\documentclass{vldb}

\usepackage{eurosym}
\usepackage{framed}
\usepackage{kpfonts}
\usepackage{xspace}
\usepackage{colortbl}
\usepackage[T1]{fontenc}
\usepackage{fancyvrb}
\usepackage[bottom]{footmisc}
\usepackage{mathtools}
\usepackage{listings}
\usepackage{enumitem}
\usepackage{accents}
\usepackage{booktabs}
\usepackage{graphicx}
\usepackage{caption}
\usepackage{subcaption}
\usepackage{fancyvrb}
\usepackage{color}
\usepackage{balance}
\usepackage[]{algorithm2e}
\usepackage[usenames,dvipsnames]{xcolor}
\usepackage[normalem]{ulem}
\usepackage{alltt} 
\usepackage[nocompress]{cite}

\newtheorem{defi}{Definition}

\newtheorem{example}[defi]{Example}

\makeatletter
\def\@copyrightspace{\relax}
\makeatother

\def\pprw{8.5in}
\def\pprh{11in}

\usepackage[bookmarks,backref=true,linkcolor=black]{hyperref} 
\hypersetup{
  pdfauthor = {},
  pdftitle = {},
  pdfsubject = {},
  pdfkeywords = {},
  colorlinks=true,
  linkcolor= black,
  citecolor= black,
  pageanchor=true,
  urlcolor = black,
  plainpages = false,
  linktocpage
}

\usepackage{cleveref} 

\definecolor{mred}{rgb}{.80,.12,.30}
\definecolor{grey}{rgb}{0.5,0.5,0.5}
\definecolor{Purple}{rgb}{.75,0,.85}
\definecolor{light-gray}{gray}{0.95}
\definecolor{mid-gray}{gray}{0.85}
\definecolor{darkred}{rgb}{0.7,0.25,0.25}
\definecolor{darkgreen}{rgb}{0.15,0.55,0.15}
\definecolor{darkblue}{rgb}{0.1,0.1,0.5}
\definecolor{blue}{rgb}{0.2,0.58,.9}

\newcommand{\red}[1]{\textcolor{red}{#1}}

\newcommand{\blue}[1]{\textcolor{blue}{#1}}

\newcommand{\stitle}[1]{\smallskip\noindent\textbf{#1}}
\newcommand{\sstitle}[1]{\noindent\textbf{#1}}

\newif\ifnotes
\notestrue

\newcommand{\eat}[1]{}

\newcommand{\sys}{\textsc{Smoke}\xspace}
\newcommand{\typej}{\textsc{Defer}\xspace}
\newcommand{\typei}{\textsc{Inject}\xspace}

\newcommand{\btxfilter}{\textsc{BT}\xspace}
\newcommand{\btftxfilter}{\textsc{BT+FT}\xspace}
\newcommand{\lazyxfilter}{\textsc{Lazy}\xspace}

\newcommand{\perm}{\textsc{Perm}\xspace}
\newcommand{\gprom}{\textsc{GProm}\xspace}

\newcommand{\sysl}{\textsc{Smoke-L}\xspace}
\newcommand{\sysd}{\textsc{Smoke-D}\xspace}
\newcommand{\sysi}{\textsc{Smoke-I}\xspace}
\newcommand{\sysic}{\textsc{Smoke-I-TC}\xspace}
\newcommand{\sysec}{\textsc{Smoke-I-EC}\xspace}
\newcommand{\sysm}{\textsc{Smoke-D-DeferForw}\xspace}

\newcommand{\sysn}{\textsc{Lazy}\xspace}
\newcommand{\lazy}{\textsc{Lazy}\xspace}

\newcommand{\lgerids}{\textsc{Logic-Rid}\xspace}
\newcommand{\lgefull}{\textsc{Logic-Tup}\xspace}
\newcommand{\lgei}{\textsc{Logic-Idx}\xspace}

\newcommand{\phvfdp}{\textsc{Phys-Mem}\xspace}
\newcommand{\phbdb}{\textsc{Phys-Bdb}\xspace}

\newcommand{\syscd}{\textsc{Smoke-CD}\xspace}
\newcommand{\metanomeuguide}{\textsc{Metanome-UG}\xspace}
\newcommand{\sysug}{\textsc{Smoke-UG}\xspace}

\newcommand{\uguide}{\textsc{UGuide}\xspace}
\newcommand{\metanome}{\textsc{Metanome}\xspace}

\newcommand{\cube}{\textbf{\textsc{Data Cube}}\xspace}
\newcommand{\immens}{\textsc{imMens}\xspace}
\newcommand{\nanocubes}{\textsc{NanoCubes}\xspace}
\newcommand{\hashedcubes}{\textsc{hashedcubes}\xspace}

\definecolor{myblack}{RGB}{20, 20, 20}
\definecolor{Gray}{gray}{0.9}

\usepackage{microtype}

\newlength{\listingindent}                
\usepackage{tikz}
\usetikzlibrary{circuits.ee.IEC}

\makeatletter
\def\@maketitle{\newpage
 \null
 \setbox\@acmtitlebox\vbox{%
\baselineskip 20pt
\vskip 2em                   
   \begin{center}
    {\ttlfnt \@title\par}       
    \vskip 1.5em                
{\subttlfnt \the\subtitletext\par}\vskip 1.25em
    {\baselineskip 16pt\aufnt   
     \lineskip .5em             
     \begin{tabular}[t]{c}\@author
     \end{tabular}\par}
    \vskip 2.5em               
   \end{center}}
 \dimen0=\ht\@acmtitlebox
 \unvbox\@acmtitlebox
 \ifdim\dimen0<0.0pt\relax\vskip-\dimen0\fi}
\makeatother

\newcommand\Ground{%
\mathbin{\text{\begin{tikzpicture}[circuit ee IEC,yscale=0.4,xscale=0.4]
\draw (0,2ex) to (0,0) node[ground,rotate=-90,xshift=.65ex] {};
\end{tikzpicture}}}%
}

\def\techr{1}
\def\subm{1}

\begin{document}

\sloppy 

\color{myblack}

\if\techr1
\title{\textsc{\huge Smoke}: Fine-grained Lineage at Interactive Speed\titlenote{A version of this paper has been accepted to VLDB 2018 (camera ready pending). This document is its associated technical report and has not been peer-reviewed in its entirety.}\vspace{-1.2em}}
\else
  \if\subm1
     \title{\textsc{\huge Smoke}: Fine-grained Lineage at Interactive Speed}
  \else
      \title{\textsc{\huge Smoke}: Fine-grained Lineage at Interactive Speed [Cover~ Letter]}
  \fi
\fi

\numberofauthors{2}
\author{
\alignauthor
Fotis Psallidas\\
		\affaddr{Computer Science Department}\\
       	\affaddr{Columbia University}\\
       	\email{fotis@cs.columbia.edu}
\alignauthor
Eugene Wu\\
       \affaddr{Computer Science Department}\\
       \affaddr{Columbia University}\\
       \email{ewu@cs.columbia.edu}
}

\maketitle

\begin{abstract}
Data lineage describes the relationship between individual input and output data items of a workflow, and has served as an integral ingredient for both traditional (e.g., debugging, auditing, data integration, and security) and emergent (e.g., interactive visualizations, iterative analytics, explanations, and cleaning) applications. The core, long-standing problem that lineage systems need to address---and the main focus of this paper---is to capture the relationships between input and output data items across a workflow with the goal to streamline queries over lineage. Unfortunately, current lineage systems either incur high lineage capture overheads, or lineage query processing costs, or both. As a result, applications, that in principle can express their logic declaratively in lineage terms, resort to hand-tuned implementations.  To this end, we introduce \sys, an in-memory database engine that neither lineage capture overhead nor lineage query processing needs to be compromised. To do so, \sys introduces tight integration of the lineage capture logic into physical database operators; efficient, write-optimized lineage representations for storage; and optimizations when future lineage queries are known up-front. Our experiments on microbenchmarks and realistic workloads show that \sys reduces the lineage capture overhead and streamlines lineage queries by multiple orders of magnitude compared to state-of-the-art alternatives. Our experiments on real-world applications highlight that \sys can meet the latency requirements of interactive visualizations  (e.g., <150ms) and outperform hand-written implementations of data profiling primitives. \eat{As such, \sys does not only enable declarative lineage constructs for applications, but also contributes to query optimization.}
\end{abstract}

\setcounter{page}{1}
\setcounter{section}{0}
\section{Introduction}
\label{s:intro}

Data lineage describes the relationship between individual input and output data items of a computation.  For instance, given an erroneous result record of a workflow, it is helpful to retrieve the intermediate or base records to investigate for errors; similarly, identifying output records that were affected by corrupted input records can help prevent erroneous conclusions.  These operations are expressed as lineage queries over the workflow: backward queries return the subset of input records that contributed to a given subset of output records; forward queries return the subset of output records that depend on a given subset of input records. 

Virtually, any application that requires an understanding over the input-output derivation process can be expressed in lineage terms. As such, data lineage has been an integral ingredient for applications such as debugging~\cite{wu2013subzero,karvounarakis:2010:proql,interlandi2015sparkprovenance,logothetis2013scalable,dbnotes2005chiticariu},  data integration~\cite{cui2000linagetrace}, auditing~\cite{eugdpr}, security~\cite{chen2017data,karvounarakis:2010:proql}, explaining query results~\cite{scorpion,wu2012demonstration,roy2015explain,pnl:2017:deutch}, data cleaning~\cite{chalamalla2014descriptive,haas2015wisteria}, iterative analytics~\cite{chothia:2016:explaining}, and interactive visualizations~\cite{wu2017dvms} that highlight the importance of lineage-enabled systems.

Lineage-enabled systems answer lineage queries by automatically capturing record-level relationships throughout a workflow.  A naive approach materializes pointers between input and output records for each operator during workflow execution, and follows these pointers to answer lineage queries.  Existing systems primarily differ based on when the relationships are materialized (e.g., {\it eagerly} during workflow execution or {\it lazily} reconstructed when executing a lineage query), and how they are represented (e.g., tuple annotations~\cite{agrawal2006trio,dbnotes2004bhagwat,glavic:2009:perm,ramp} or explicit pointers~\cite{wu2013subzero,logothetis2013scalable}).  Each design trades off between the time and storage overhead to capture lineage and lineage query performance.  For instance, a query execution engine may augment each operator to materialize a hash index that looks up input records for a given output record, thus speeding up backward lineage query execution.  However, the overhead of constructing the index can dwarf the operator execution cost by $100\times$ or more~\cite{wu2013subzero}---particularly if the operator is heavily optimized for latency or throughput.

\begin{figure}
  \centering
  \includegraphics[width=.68\columnwidth]{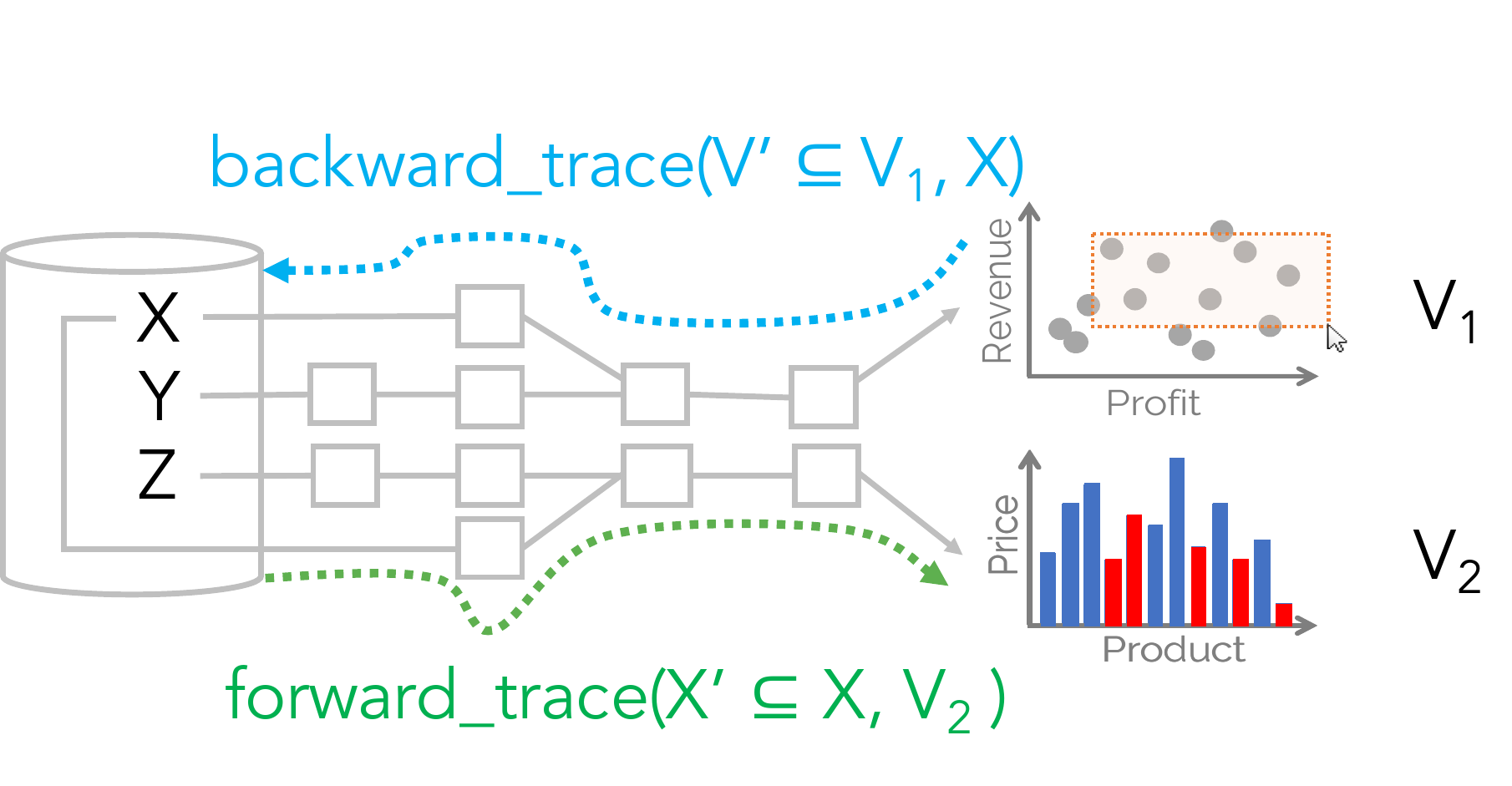} 
  \caption{\small Two workflows generate visualization views $V_1$ and $V_2$.  Then, a linked brushing interaction highlights in \red{red} marks in $V_2$ that share the same input records with selected marks of $V_1$.  This interaction can be expressed as a backward query from selected circles in $V_1$ followed by a forward query to  highlight the bars in $V_2$.  }
  \vspace*{-.2in}
  \label{f:vizworkflow}
\end{figure}

As data processing engines become faster, an important question---and the main focus of this paper---is whether it is possible to achieve the best of both worlds: negligible lineage capture overhead {\it as well as} fast lineage query execution. 

Unfortunately, current lineage systems incur either high lineage capture overhead, or high lineage query processing costs, or both. As a result, applications that could be expressed in lineage terms resort to manual implementations: 

\vspace*{-10pt}
\begin{example}
Figure~\ref{f:vizworkflow} shows two views $V_1$ and $V_2$ generated from queries over a database.  Linked brushing is an interaction technique where users select a set of marks (e.g., circles) in one view, and marks derived from the same records are highlighted in the other views.  Although this functionality is typically implemented manually, it can be logically expressed as a backward lineage query from selected points in $V_1$ to input records followed by a forward query to highlight the corresponding bars in $V_2$.
\end{example}

To avoid the shortcomings of current lineage systems and tackle the competing requirements of lineage applications, we employ a careful combination of four design principles:

\stitle{P1. Tight integration.} In high throughput query processing systems, per-tuple overheads incurred within a tight loop---even a single virtual function call to write lineage metadata to a separate lineage subsystem~\cite{wu2013subzero,interlandi2015sparkprovenance,logothetis2013scalable}---can slow down operator execution by more than an order of magnitude. In response, we introduce a physical algebra that tightly integrates lineage capture into query execution, and we design simple, write-efficient data structures for lineage capture to avoid the overhead of crossing system boundaries.

\stitle{P2. Apriori knowledge.} Lineage applications such as debugging need to capture lineage to answer ad-hoc lineage queries that can trace back to any base or intermediate table. In applications such as interactive visualizations or data profiling we typically know the possible set of lineage queries up front. Having apriori knowledge enables us to avoid materializing lineage that does not contribute to these lineage queries.

\stitle{P3. Lineage consumption.} Lineage applications rarely require all results of a lineage query (e.g., all records that contributed to an aggregation result), unless the results have low cardinality.  Instead, the results are filtered, transformed, and aggregated by additional SQL queries. We term these queries {\it lineage consuming queries}. If such queries are known up-front, as is typically the case for applications based on templated analysis (e.g., Tableau or Power BI), physical design logic based on these templates can be pushed into the lineage capture phase. Such physical design logic may include materialization of aggregate statistics or prune/re-partition lineage indexes to speed up future lineage consuming queries.

\stitle{P4. Reuse.} Finally, we have found that significant lineage capture costs arise from generating and storing unnecessary amounts of lineage data such as expensive annotations and denormalized forms of lineage. Following the concept of reusing data structures~\cite{dursun2016revisiting}, we identify cases where data structures constructed during normal operator execution can be augmented and reused, but this time with the goal set to capture lineage with low overhead.

\smallskip
This paper presents \sys, a lineage-enabled system that embodies the above four principles to support lineage capture and querying with low latency. More specifically, \sys is an in-memory query compilation database engine that tightly integrates the lineage capture logic within query execution and uses simple, write-efficient lineage indexes for low-overhead lineage capture (\textbf{P1}).  In addition, \sys enables workload-aware optimizations that prune captured lineage and push the logic of lineage consuming queries down into the lineage capture phase (\textbf{P2,P3}). Finally, \sys identifies data structures, which are constructed during normal operator execution (i.e., hash tables), and (re)uses them, whenever possible, for low-overhead lineage capture (\textbf{P4}).

In the rest of the paper, we start by discussing necessary background and related work (\Cref{s:bg}). Then, we present our techniques and contributions as follows:
\begin{itemize}[leftmargin=*]  
\item We introduce a physical algebra that tightly integrates the lineage capture logic within the processing of single and multi-operator plans. Each logical operator has a dual form to both execute its logic and generate lineage. In turn, physical
operators are combined to implement this extended logic. Furthermore, we design write-efficient data structures to materialize lineage with low overhead.(\Cref{s:instr})

\item We design a suite of simple optimizations based on the availability of future lineage consuming queries to 1) prune lineage that will not be used and 2) materialize aggregates and prune or re-partition our lineage data structures to
answer lineage consuming queries faster. (\Cref{s:optf})

\item We conduct experiments to 1) compare Smoke with state-of-the-art lineage systems and 2) show how it can enable real-world applications (i.e., interactive visualizations and data profiling). The former experiments show that Smoke reduces the lineage capture overhead and streamlines lineage queries by multiple orders of magnitude compared to state-of-the-art lineage systems. The latter suggest that \sys lets applications express their logic using declarative lineage constructs {\it and also} speeds up these applications---to the extent that \sys is  on par with or outperforms hand-written implementations. (Sections \ref{s:settings} and \ref{s:results})
\end{itemize}
\vspace{-0.5em}
\noindent We conclude with open questions for future work (\Cref{s:future}).

\section{Background}
\label{s:bg}

In this section, we provide background on the fine-grained lineage capture problem, the approach of \sys for this problem, and applications of focus in this paper. 

\subsection{Fine-Grained Lineage Capture}
\label{s:bglin}
Our lineage semantics adhere to the transformational provenance semantics of~\cite{chothia:2016:explaining,glavic:2009:perm,ikedathesis} over relational queries.

\sstitle{Base queries.} Formally, let the {\it base query} $Q_{\Ground}(D) = O$ be a relational query over a database of relations $D = \{R_1,\cdots, R_n\}$ that generates an output relation $O$.  An application can initially execute multiple base queries $\mathbb{Q}_{\Ground} = \{Q_{{\Ground}1},\cdots, Q_{{\Ground}m}\}$.  For instance, $\mathbb{Q}_{\Ground}$ in Figure~\ref{f:vizworkflow} consists of two queries that generate two output relations rendered as visualization views. 

\sstitle{Lineage and lineage consuming queries.} After a base query runs, the user may issue a backward lineage query $L_b(O',R_i)$ that traces from a subset of an output relation $O' \subseteq O$ to a base table $R_i$, or a forward lineage query $L_f(R_i',O)$ that traces from a subset of an input relation $R' \subseteq R_i$ to the query's output relation $O$.   A lineage query $L(\bullet)$ results in a relation that can be used is another query $C(D \cup \{L(\bullet)\})$ which we term a {\it lineage consuming query}; a lineage query is a special case of lineage consuming queries: $C = \texttt{SELECT * FROM L(}\bullet\texttt{)}$. Finally, $C$ itself can be used as a base query, meaning that another lineage consuming query $C'$ can use $C$ as a base query.\footnote{\sys's query model includes multi-backward and multi-forward queries as well as refresh and forward propagation~\cite{ikedathesis}. We limit the discussion to $L_b$, $L_f$, and $C$ in that they form the basis to express general query constructs.} 

\begin{example}
  Let $Q_{{\Ground}1}(\{X, Y\}) = V_1$ and $Q_{{\Ground}2}(\{X, Z\}) = V_2$ be the base queries in Figure~\ref{f:vizworkflow}.  The linked brushing interaction is expressed as a backward query $L_b(V'_1, X)$ from the selected circles $V'_1 \subseteq V_1$ back to the input records in $X$ that generated them. The forward lineage query $F = L_f(L_b(V'_1, X), V_2)$ retrieves the linked bars in $V_2$. A lineage consuming query $C(D \cup F)$ can then be used to change the color of the bars to \red{red}, similarly to the ones in~\cite{wu2017dvms}.
\end{example}
\noindent We can model interactive visualization applications as base queries (e.g., $Q_{{\Ground}1}$, $Q_{{\Ground}2}$ above) that load the initial visualization, followed by lineage consuming queries $W$ that express user interactions~\cite{wu2017dvms}. Therefore, optimizing the visualization responsiveness corresponds to quickly executing the base queries followed by streamlining lineage consuming queries.

\sstitle{Lazy and Eager lineage query evaluation.} 
How can we answer lineage queries quickly?  {\it Lazy} approaches rewrite lineage queries as relational queries over the input relations---the base queries do not incur overhead at the cost of potentially slower lineage query execution costs~\cite{ikedathesis,cui2000linagetrace,provdb}.  In contrast, we might {\it Eagerly} materialize data structures during base query execution to speed up future lineage queries~\cite{provdb,ikedathesis}.  We refer to this as lineage capture, and we seek to minimize capture overhead on the base query to speed up lineage queries.

\sstitle{Lineage capture overview.} The eager approach incurs overhead to capture the base query's {\it lineage graph}.  Logically, each edge $a\xleftrightarrow{op} b$ maps an operator $op$'s input record $a$ to $op$'s output record $b$ that is derived from $a$. Backward lineage connects tuples in the query output $o\in O$ with tuples in each input base relation $r\in R_i$ by identifying all end-to-end edges $o\leadsto r$ for which a path exists between the two records. Forward lineage reverses these arrows.  Materializing such end-to-end forward and backward {\it lineage indexes} can help speed up lineage consuming queries.    

We will present techniques that can efficiently capture lineage indexes in a {\it workload-agnostic} setting by carefully instrumenting operator implementations, and in a {\it workload-aware} setting by tailoring the indexes for future lineage consuming queries if they are known up-front.  In general, lineage capture techniques fall into two categories: logical and physical.

\sstitle{Logical lineage capture.} 
This class of approaches stay within the relational model by rewriting the base query into $Q'_{\Ground}(\{R_1,\cdots,R_n\}) = O'$, so that its output is annotated with additional attributes of input tuples.  Some systems~\cite{agrawal2006trio,dbnotes2005chiticariu} generate a normalized representation of the lineage graph such that a join query between $O'$ and each base relation $R_i$ can create the lineage edges between $O'$ and $R_i$. The correct output relation $O$ can be retrieved by projecting away the annotation attributes from $O'$.  Alternative approaches~\cite{glavic:2009:perm,dbnotes2005chiticariu} output a single denormalized representation that extends $O'$ with attributes of the input relations.  Recent work has shown that the latter rewrite rules (\perm~\cite{glavic:2009:perm}) and optimizations leveraging the database optimizer (\gprom~\cite{gprom}) incurs lower capture overheads than the former normalized approach.

Although these approaches can run on any relational database and benefit from the database optimizer, they suffer from several performance drawbacks.  The normalized representation requires expensive independent joins when running lineage queries. The denormalized representation can incur significant data duplication---an aggregation output $o$ computed over $k$ input records will be duplicated $k\times$---and require further projections to derive $O$ from $O'$.  Furthermore, indexes are needed to speed up lineage queries.

\sstitle{Physical lineage capture.} This approach directly instruments physical operators write lineage edges to a lineage subsystem through an API; the subsystem stores and indexes the edges and answers lineage queries~\cite{logothetis2013scalable,interlandi2015sparkprovenance,wu2013subzero,ramp,panda}.  This can support black-box operators and decouples lineage capture from its physical representation. However, we find that virtual function calls alone (ignoring the cross-process overheads) can slow data-intensive operators by up to $2\times$.  Further, lineage capture cannot easily leverage and be co-optimized with base query execution.

\begin{figure}[t]
\centering
\includegraphics[width=.9\columnwidth]{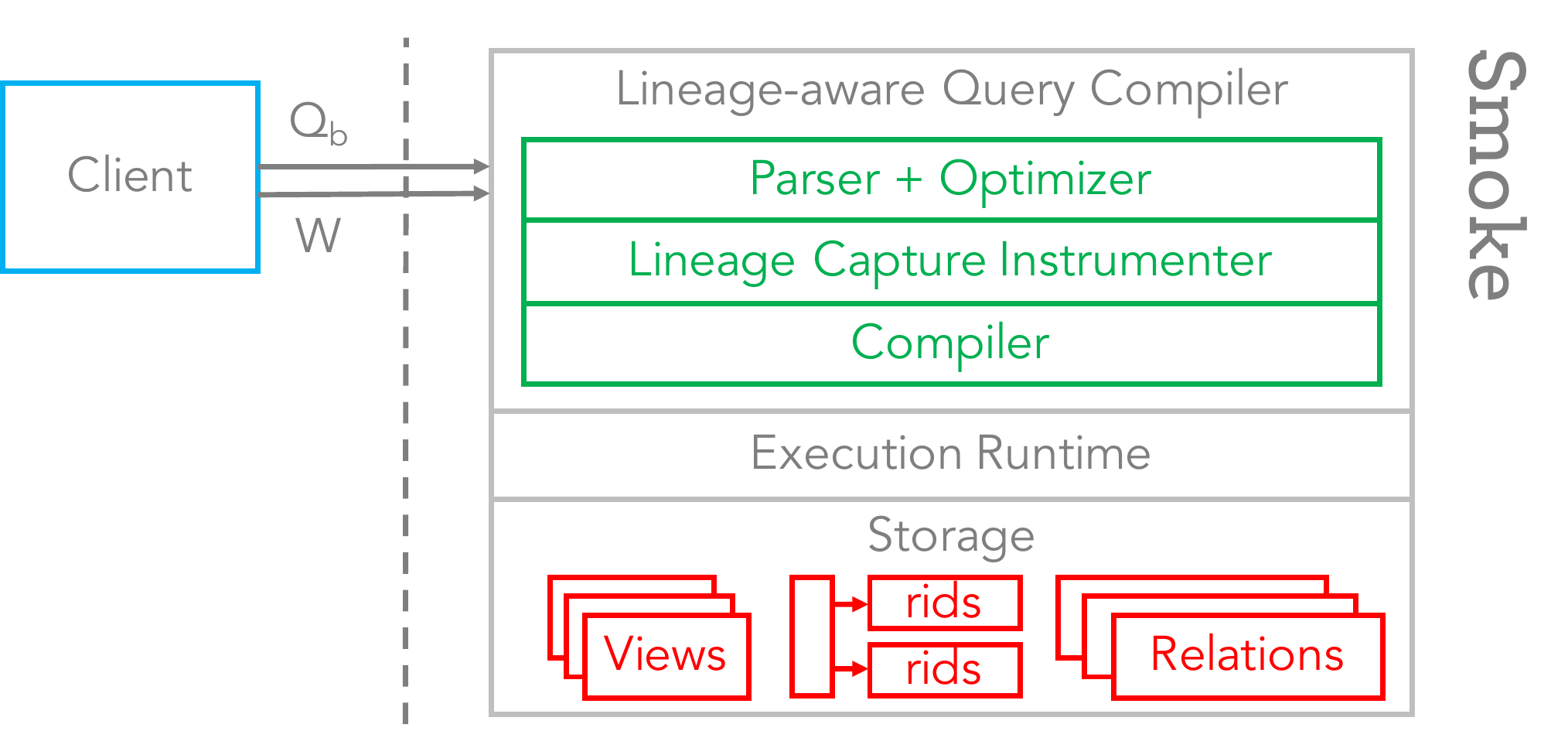}
\caption{\small \sys is a query compilation engine that instruments physical query plans to capture lineage efficiently:  Query execution generates lineage indexes that map input and output record ids (rids) as well as materialized views.  }
\label{fig:sys}
\vspace{-1em}
\end{figure}

\subsection{Approach of Smoke}
\label{s:arch}
To this end, we introduce \sys, an in-memory database engine that avoids the drawbacks of logical and physical approaches. \sys improves upon logical approaches by physically representing the lineage edges as read- and write-efficient indexes instead of relationally-encoded annotations.   We improve upon physical approaches by introducing a physical algebra that tightly integrates lineage capture and relational operator logic to avoid API calls and in a way amenable to co-optimization. Finally, \sys can exploit knowledge of lineage consuming queries to (a) prune or partition lineage indexes or (b)  materialize views {\it during the base query execution} that will benefit the lineage consuming queries.

The primary focus of this work is to explore mechanisms to instrument physical operator plans with lineage capture logic.  To do so, we have implemented \sys as a query compilation execution engine using the produce-consumer model~\cite{neumann2014compiling} (\Cref{fig:sys}). It takes as input the base query $Q_{\Ground}$ and an optional workload of lineage consuming queries $W$; parses and optimizes $Q_{\Ground}$ to generate a physical query plan; instruments the plan to directly generate indexes to speed up backward and forward lineage queries; and compiles the instrumented plan into machine code that, when executed, generates $Q_{\Ground}(D)$ as well as lineage indexes.   Internally, \sys uses a single-threaded, row-oriented execution model, and leverages hash-based operator implementations that are widely used in fast query engines and are amenable to low-overhead lineage capture. Execution models that use advanced features such as compression or vectorization are interesting future work.

\subsection{Lineage Applications}
\label{s:lapps}

Many applications logically rely on lineage (and generally provenance), including but not limited to:
debugging~\cite{wu2013subzero,karvounarakis:2010:proql,interlandi2015sparkprovenance,logothetis2013scalable,dbnotes2005chiticariu}, diagnostics~\cite{uguide:2017:thirumuruganathan}, data integration~\cite{cui2000linagetrace}, security~\cite{chen2017data,karvounarakis:2010:proql}, auditing~\cite{eugdpr} (the recent EU GDP regulation~\cite{eugdpr} mandates tracking lineage), data cleaning~\cite{chalamalla2014descriptive,haas2015wisteria}, explaining query results~\cite{scorpion,wu2012demonstration,roy2015explain,pnl:2017:deutch}, debugging machine learning pipelines~\cite{zhang2017diagnosing,kumar2017data}, and interactive visualizations~\cite{wu2017dvms}.

Unfortunately, there is a disconnect between modeling applications in terms of lineage, and the performance of existing lineage capture mechanisms---the overhead is enough that applications resort to manual implementations instead.  For this reason,  we center the paper around interactive visualizations:  it is a domain that can directly translate to lineage~\cite{heer:2008:gs,wu2017dvms}, yet is dominated by hand-written implementations.  Furthermore, it imposes strict latency requirements on lineage capture (to show the initial visualization) and lineage consuming queries (to respond to user interactions). Finally, our experiments seek to argue, using visualization and data profiling applications, that lineage is not only an elegant logical description of many use cases, but can be on a par with or even {\it improve} on performance compared to hand-tuned implementations.

\section{Fast Lineage Capture} 
\label{s:instr}

This section describes lineage capture without knowledge of the future workload. We present (a) lineage index representations to map output-to-input or input-to-output record ids ({\it rids}) that are read- and write-efficient, and (b) a physical algebra that tightly integrates the lineage capture logic with the base query execution.  To do so, we will describe how to instrument individual as well as multiple operators to capture lineage with low overhead.

\subsection{Lineage Index Representations} 
\label{s:storage}

\begin{figure}
\centering
\includegraphics[height=1in]{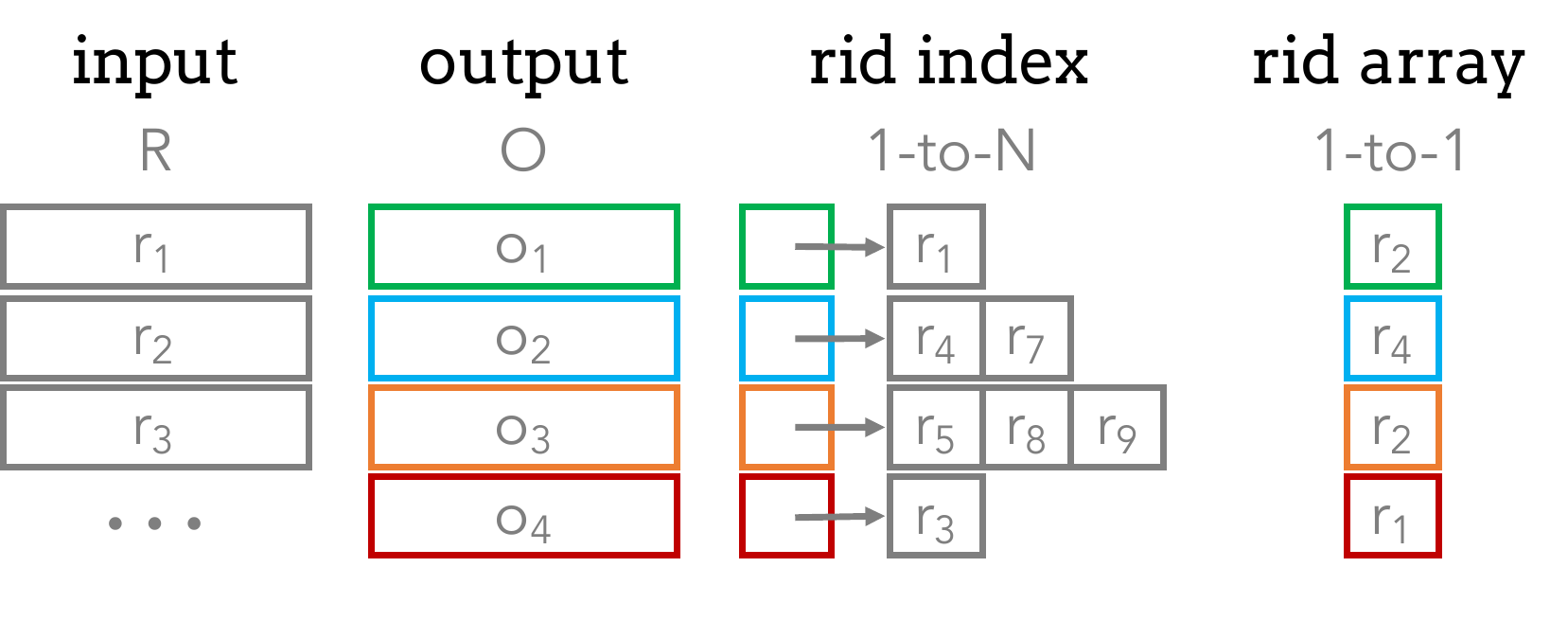}
\caption{\small Lineage index representations: rid index for 1-to-N operators (e.g., GROUPBY); rid array for 1-to-1 operators (e.g., SELECT).  }
\label{fig:ridindexes}
\vspace{-1em}
\end{figure}

\sys uses two main lineage index representations. Figure~\ref{fig:ridindexes} illustrates the input and output relations $R$ and $O$, respectively, and the two $rid$-based representations for 1-to-N and 1-to-1 operators.  We index rids because the lineage indexes are cheap to write, and lookups---which simply index into the relation's array---are fast.   In contrast, indexing full tuples incurs high write costs, while indexing primary keys is not beneficial without building primary key indexes apriori or when the primary keys are wide.  Furthermore, in-memory columnar engines~\cite{abadi2013design,mmdbs} already create rid lists as part of query processing that resemble our lineage indexes, and enable reuse opportunities to reduce lineage capture costs.

\sstitle{Rid Index.} 1-to-N relationships  are represented as inverted indexes. Consider the backward lineage of GROUPBY. The index's $i^{th}$ entry corresponds to the $i^{th}$ output group, and points to an rid array containing rids of the input records that belong to the group.  The Rid Index is used for 1-to-N forward lineage relationships as well, such as the JOIN operator.   Following high performance libraries~\cite{folly}, the index and rid arrays are initialized to 10 elements and grow by a factor of $1.5\times$ on overflow.  Our experiments show that array resizing dominates lineage capture costs, and statistics that allow \sys to pre-allocate appropriate sized arrays can reduce lineage capture costs by up to $60\%$.   To avoid offline statistics computation, we show how useful statistics can be collected during query processing.

\sstitle{Rid Array. } 1-to-1 relationships between output and input are represented as a single array. Each entry is an input record rid rather than a pointer to an rid array.

\subsection{Single Operator Instrumentation}
\label{ss:instr}
We now introduce instrumentation techniques to generate lineage indexes when executing single and multi-operator plans. Our designs are based on two paradigms: \typej defers portions of the lineage capture until after operator execution while \typei incurs the full cost during execution. \typej is preferable when the base query execution overhead {\it must} be minimized, or when it is possible to collect cardinality statistics during base query execution to allocate appropriately sized lineage indexes and avoid resizing costs. In contrast, \typei typically incurs lower overall overhead, but the client needs to wait longer to retrieve the results of the base query.

We now describe how both paradigms illustrate the application of the tight integration and reuse principles from the Introduction for core relational operators.  Our focus is on the mechanisms and~\Cref{s:future} discusses future work on automatically choosing between these paradigms.  Code snippets of the compiled code and details for additional operators (including $\cup$, $\cap$, $-$, $/$, $\times$, $\bowtie_\theta$) are covered in our technical report~\cite{psallidas2017extendedlineage}. \Cref{ss:instr_multi} extends our support to multi-operator plans.

\begin{figure}[tb]
  \centering
  \includegraphics[width=.9\columnwidth]{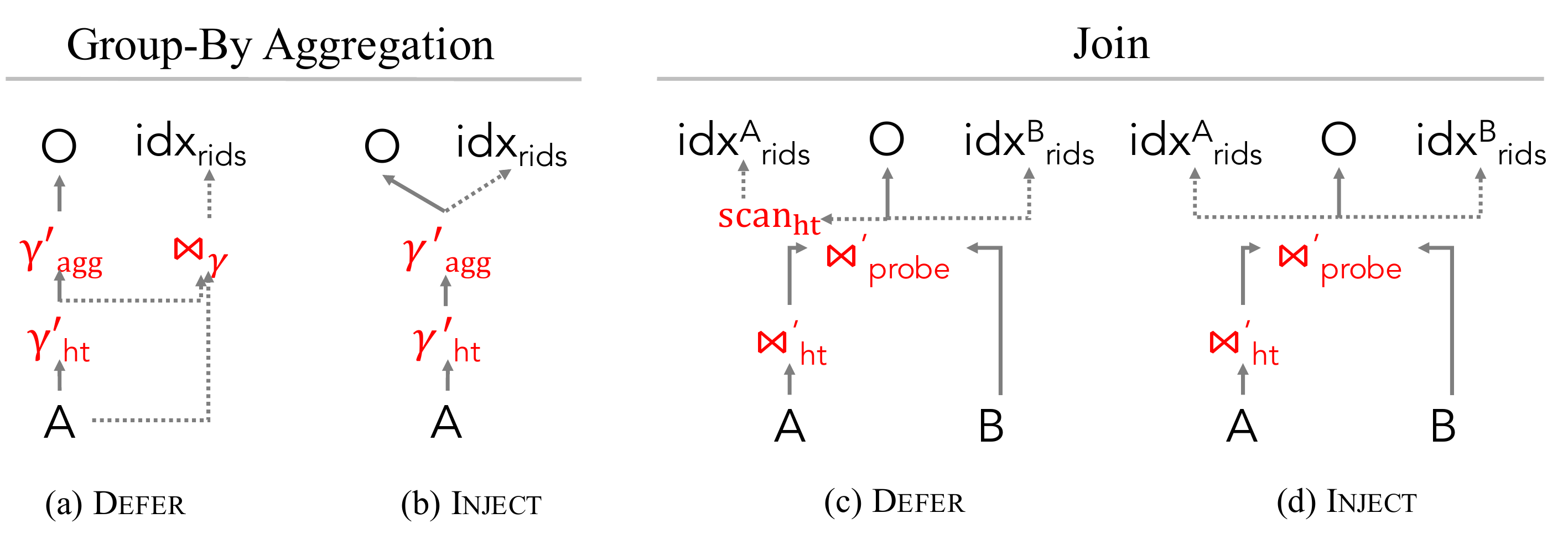}
  \caption{{\small\typei and \typej plans for group-by aggregation and join. Dotted arrows are only necessary for lineage capture.}}
  \label{fig:smoke_inst}
\vspace{-1em}
\end{figure}

\subsubsection{Projection}
Projection under bag semantics does not need lineage capture because the input and output order and cardinalities are identical---the rid of an output (input) record is its backward (forward) lineage. Projection with set semantics is implemented using grouping, and we use the same mechanism as that for group-by aggregation below.

\subsubsection{Selection}
Selection is an \texttt{if} condition in a \texttt{for} loop over the input relation, and emits a record if the predicate evaluates to true~\cite{neumann2011efficiently}. Both forward and backward lineage use rid arrays; the forward rid array can be pre-allocated based on the cardinality of the input relation.  \typei adds two counters, $ctr_i$ and $ctr_o$, to track the rids of the current input and output records, respectively.  If a record is emitted, we set the $ctr_i^{th}$ element of the forward rid array to $ctr_o$, and append $ctr_i$ to the backward rid array.  Available selectivity estimates can be used to pre-allocate the backward rid array and avoid reallocation costs during the append operation.  We don't implement \typej because it is strictly inferior to \typei.

\subsubsection{Group-By Aggregation}
Query compilers decompose \texttt{GROUPBY} into two physical operators: $\gamma_{ht}$ builds the hash table that maps group-by values to the group's intermediate aggregation state; $\gamma_{agg}$ scans the hash table, finalizes aggregation results for each group, and emits output records.  Figure~\ref{fig:smoke_inst} shows the plans for both instrumentation paradigms; the lineage indexes consist of a forward rid array and a backward rid index. 

\sstitle{\typej: } Consider the \typej plan in~\Cref{fig:smoke_inst}.a. $\gamma'_{ht}$ for \typej extends $\gamma_{ht}$ to store an $oid$ number to each group's intermediate aggregation state.  When $\gamma'_{agg}$ scans the hash table to construct the output records, it uses a counter to track the output record's rid and assign it to the group's $oid$ value (i.e., $oid$ tracks the output rid of the group in the result).  \sys then pins the hash table in memory.   At a later time, $\Join_\gamma$ can scan each record in $A$, reuse the hash table to probe and retrieve the associated group's $oid$, and populate the backward rid index and forward rid array.  

Although \typej must scan $A$ twice, the operator's input and output cardinalities are used to avoid resizing costs during $\Join_\gamma$. Also, $\Join_\gamma$ can be freely scheduled (e.g., immediately after $\gamma'_{ht}$ or during user think time when system resources are free).

\sstitle{\typei: } Consider the \typei plan in~\Cref{fig:smoke_inst}.b.  $\gamma'_{ht}$ this time augments each group's intermediate state with an rid array $i\_rids$ that contains the rids of the group's input records (i.e., backward lineage).   $\gamma'_{agg}$ tracks the current output record id $oid$ in order to set the pointer in the backward index to the bucket's rid list and the values in the forward rid list.   Since $\gamma'_{agg}$ knows the input and output cardinalities, it can correctly allocate arrays for the backward and forward indexes. 
The primary overhead is due to reallocations of $i\_rids$ during the build phase in $\gamma'_{ht}$. We find that knowing group cardinalities can decrease the lineage capture overhead up to 60\%.

\subsubsection{Join}  
\label{ss:join}

\sys instruments hash joins in a similar way as hash aggregation. A hash join is split into two physical operators: $\Join_{ht}$ builds the hash table on the left relation $A$, and $\Join_{probe}$ uses each record of the right relation $B$ to probe the hash table. We now introduce \typei and \typej techniques for lineage capture that can be used for general M:N joins, and further optimizations for primary key-foreign key (pk-fk) joins. \sys generates backward rid arrays and forward rid indexes (an input record can generate multiple join results).

\sstitle{\typei: } Consider the \typei plan for joins in~\Cref{fig:smoke_inst}.d. The build phase $\Join'_{ht}$ augments each hash table entry with an rid array $i\_rids$ that contains the input rids from $A$ for that entry's join key.  The probe phase $\Join'_{probe}$ tracks the $rid$ for each output record, and populates the forward and backward indexes as expected.  Note that output cardinalities are not yet known within the $\Join'_{probe}$ phase and we cannot pre-allocate our lineage indexes. As a result, although the backward rid array is relatively cheap to resize, the forward rid indexes can potentially trigger multiple reallocations if an input record has many matches and penalize the performance.

\sstitle{\typej}: Our main observation is that exact cardinalities needed to pre-allocate the forward rid indexes are known {\it after} the probe phase and can be used by \typej. Note that deferring the whole construction after the probe phase is similar to the logical approaches, and incurs the cost of re-running the join, or annotating, indexing and projecting the join output.   \typej instead augments \typei by partially deferring index construction for the left input relation $A$  (see~\Cref{fig:smoke_inst}.c).

The build phase adds a second rid list $o\_rids$ to the hash table entry, in addition to $i\_rids$ from \typei. When $B$ is scanned during the probe phase, its output records are emitted contiguously, thus $o\_rids$ need only store the rid of the first output record for each match with a $B$ record.  After the $\Join'_{\text{probe}}$ phase, the forward and backward indexes for the left relation $A$ can then be pre-allocated and populated in a final scan of the hash table ($\text{scan}_{\text{ht}}$ in~\Cref{fig:smoke_inst}.c). Deferring for $B$ is also possible, however the benefits are minimal because we need to partition the output records for each hash table entry by the $B$ records that it matches, which we found to be costly.

\sstitle{Further optimizations. } If the hash table is constructed on a unique key, then the $i\_rids$ do not need to be arrays and can be replaced with a single integer.  Also, if the join is a primary-key foreign-key join, the forward index of the foreign-key table is an rid array; since the join cardinality is the same as the foreign-key table cardinality, backward indexes are pre-allocated. Finally, join selectivity estimates can help pre-allocate the forward rid indexes.

\pagebreak

\subsection{Multi-Operator Instrumentation}
\label{ss:instr_multi}
The na\"ive way to support multi-operator plans is to individually instrument each operator to generate its lineage indexes; lineage queries can use the indexes to trace backward or forward through the plan.  This approach is correct and can be used to support any DAG workflow composed of our physical operators.  However, it unnecessarily materializes all intermediate lineage indexes even though only the lineage between output and input records are strictly needed.  

We address this issue with a technique that 1) propagates lineage information throughout plan execution so that only a single set of lineage indexes connecting input and final output relations are emitted, and 2) reduces the number of lineage index materialization points in the query plan.

To propagate lineage throughout plan execution, consider a two-operator plan $op_p(op_c(R))=O$  with input relation $R$. When $op_p$ runs, it will use $op_c$'s backward lineage index to populate its own lineage index with rids that point to $R$ rather than the intermediate relation $op_c(R)$;  $op_c$'s lineage indexes can be garbage collected when not needed further. 

To reduce lineage index materialization points, recall that database engines pipeline operators to reduce intermediate results by merging multiple operators into a single pipeline~\cite{neumann2011efficiently}.  Operators such as building hash tables are {\it pipeline breakers} because the input needs to be fully read before the parent operator can run. Within a pipeline there is no need for lineage capture, but pipeline breakers need to generate lineage along with the intermediate result. In~\Cref{ss:instr}, we showed how pipeline breakers (e.g., hash table construction for the left-side of joins and group-by aggregations) can augment the hash tables with lineage.  Parent pipelines that use the same hash-tables for query evaluation (e.g., cascading joins) can also use the lineage indexes embedded in the hash tables to implement the lineage propagation above.

\sstitle{Implementation Details}  Our engine supports naive instrumentation for arbitrary relational DAG workflows, and we focused our optimizations for SPJA query blocks composed of pk-fk joins.  This was to simplify our engineering, and because fast capture for SPJA blocks can be extended to nested blocks by using the propagation technique above.  We focus on pk-fk joins due to their prevalence in benchmarks and real-world applications, and because the \typei and \typej instrumentation for pk-fk joins are identical (\Cref{ss:instr}).  Thus, the main distinction between \typei and \typej for SPJA blocks is how the final aggregation operator in the block is instrumented---the joins are instrumented identically, while select and project are pipelined.   Details are in our technical report~\cite{psallidas2017extendedlineage}.

\section{\mbox{Workload-Aware Optimizations}}
\label{s:optf}

Lineage applications such as interactive visualizations will often support a pre-defined set of interactions (e.g., filter, pan, tooltip, cross-filter~\cite{crossfilter}) that amount to a pre-declared lineage consuming query workload $W$.   This section describes simple but effective optimizations that exploit knowledge of $W$ to avoid capturing lineage that is not queried, and generate lineage representations that directly speed up queries in $W$.  To simplify the discussion, we will center each optimization around different classes of lineage consuming queries over the base query $Q_{\Ground} = \sigma_{\textit{o\_orderdate}>\text{`2017-08-01'}} (\textit{orders}\bowtie \textit{lineitem})$. 

\subsection{Instrumentation Pruning}
\label{ss:optsf::prune}
Instrumentation pruning disables lineage capture if the lineage indexes will not be used by $W$.  We present two types of pruning that do not generate lineage for specific input relations, and for backward/forward lineage. 

\sstitle{Pruning input relations.} Suppose the visualization only supports a tooltip interaction that shows detailed lineitem information when the user hovers over a visualization mark. This is expressed as a backward lineage query to \texttt{lineitem}.  In this case, we can avoid capturing lineage for the \texttt{orders} table.  In general, \sys does not capture lineage for any relation not referenced in $W$.

\sstitle{Pruning lineage direction.} Extending the previous example, it is clear that $W$ will only execute a backward lineage query to \texttt{lineitem} and not vice versa.  Thus, \sys can also avoid generating the forward lineage index from \texttt{lineitem} to the base query output.  The lineage indexes that can be pruned is evident from the lineage consuming queries in $W$.

\subsection{Push-Down Optimizations}
\label{ss:optsf::push}
User facing applications rarely present a large set of query results to the users---instead they will {\it reduce} the result cardinality with further filter, transform, and/or aggregation operations. This is also the case for lineage consuming queries, and presents opportunities to push these reduction operations into the lineage capture logic.  We present three simple push-down optimizations for fixed filter predicates, templated predicates, and aggregation operations, and then discuss the relationship between push-down optimizations and common provenance semantics.

\sstitle{Selection push-down.} Visualizations often update metrics that summarize data that the user selects.  For instance, the following query retrieves Christmas shipment order information for parts of the visualization that the user interacts with:  $C=\sigma_{shipdate=`xmas'}(L_B(O' \subseteq Q_{\Ground}(D), \textrm{orders}))$.   This optimization pushes the predicate \texttt{shipdate=`xmas'} into the instrumented base query; before \sys populates the backward lineage indexes, it checks whether the input tuple satisfies the predicate.  If the predicate is on a GROUPBY key, \sys can also avoid any lineage capture overhead for all other groups.  This technique reduces lineage space requirements, and typically reduces capture overhead.  However, if the predicate is expensive to evaluate (e.g., slow UDF), it is possible to introduce more capture overhead.

\sstitle{Data skipping using lineage.} Push down selections require a static predicate, however interactive visualizations also use parameterized predicates. For instance, the user may use a slider to dynamically specify the filtered shipping date:
$C=\sigma_{\textit{shipdate}=:p1}(L_B(O' \subseteq Q_{\Ground}(D), \textrm{orders}))$.  This pattern is ubiquitous in interactive visualizations and applies to faceted search, cross-filtering, zooming, or panning.  \sys pushes the parameterized predicate into lineage capture by partitioning the rid arrays (standalone and part of rid indexes) by the predicate attribute.  For instance, \sys would partition the rid arrays in the backward index for \texttt{orders} by the \texttt{shipdate} attribute, so that $C$ only reads the rid partition matching the parameter \texttt{:p1}.    This technique is applicable to categorical attributes and continuous attributes that can be discretized.  This is almost always possible because user-facing output is ultimately discretized at pixel granularity~\cite{m4}.

\sstitle{Group-by push-down.} Interactions such as cross-filtering let users select marks in one view, trace those marks to the input records that generated them, and recompute the aggregation queries in other views based on the selected subset of input records.  This pattern is precisely an aggregation query over the backward lineage of the user's selection.  \sys pushes the group-by aggregation into lineage capture by partitioning the rid arrays on the group-by attributes, and incrementally computing the intermediate aggregation state. This technique works if the main difference between the base and lineage consuming query is the addition of grouping attributes.   In effect, lineage capture generates data cubes to answer the linage consuming aggregation query.  In contrast to building data cubes offline, which requires separate scans of the database, this approach piggy-backs on top of the base query's existing table scans. As with prior work~\cite{datacube,liu2013immens,hanusse2011revisiting}, this technique supports algebraic and distributive functions (e.g., \texttt{SUM}, \texttt{COUNT}, and \texttt{AVG}), and we evaluate this optimization extensively in synthetic (\Cref{ss:exp:opts}) and real-world settings (\Cref{sss:exp:xfilter}).

\sstitle{Relationships with Provenance Semantics.} We observe that popular provenance semantics (e.g., which\cite{cui2000linagetrace,tannen2017pods} and why\cite{buneman:2001:wwp} provenance) can be expressed as lineage consuming queries and pushed down using the above optimizations. In other words, \sys can operate as a system with alternative provenance semantics depending on the given lineage consuming query. For space reasons, we include a brief discussion in our technical report~\cite{psallidas2017extendedlineage}.

\sstitle{Applying Optimizations.} Choosing the appropriate optimizations, manually or automatically, each poses challenges that we leave to future work. ``What language extensions (e.g., {\small\texttt{CREATE BACKWARD INDEX ON (SELECT \ldots)}}) are needed to capture lineage manually?''  and ``What cost models are needed to model the trade-offs for each capture and optimization choice?'' constitute interesting research questions.

\section{Experimental Settings}
\label{s:settings}

Our experiments seek to show that \sys (1) incurs significantly lower lineage capture overhead than logical and physical lineage capture approaches, (2) can execute lineage queries faster than lazy, logical, and physical lineage query approaches, and (3) can leverage lineage indexes and workload-aware optimizations to speed up real-world applications as compared to current manual implementations.

To this end, we compare \sys to state-of-the-art logical and physical lineage capture and query approaches using microbenchmarks on single operator plans, as well as  end-to-end evaluations over a subset of TPC-H queries. Using TPC-H, we further show that our workload-aware optimizations can provide further lineage query speedups on the ``Overview first, zoom and filter, and details on demand'' interaction paradigm and respond within interactive latencies of $<150ms$~\cite{liu2014effects,card91,miller1968response}. Finally, we express two real-world applications (cross-filter~\cite{crossfilter} and data profiling~\cite{uguide:2017:thirumuruganathan}) in lineage terms and show that {\it \sys can match or outperform hand-optimized implementations of the same applications}.


\sstitle{Data.} The microbenchmarks use a synthetic dataset of tables \texttt{zipf}$_{\theta,n,g}$\texttt{(id,z,v)} containing zipfian distributions of varying skew.  \texttt{z} is an integer that follows a zipfian distribution and \texttt{v} is a double that follows a uniform distribution in $[0,100]$. $\theta$ controls the zipfian skew, $n$ is the table size, and $g$ specifies the number of distinct $z$ values (i.e., groups).  Tuple sizes are small to emphasize worst-case lineage overheads.  End-to-end and workload-aware experiments use the TPC-H data generator and vary the scale factor. Our experiments on real-world applications use the Ontime~\cite{ontimeoriginal,ontime} (123.5m tuples, 12GB) and Physician~\cite{physicians} (2.2m tuples, 0.6GB) datasets.

{
\begin{table}[t]
\fontsize{8}{8}\selectfont
\centering
\begin{tabular}{ |c|l| } 
 \hline
 \textbf{Abbreviation} & \multicolumn{1}{c|}{\textbf{Description}}  \\
 \hline

 \rowcolor{Gray}
	\multicolumn{2}{c}{\textbf{Smoke}}\\
 \hline
 \textsc{Baseline} & \sys without lineage capture  \\

 \sysd & \sys with defer lineage capture \\

 \sysi & \sys with inject lineage capture  \\

  \hline

\rowcolor{Gray}
	\multicolumn{2}{c}{\textbf{Logical}}\\	
\hline
\lgerids & Rid-based annotation\\
\lgefull & Tuple-based annotation \\
\lgei    & Indexing input-output relations\\

	\hline

\rowcolor{Gray}
	\multicolumn{2}{c}{\textbf{Physical}}\\	
\hline
\phvfdp & Virtual emit function calls and no reuse   \\

\phbdb & Lineage capture using BerkeleyDB \\

\hline

\end{tabular}
\caption{Lineage capture techniques used in our evaluation.}

\label{tab:techniques}
\vspace{-2em}
\end{table}
}

To ensure a fair comparison, we implement and optimize alternative, state-of-the-art  techniques in our query engine.  Our implementation reduces the capture overheads (by several orders of magnitude) as compared to their original implementations, and is detailed in our extended report~\cite{psallidas2017extendedlineage}. 

First, we describe the compared lineage capture techniques (see also~\Cref{tab:techniques} for a minimal description):

\sstitle{\sys techniques.} \textbf{\sysi} and \textbf{\sysd} instrument the plan using \typei and \typej instrumentation (\Cref{s:instr}). Unless otherwise noted, \sysi and \sysd don't use optimizations from~\Cref{s:instr}. \textbf{\textsc{Baseline}} evaluates base queries on \sys without capturing lineage.

\sstitle{Baseline logical techniques.} State-of-the-art logical approaches (\perm~\cite{glavic:2009:perm}, \gprom~\cite{gprom}) use query rewrites to annotate the base query output with lineage. However, they are built on production databases that incur unneeded capture overheads from e.g., transaction and buffer managers, lack of hash-table reuse, no query compilation.   For this reason, we used \perm's rewrite rules (and whenever applicable, \gprom's optimizations) to generate physical plans that annotate the output with either rids (\textbf{\lgerids}) or full input tuples (\textbf{\lgefull}), and implemented the plans in \sys.  Note that the output relation needs to be indexed to support fast lineage lookups.  To this end, \textbf{\lgei} scans the annotated output relation to construct the same end-to-end lineage indexes as those created by \sys.    Ultimately, our implementations are two orders of magnitude faster than \perm and \gprom but still incur capture overheads higher than \sys.

\sstitle{Baseline physical techniques.} To highlight the importance of tightly integrating lineage capture and operator logic, we use two baseline physical techniques.  \textbf{\phvfdp} instruments each operator to make virtual function calls to store input-output rid pairs in \sys lineage indexes from~\Cref{s:instr}, which highlights the overhead of making a virtual function call for each lineage edge.  \textbf{\phbdb} instead indexes lineage data in BerkeleyDB to showcase the drawbacks of using a separate storage subsystem~\cite{wu2013subzero}.


Moreover, we compare lineage querying techniques based on data models and indexes induced during lineage capture:

\sstitle{Lineage consuming queries.} \sysi, \sysd, \lgei, and \phvfdp all capture the same lineage indexes from \Cref{s:storage}, thus their lineage consuming query performance will be identical.  We call this group \textbf{\sysl}.  We compare with a baseline lazy approach, \textbf{\sysn}, which uses standard rules~\cite{cui2000linagetrace,ikedathesis} to rewrite lineage consuming queries into relational queries that scan the input relations. We also compare with the data model that \lgerids and \lgefull produce and the indexes that \phbdb generate.  Finally, we consider \sysn and \sys without optimizations as baselines to our workload-aware optimizations. 

Settings for the real-world applications are provided inline.

\sstitle{Measures.} For lineage capture, we report the absolute base query latency and relative overhead compared to not capturing lineage. For lineage and lineage consuming queries, we report absolute latency and speedup over baselines. All numbers are averaged over 15 runs, after 3 warm-up runs.

\sstitle{Platforms.} We ran experiments on a MacBook Pro (macOS Sierra 10.12.3, 8GiB 1600MHz DDR3, 2.9GHz Intel Core i7), and a server-class machine (Ubuntu 14.04, 64GiB 2133MHz DDR4, 3.1GHz Intel Xeon E5-1607 v4). Both architectures have caches sizes 32KiB L1d, 32KiB L1i, and 256KiB L2---the MacBook has 4MiB L3 and the server-class has 10MiB L3.  Our overall findings for lineage capture are consistent across the two architectures.  Since lineage capture is write-intensive, we report results using the lower memory bandwidth setting (MacBook).  For the crossfilter application, we report the server-class results because the Ontime dataset doesn't fit in the MacBook memory.

\section{Experimental Results}
\label{s:results}

In this section, we first compare lineage capture techniques on microbenchmarks (\Cref{ss:exp:lcapture}) and TPC-H queries (\Cref{ss:exp:multi}). Then, we compare techniques on lineage query evaluation (\Cref{ss:exp:lquery}) and showcase the impact of our workload-aware optimizations (\Cref{ss:exp:opts}). We conclude with experiments on real-world applications (\Cref{ss:lapps}).

\begin{figure}[t]
\centering
\includegraphics[width=\columnwidth]{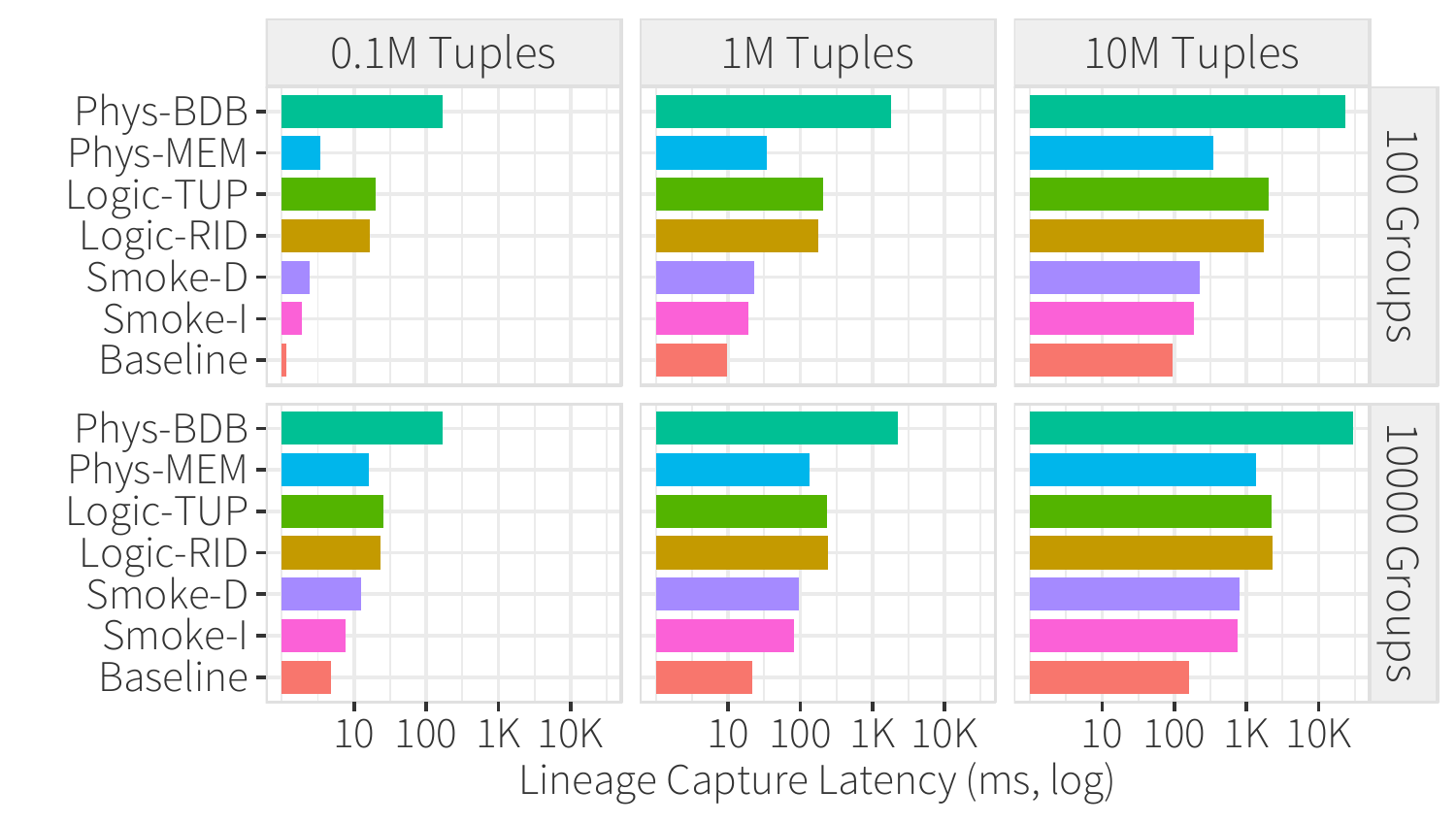}
\caption{\small Comparison of lineage capture costs for the group-by aggregation operator for different relation cardinalities (columns) and number of distinct groups (rows).  \sysi and \sysd slow down the non-instrumented Baseline the least as compared to alternative logical and physical capture methods.}
\label{fig:gagg_overhead}
\vspace*{-.2in}
\end{figure}

\subsection{Single Operator Lineage Capture}
\label{ss:exp:lcapture}
We first evaluate lineage capture with a set of single operator microbenchmarks for group-by (\Cref{sss:exp:micro:gagg}), pk-fk joins (\Cref{sss:exp:micro:pkfk}), and m:n joins (\Cref{sss:exp:micro:mnjoins}).

\subsubsection{Group-by Aggregation}
\label{sss:exp:micro:gagg}
We use the following base query, which groups by $z$ drawn from a zipfian distribution so that cardinalities are skewed ($\theta=1$).  Visualizations often compute multiple statistics to avoid redundant scans when users ask for new statistics~\cite{tableau:2015:improving}:
{
\small
\begin{lstlisting}[
	language = SQL,
	showspaces=false,
	basicstyle=\ttfamily,
	commentstyle=\color{gray},
	mathescape=true,
	numbers=none,
	frame = none,
    escapeinside={<}{>},
    label={dl:lsql1},
    aboveskip=-0.1em,
    belowskip=-0.1em
 		]
  <$Q_{\Ground}$>= SELECT    z, COUNT(*), SUM(v), SUM(v*v),
                 SUM(sqrt(v)), MIN(v), MAX(v) 
       FROM      zipf 
       GROUP BY  z -- #groups follow a zipfian
\end{lstlisting}
}
\noindent \Cref{fig:gagg_overhead} reports the lineage capture latency (base query cost + overhead) for each instrumentation technique, and varies the input size (columns) and the number of groups (rows).  

\sstitle{Smoke.}  \sysi incurs the lowest overhead among techniques ($0.7\times$ on average).  \sysd is slightly slower ($1.2\times$ on average) due to the cost of join to construct lineage indexes.  

\sstitle{Comparison with \texttt{Logical} systems. } \lgerids and \lgefull use \perm's aggregation rewrite rule, which computes $Q_{\Ground}\Join_{z} \texttt{zipf}$ to derive the denormalized lineage graph as a single relation. The cost of computing and writing the denormalized lineage graph is costly and can slow the base query by multiple orders of magnitude.   Since \texttt{zipf} is narrow, \lgefull performs similarly to \lgerids, however we expect the cost to increase for wider input relations. \lgei has extra indexing costs over \lgerids and is not plotted.

\sstitle{Comparison with \texttt{Physical} systems. } The primary overhead for \phvfdp is the cost of a virtual function call for each written lineage edge. The cost of building index data structures is comparable to \sys's write costs, however \sys can reuse the hash table built by $\gamma'_{ht}$ and incur lower costs for building the backward lineage rid index.  \phbdb incurs by far the highest overhead (up to $250\times$ slowdown), due to the overhead of communicating with BerkeleyDB.  The same trends hold for the other operators, and we have not found physical approaches to be competitive.   {\it As such, we do not report physical approaches in the rest of the experiments.  }

\sstitle{Varying dataset size, skew, and groups.}
In general, the lineage capture techniques all incur a constant per input tuple overhead, and differ on the constant value.  This is why increasing the input relation size increases costs linearly for all techniques.  Increasing the number of groups increases the costs of building and scanning the group-by hash table as well as the output cardinality, and affects all techniques including the baseline.    We find that the overhead is independent of the zipfian skew because it does not change the number of lineage edges that need to be written; it does affect querying lineage as we will see in~\Cref{ss:exp:lquery}.

\sstitle{Complexity of group-by keys and aggregate functions.} 
We find that the techniques differ in their sensitivity to the size of the group-by keys and the number of aggregation functions in the project clause of the query.  \sysi simply generates rid index and rid arrays, and is not affected by these characteristics of the base query.  In contrast, \sysd and both logical approaches are sensitive to the size of the group-by keys, since they are used to join the output and input relations.  Finally, the logical approaches are also affected by the number of aggregation functions because they affect the cost of the final projection.  In short, we believe our setup is favorable to alternative approaches, and find that \sys still shows substantial lineage capture benefits.

\sstitle{Cardinality Statistics. }  If \sys knows the cardinality statistics for each group, then it can allocate correctly sized arrays in the lineage indexes (\Cref{s:instr}).   This further reduces the lineage capture overhead by 52\% on average and leads to overhead reduction from $0.7\times$ to $0.3\times$ for \sysi (not plotted).

\subsubsection{Primary-Foreign Key (Pk-Fk) Joins}
\label{sss:exp:micro:pkfk}

We use the following primary-foreign key join query:
\texttt{\lstinline|SELECT * FROM gids,zipf WHERE gids.id=zipf.z|}. 
\texttt{zipf.z} is a foreign key that references \texttt{gids.id} and drawn from a zipfian distribution so that some keys are more popular than others.  We vary the number of join matches (i.e., groups) by varying the number of unique values for \texttt{gids.id}.  In addition to \textsc{Baseline} and \sysi, we evaluate \sysic, which assumes that we know the number of matches for each join attribute value---this is to highlight the costs of array resizing. Note that \sysd is equivalent to \sysi due to the pkfk optimization (\Cref{ss:join}).  We compare against \lgei because \lgerids and \lgefull do not support forward queries without additional indexes.  

\begin{figure}[t]
\centering
\includegraphics[width=\columnwidth]{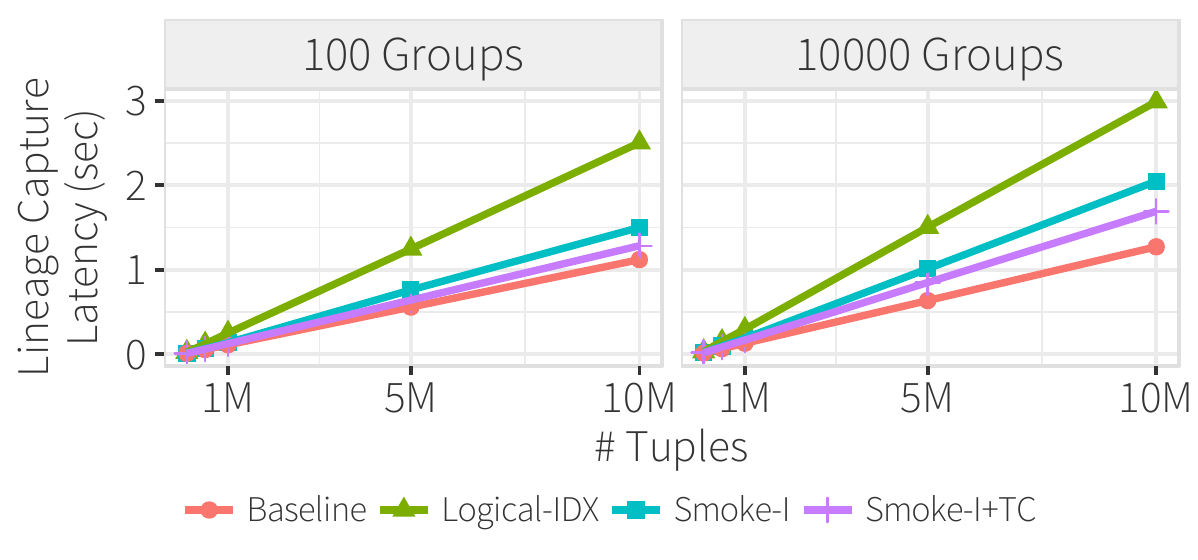}
\caption{\small \sysi reduces the instrumented pk-fk join latency from 1.4$\times$ (\lgei) to 0.41$\times$.  Knowing the join cardinalities further reduces the overhead to 0.23$\times$ (\sysic).  \sysd is equivalent to \sysi for pk-fk joins.}
\label{fig:joinpkfk_time_overhead}
\vspace{-2em}
\end{figure}

\sstitle{Comparison with logical techniques.} \lgei incurs 1.4$\times$ relative overhead on average due to the costs of computing and materializing the denormalized lineage graph in the form of the annotated output relation, and scanning the annotated table to build backward and forward lineage indexes for both input relations.  In contrast, \sysi incurs on average $0.41\times$ overhead; knowing join cardinalities reduces the overhead to $0.23\times$ on average.   Finally, note that \sysi already knows the cardinalities for the backward indexes and the forward index of the right table for pkfk joins (\Cref{ss:join}), thus \sysic's lower overhead can be attributed to lower reallocation costs to build the forward index for the left table.

\subsubsection{Many-to-Many Joins}
\label{sss:exp:micro:mnjoins}
For M:N joins we use the following base query that performs a join over two \texttt{z} attributes with zipfian distributions: 
{\centering\small\texttt{\lstinline|SELECT * FROM zipf1,zipf2 WHERE zipf1.z=zipf2.z|}}.   The join is highly skewed to stress lineage capture.  Both \texttt{z} integer attributes are drawn from a zipfian distribution, where \texttt{zipf1.z} is within $[1, 10]$ or $[1, 100]$, while \texttt{zipf2.z}$\in [1,100]$.  This means that tuples with $z=1$ have a disproportionate number of matches, whereas larger $z$ will have few matches.  Furthermore, we fix the size of the left table to 1000 records and  vary the right from $10^3$ to $10^5$. 

\Cref{ss:join} described the \typei approach, which populates the lineage indexes within the join's probe phase ($\Join_{probe}$), and the \typej approach, which computes cardinality statistics during the probe phase to correctly allocate and populate the lineage indexes for the left table (\texttt{a\_fw},\texttt{a\_bw}) after the probe phase and avoid array resizing costs.  Finally, to break down the benefits of \typej, we also evaluate \sysm which still defers \texttt{a\_fw} but populates \texttt{a\_bw} within the $\Join_{probe}$ phase. To simplify the presentation, we only report \sys-based techniques since the relationship with alternatives is consistent with the previous results.

\begin{figure}[t]
\centering
\includegraphics[width=\columnwidth]{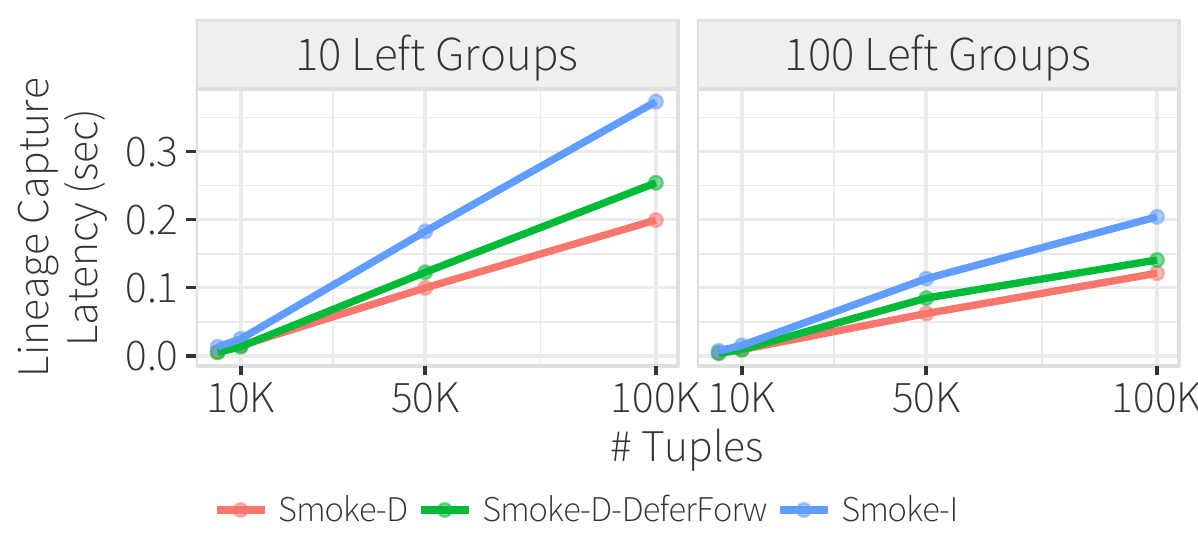}
\caption{\small M:N join latency when all indexes are populated during the probe phase (\sysi), only forward indexes for the left table are deferred (\sysm), and when both lineage indexes are deferred (\sysd).  These graphs highlight rid array resizing costs.}
\vspace{-.1in}
\label{fig:mnjoin_overhead}
\end{figure}

\sstitle{Comparison of \sys techniques.} In contrast to the other operators, the MN join over skewed inputs is similar to a cross-product, and generates $>1$billion results.  Materializing the result set renders the capture overheads non-informative so we do not materialize the output.  In this case, the MN execution is nearly 0, so~\Cref{fig:mnjoin_overhead} primarily reports instrumentation overhead for the three techniques.  The overhead is predominantly due to rid array resizing, which is why deferring the forward and backward lineage index construction for the left table reduces overhead by up to $2.65\times$. Finally, increasing the number of groups for \texttt{zipf1.z} reduces the costs of all techniques because the output cardinality is smaller, but their relative overheads are the same.

\sstitle{Other Operators.} Our technical report~\cite{psallidas2017extendedlineage} describes additional results and operators.  The main additional finding is that it is preferable to overestimate selection cardinality estimates to avoid array resizings for the selection operator.  

\subsection{Multi-Operator Lineage Capture}
\label{ss:exp:multi}

\begin{figure}[t]
\centering
\includegraphics[width=\columnwidth]{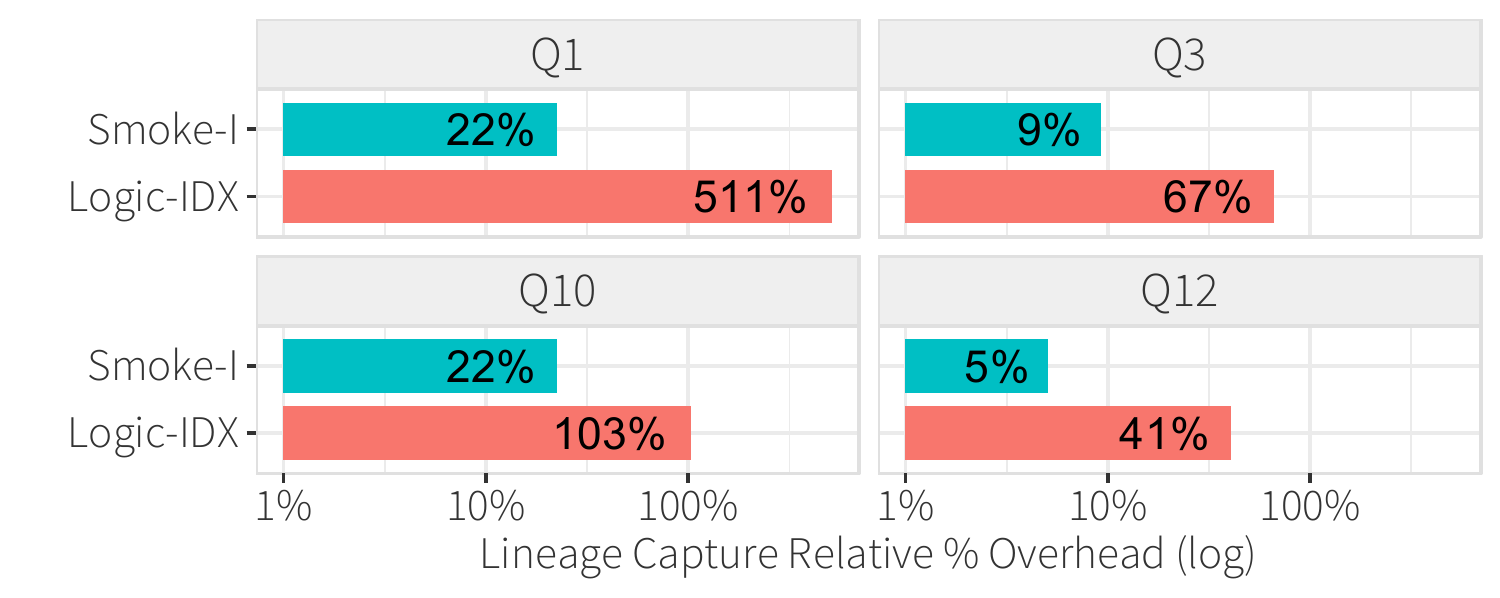}
\caption{\small Relative overhead of \sys and logical lineage capture techniques for TPC-H queries Q1, Q3, Q10, and Q12.  Scale factor 1.}
\label{fig:tpch_overhead}
\vspace{-1em}
\end{figure}

We used four queries from TPC-H---Q1, Q3, Q10, and Q12. Their physical query plans contain group by aggregation as the root operator, selections that vary in predicate complexity and selectivity, and up to three pk-fk joins. (Our hash-based execution precludes sort operations.)  ~\Cref{fig:tpch_overhead} summarizes the relative overhead of the best performing \sys (i.e., \sysi) and logical (i.e., \lgei)  techniques for the four TPC-H queries.  We use scale factor 1.

\sstitle{Overall Results.} \sysi reduces the lineage capture overhead as compared to \lgei by up to $22\times$.     In addition, \sysi incurs at most $22\%$ overhead across the four queries.  To make the overhead results meaningful, we have tried to ensure that \sys query engine has respectable performance---despite row-oriented execution it matches the performance of single-threaded MonetDB---non-instrumented Q1 runs in $176$ms, while the slowest query Q12 runs in $306$ms.\footnote{\small The purpose is {\it not} to compare \sys with MonetDB, but to suggest that the reported overheads are with respect to reasonable baseline performance.} \sysd (not shown) is slower than \sysi due to the join cost between the input and output relations, however it is still faster than the logical approaches. Finally, although Q1 is simple (e.g., it has no joins), its results are arguably the most informative because the plan is simple and has the highest selectivity, which most stresses lineage capture.

\sstitle{Impact of selections in lineage capture. } We found that the selectivity of the query predicate has a large impact on the overhead of the logical approaches.   Q1 shows a setting where the predicate has a high selectivity, thus the input to the subsequent aggregation operator has a high cardinality, each output group depends on a larger set of input records, and ultimately leads a large amount of data duplication to create the denormalized lineage graphs.  In contrast, the other queries have low predicate selectivity, thus the cardinality of the subsequent aggregation operator is small and leads to a substantially smaller lineage graph.  \sys is less sensitive to this effect because its lineage indexes represent a normalized lineage graph and avoids this duplication.

{\it \stitle{Lineage Capture Takeaways (Sections~\ref{ss:exp:lcapture} and~\ref{ss:exp:multi}):} \sys-based techniques outperform both logical and physical approaches by up to two orders of magnitude: Logical approaches that adhere to the relational model are affected by the denormalized lineage graph representation, extra indexing steps, and expensive joins. The physical approaches are affected by virtual function calls and write-inefficient lineage indexes. We find that array resizing contributes to a large portion of \sys overheads, and accurate or overestimated cardinality estimates can reduce resizing costs.}

\subsection{Lineage Query Performance}
\label{ss:exp:lquery}

\begin{figure}[t]
\centering
\includegraphics[width=\columnwidth]{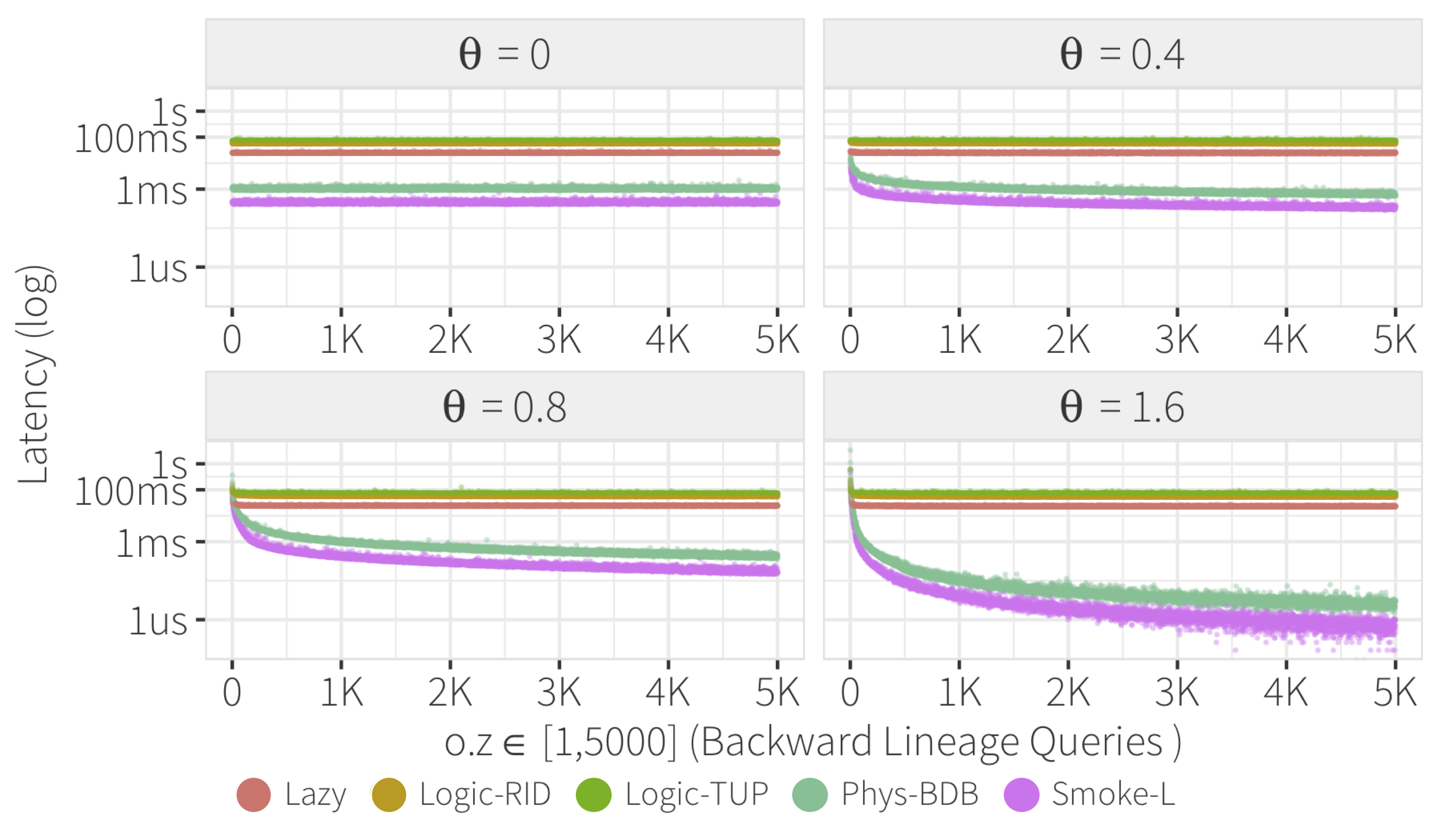}
\caption{\small Lineage query latency for varying data skew ($\theta$).  \sysn has a fixed cost to scan the input relation and evaluates the simple selection predicate on the group-by key \texttt{z=?}. \lgerids and \lgefull performs the same selection but on annotated output relations.  \sysl is mainly around $1ms$, and outperforms \sysn, \lgerids, and \lgefull by up to five orders of magnitude for low selectivity lineage queries.  The crossover point at high selectivities is due to the overhead of \sysl's index scan. \sysl is a lower bound for \phbdb that incurs extra costs for reading lineage indexes. }
\label{fig:lquery}
\vspace{-1em}
\end{figure}

We now evaluate the performance of different lineage query techniques.  Recall that lineage queries are a special case of lineage consuming queries.  We evaluate the query: \texttt{\textbf{SELECT} * \textbf{FROM} $\text{L}_\text{b}($o$\in Q_{\Ground}($zipf$)$, zipf$)$}, where $Q_{\Ground}(zipf)$ is the query used in the group-by microbenchmark (\Cref{sss:exp:micro:gagg}). $o$ is an output record (a group).  \texttt{zipf} contains 5000 groups, $10M$ records, and we vary its skew $\theta$.   Varying $\theta$ highlights the query performance with respect to the cardinality of the backward lineage query. ~\Cref{fig:lquery} reports lineage query latency for all $5000$ possible $o$ assignments and different $\theta$ values.

\sysl evaluates the lineage query using a secondary index scan---it probes the backward index, and uses the input rids as array offsets into \texttt{zipf}. Recall that \sysl refers to any of \sysi, \sysd, \lgei, or \phvfdp (\Cref{s:settings}).

\sstitle{\sysl vs \sysn.} In contrast to \sysl, \sysn performs a table scan of the input relation and evaluates an equality predicate on the integer group key. This is arguably the cheapest predicate to evaluate and constitutes a strong comparison baseline. We find that \sysl outperforms \sysn up to five orders of magnitude, particularly when the cardinality of the output group is small.  We expect the performance differences to grow when the base query uses more complex group keys that increase the predicate evaluation cost~\cite{johnson2008row,chaudhuri2003factorizing,aps:2017:kester:2017}, or when the input relation is wide, which increases scan costs. There is a cross over point when the input relation is highly skewed ($\theta \in \{0.8,1.6\}$) and the backward lineage of some groups have high cardinality.  This increases \sysl's secondary index scan cost as compared to a linear table scan that can benefit from scan pre-fetching. 

\sstitle{\sysl vs Logical Approaches.} We also report the cost of scanning the annotated relations generated by \lgerids and \lgefull (highest two lines).  Scanning these relations to answer lineage queries is worse than \sysn because the annotated relation is wider than the input relation, yet they have the same cardinality. Note that indexing the annotated relation is \lgei, and represented by \sysl.

\sstitle{\sysl vs Physical Approaches.} \phvfdp is included as part of \sysl, so we report \phbdb. Using an external lineage subsystem to perform a lineage query, we need to perform function calls to the external system to fetch the input rids for an output. As long as we have the input rids, we can perform a secondary index scan to evaluate the lineage query similarly to \sysl. In our experiments, we compare both fetching all input rids in a single function call, and with consecutive function calls to fetch the rids in a cursor-like fashion. The cursor-like approach outperformed the bulk approach since it avoids allocation costs for input rids. \sysl provides a lower bound for \phbdb: both perform the same secondary index scan but \phbdb pays the cost of function calls to the external lineage subsystem, and has worse lineage index read performance due to the B-Tree of BerkeleyDB.

{\it \stitle{Lineage Query Takeaways:} \sys outperforms logical and lazy lineage query evaluation strategies by multiple orders of magnitude, especially for low-selectivity lineage queries. We believe \sys is a lower bound for physical approaches by avoiding functions calls and using read-efficient indexes.}

\subsection{Workload-Aware Optimizations}
\label{ss:exp:opts}

We explore the effectiveness of the data skipping and group-by push-down optimizations by incrementally building up an example motivated by the ``Overview first, zoom and filter, details on demand''~\cite{shneiderman1996eyes} interaction paradigm. We focus only on zoom and filter because the base query generates the initial overview, while details on demand is the simple backward lineage query evaluated in~\Cref{ss:exp:lquery}. We report selection push-down and pruning in our technical report~\cite{psallidas2017extendedlineage}.

We use TPC-H Q1 as the initial ``Overview'' base query, and we render its output groups as a bar chart; there are four bars each generated from 48\%, 24\%, 24\%, and 0.06\% of the \texttt{Lineitem} input relation.  Subsequent user interactions (zoom by drilling down, filter by adding predicates) will be expressed as lineage consuming queries that incrementally modify its preceding lineage consuming query.

\sstitle{No Optimization. }  We start off by evaluating the effectiveness of using lineage indexes (without optimizations) as compared to the lazy approach for lineage consuming queries (not plotted).   Suppose the user will be interested in drilling into a particular bar to see the statistics broken down by month and year of the shipping date.  This can be modeled as a lineage consuming query $Q1_a$ that extends Q1 in two ways: (1) replace the input relation with the backward lineage of the bar that the user is interested in (i.e., $L_b(o_a \in Q1($\texttt{Lineitem}$),$ \texttt{Lineitem}$)$) and (2) add \texttt{Month, Year} to the GROUP BY.

We evaluate $Q1_a$ for every value of $o_a$.  \sysn runs Q1$_a$ as a table scan followed by filtering on Q1's group by keys,  grouping on year and month, and computing the same aggregates as Q1.  \sysi is best when the group cardinality is low ($0.06\%$ selectivity) and outperforms \sysn by 6.2$\times$. Higher cardinality groups incur random seek overheads.  The performance converges for high cardinality because the performance is dominated by the query processing costs (i.e., aggregation in this case).  To address the overhead of high cardinality lineage queries, we next evaluate workload-aware optimizations.

\begin{figure}[t]
\centering
\includegraphics[width=\columnwidth]{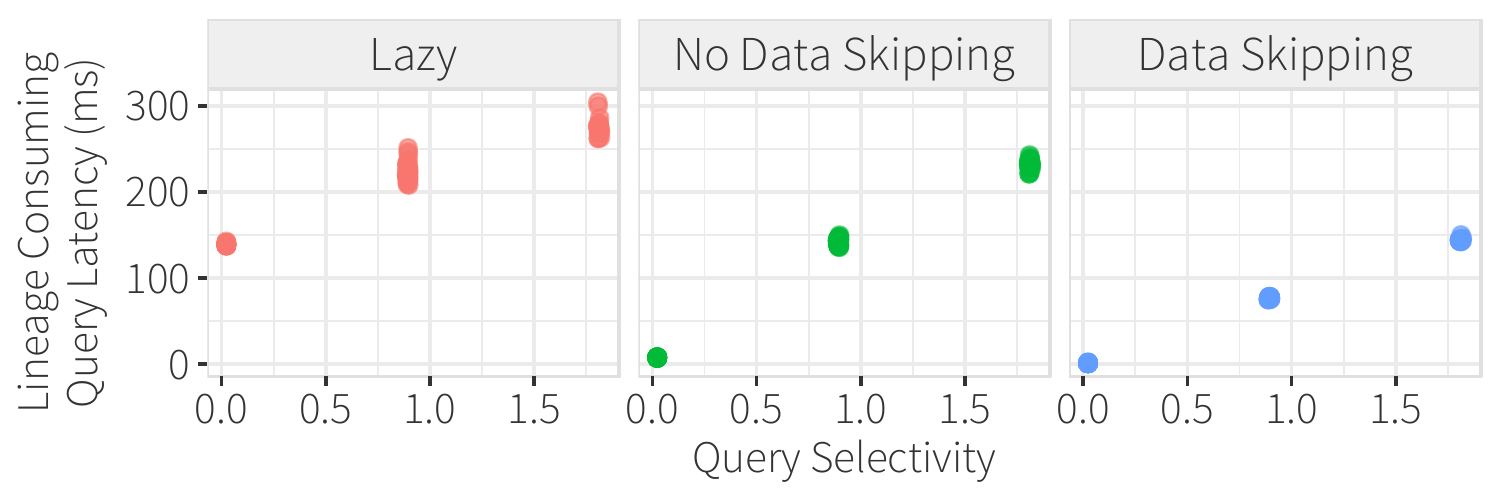}
\caption{\small Lineage consuming query latency for different instrumentation approaches as the lineage consuming query's selectivity varies.   Lazy requires full table scans, No Data Skipping performs more efficient secondary index scans, and Data Skipping is $\le150$ms because it only scans the relevant partition of the lineage index.}
\label{fig:dskip}
\vspace{-2em}
\end{figure}

\sstitle{Data skipping.}  Suppose we know that the user wants to filter the result of $Q1_a$ (say, based on interactive filter widgets), then we can push this logic into lineage capture using the data skipping optimization.  We evaluate Q1$_b$, which extends Q1$_a$ with two parameterized predicates: \texttt{l\_shipmode = :p1 AND l\_shipinstruct = :p2}. Q1 is the base query for Q1$_b$.   
To exercise push-down overheads, both are text attributes and thus more expensive to evaluate than for numeric attributes.
The lineage capture overhead was 0.22$\times$ for \sysi, and 1.65$\times$ with the data skipping optimization due to the additional cost of partitioning the rid arrays on the text attributes, but still lower than logical approaches (Figure~\ref{fig:tpch_overhead}).

\Cref{fig:dskip} plots the lineage consuming query latency vs the selectivity of every possible combination of the predicate parameters.  The \sysn baseline executes the lineage consuming query as a filter-groupby query over a table scan of \texttt{Lineitem}.  Although lineage indexes substantially reduce query latency (No Data Skipping in~\Cref{fig:dskip})---particularly for low predicate selectivities---it is bottlenecked by the secondary scan costs of backward lineage for high cardinality groups.  In contrast, data skipping reduces even high selectivity queries by at least $2\times$ compared to \sysn, and is consistently below the interactive $150ms$ threshold~\cite{liu2014effects}. This is because rid arrays are partitioned by \texttt{l\_shipmode, l\_shipinstruct}, so that the lineage query only performs an indexed scan over the rids needed to answer the query.

\sstitle{Aggregation push down. } After users filter and identify interesting statistics from the filter interactions in Q1$_b$, they may want to drill down further.  If we know this upfront, \sys may pre-compute aggregates for new dimensions during lineage capture.  To this end, we define Q1$_c$ by adding \texttt{l\_tax} to the group by clause in Q1$_b$, and setting the input relation to $L_b(o_c \in Q1_b(...),$\texttt{Lineitem}$)$.  We compare \lazy (rewrites Q1$_c$ as a table scan-based query) against \sysi with and without the group-by optimization.  In this setup, the previous lineage consuming query Q1$_b$ is the base query for Q1$_c$.  

\begin{figure}[tb]
  \centering
   \includegraphics[width=\columnwidth]{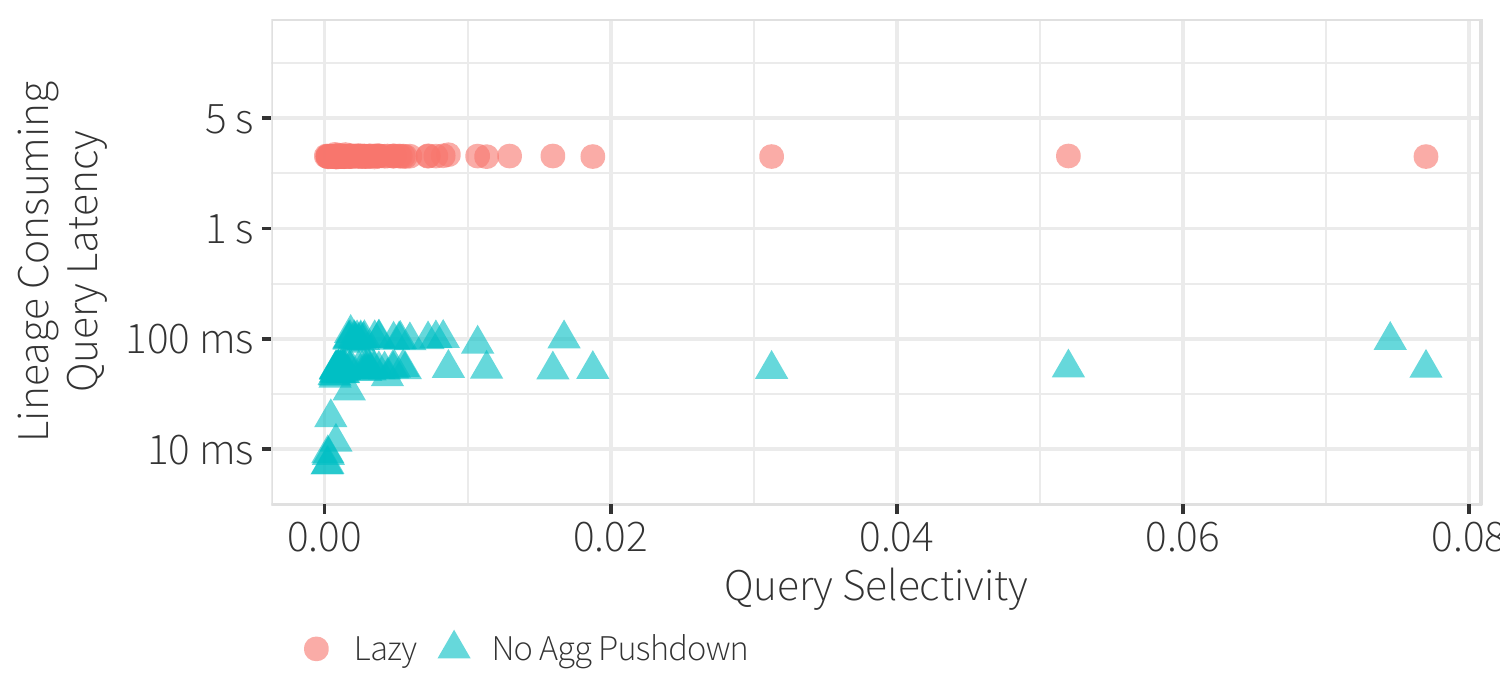}
  \caption{\small \sysi reduces the lineage consuming query latency by $72.9\times$ on average as compared to \lazy.  With aggregation push-down, the latency is $\approx 0$ms and we do not plot it. }
    \label{fig:apush}
\vspace{-1em}
\end{figure}

~\Cref{fig:apush} compares the lineage query latency under \lazy (\red{red dots}) against \sysi without the optimization (\blue{blue triangles}). The push-down optimization is not plotted because it takes $\approx 0$ms (i.e., just fetches the materialized aggregates).   For completeness we vary the parameters of the backward lineage statement $L_b()$ for Q1$_c$ ($L_b(o_c \in Q1_a,...)$) as well as for the base query Q1$_a$ ($L_b(o_a \in Q1,...)$) of Q1$_b$ and report the lineage consuming query's latency for all combinations.   Although \lazy takes $>4$ seconds per Q1$_c$ instance, \sysi's index scan takes on average $100ms$, and $10ms$ for low selectivity queries.  

\begin{figure}[tb]
  \centering
   \includegraphics[width=\columnwidth]{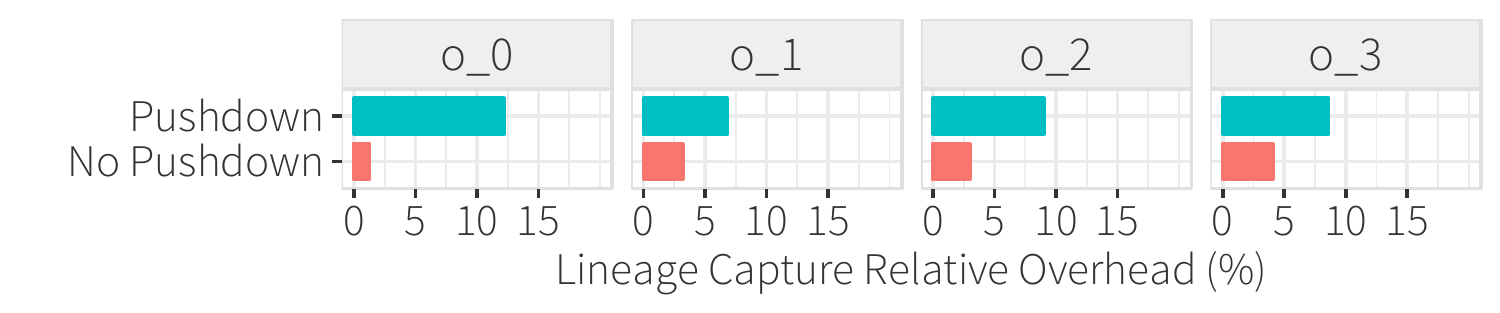}
  \caption{\small The average relative instrumentation overhead increases from $2.9\%$ without to $9.15\%$ with aggregation push-down. } 
    \label{fig:apush_lcapture}
\vspace{-1em}
\end{figure}

Pre-computing aggregation statistics is not free---\Cref{fig:apush_lcapture} plots the lineage capture latency for both \sys variants as compared to the non-instrumented lazy approach.  We report the result for all 4 parameters to the base query Q1$_a$'s backward lineage statement ($L_b(o_a \in Q1,...)$). The overhead of \sysi is low compared to the cost of partitioning the rid arrays on \texttt{l\_tax} and computing aggregates.

{\it \stitle{Push-down Takeaways:} Our experiments highlight that lineage indexes are sufficient whenever the lineage cardinality is low for the complexity of future lineage consuming queries. For higher lineage cardinalities, our workload-aware optimizations provide a principled way to push-down computation into lineage capture and optimize future lineage consuming queries. They also highlight trade-offs that future optimizers would need consider (see also open research questions in~\Cref{ss:optsf::push}). }

\subsection{{\large \textbf{\sys}}-Enabled Applications}
\label{ss:lapps}
We now present evidence that lineage can be used to optimize two real-world applications---cross-filter visualizations (\Cref{sss:exp:xfilter}) and data profiling (\Cref{sss:exp:dprepare})---enough to perform on par with or better than hand-tuned, application-specific implementations.  We highlight the main results here, and defer details to the technical report~\cite{psallidas2017extendedlineage}.

\subsubsection{Crossfilter}
\label{sss:exp:xfilter}

Crossfilter is an important interaction technique to help explore correlated statistics across multiple visualization views~\cite{crossfilter}. In the common setup, multiple group-by queries along different attributes are each rendered as e.g., bar charts. When the user highlights a bar (or set of bars) in one view, the other views update to show the group-by results over only the subset that contributed to the highlighted bar(s).   This is naturally expressed as backward lineage from the highlighted bar, followed by refreshing the other views by executing the group-by queries on the lineage subset.

Since the views are fundamentally aggregation queries, recent research have proposed variations of data cubes to accelerate the interactions~\cite{liu2013immens,lins2013nanocubes,pahins2017hashedcubes}, however it can take minutes or hours to construct the cubes.  Such offline time is not available if the user has loaded a new dataset (e.g., into Tableau) and wants to explore using cross-filter as soon as possible.  This has recently been referred to as the cold-start problem for interactive visualizations~\cite{position:2017:leilani}.

\begin{figure}[tb]
	\vspace{-0.5em}
  \centering
   \includegraphics[width=\columnwidth]{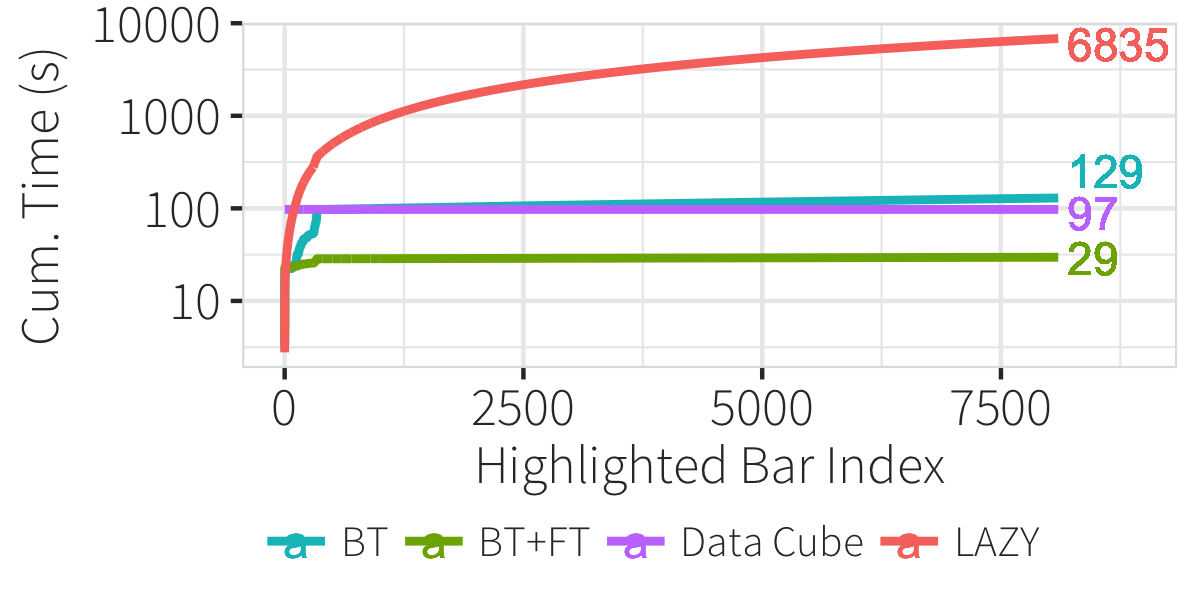}
  \caption{\small Cumulative latency of different crossfiltering techniques. \btftxfilter outperforms all approaches with the total time to perform the initial group-by aggregates, track lineage, and evaluate all interactions being thirty seconds.} 
    \label{fig:flights_performance_cum}
    \vspace{-2em}
\end{figure}

\begin{figure}[tb]
  \centering
   \includegraphics[width=\columnwidth]{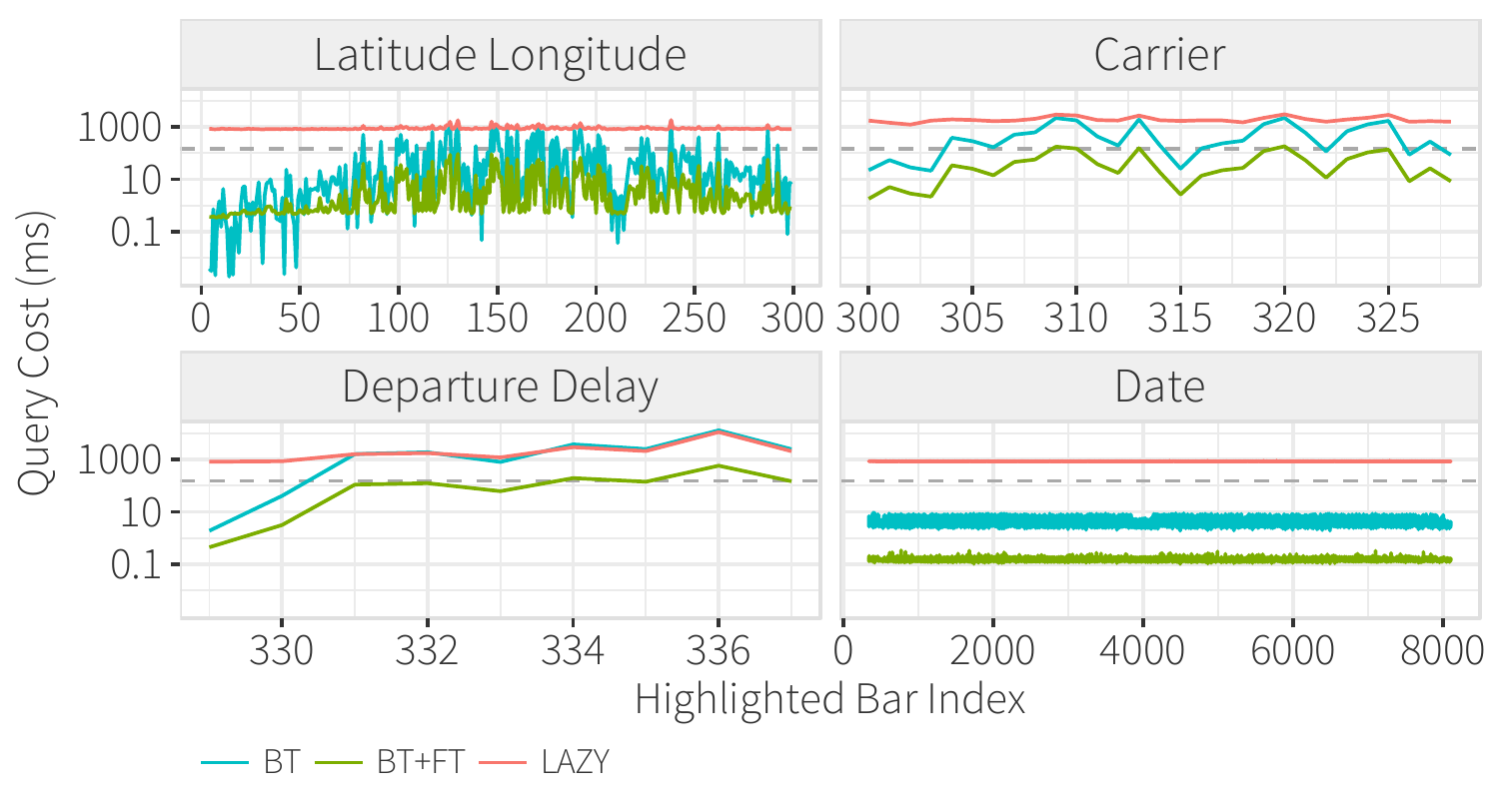}
  \caption{\small Latency for each 1D brushing crossfilter interaction. Dashed lines correspond to 150ms interaction layer. \btftxfilter performs under the 150ms interaction layer for all 8,100 but 5 interactions, with interactions on the spationatemporal dimensions to be <10ms. Data Cube has instantaneous response time and we do not plot it.} 
    \label{fig:flights_performance_per}
    \vspace{-2em}
\end{figure}

\sstitle{Setup.} Following previous studies~\cite{liu2013immens,lins2013nanocubes,pahins2017hashedcubes}, we used the Ontime dataset and four group-by \texttt{COUNT} aggregations on <lat, lon> (65,536 bins), <date> (7,762 bins), <departure delay> (8 bins), and <carrier> (29 bins); only 8,100 bins have non-zero counts because <lat, lon> is sparse.  Each group-by query corresponds to one output view. This setup favors cube construction because it involves only four views and coarse-grain binning on spatiotemporal dimensions (which decreases the number of cubes, and increases group cardinalities). We report the individual (\Cref{fig:flights_performance_cum}) and cumulative (\Cref{fig:flights_performance_per}) latency to highlight each and every bar, respectively.  The experiment was run on our server-class machine so that the dataset can fit in memory.

\sstitle{Techniques.} We compare the following:  \textbf{\lazyxfilter} uses lazy lineage capture and re-executes the group-by queries on the lineage subset.  \textbf{\btxfilter} uses \sys to capture backward lineage indices, but re-runs the group-by queries (which requires re-building group-by hashtables).    \textbf{\btftxfilter} also captures forward lineage indices that map input records to the output bars that they contribute to, which can be used to incrementally update the visualization bars without re-running the aggregation query.   Finally, we compare with \cube construction.  We first ran \immens~\cite{liu2013immens}, \nanocubes~\cite{lins2013nanocubes}, and \hashedcubes~\cite{pahins2017hashedcubes} to construct the data cubes, however \immens and \nanocubes did not finish within 30 minutes, while \hashedcubes required 4 minutes. For this reason, we implemented a custom partial cube construction based on our group-by aggregation push-down optimization that took $1.6$ minutes to construct. This construction resembles the low dimensional cube decomposition described by \immens, but using the sparse encoding recommended by \nanocubes.

\sstitle{Main Results.} We make four main observations. First, \btxfilter outperforms \lazyxfilter by leveraging the backward index to avoid table scans, and is consistent with our TPC-H benchmarks.  Despite the overhead of forward index capture, \btftxfilter outperforms \btxfilter because the forward index lets \sys directly update the associated visualization bars without the need to re-run aggregation queries (and re-build group-by hash tables).  Although the \cube response time is near-instantaneous, the offline construction cost is considerable and \btftxfilter is able to complete the benchmark before the cube is constructed (\Cref{fig:flights_performance_cum}).   Second, \btftxfilter performs best ($<10$ms) when group-by queries output many groups  (e.g., lat/lon, day) because they reduce group cardinality.  This suggests that lineage can complement cases when data cubes are expensive (a cube dimension contains many bins) by computing the results online.  Third, \Cref{fig:flights_performance_per} shows that \btftxfilter responds within $<150$ms (dotted line) for all but five bars, whose lineage depends on a large subset of the input tuples (>10\% selectivity; >13M tuples). Fourth, the capture overhead for \btftxfilter and \btxfilter on base visualization queries are relatively low ($<2\times$ using \sysi). We expect that optimizations that leverage parallelization, sampling, or deferred capture scheduled during user ``think time'' can further reduce the capture costs.

\subsubsection{Data Profiling Applications}
\label{sss:exp:dprepare}

Data profiling studies the statistics and quality of datasets, including constraint checking; data type extraction; or key identification. Recent work, such as \uguide~\cite{uguide:2017:thirumuruganathan}, proposes human-in-the-loop approaches towards mining and verifying functional dependencies, and present users with examples of potential constraint violations to double-check.  This experiment compares \uguide with \sys on a lineage-oriented specification of a data profiling problem.

\sstitle{Setup.} We evaluate the following task: given a functional dependency (FD) $A \rightarrow B$ over a table $T$ and an FD evaluation algorithm that outputs the distinct values $a \in A$ that violate the FD, our goal is to construct a bipartite graph that connects the violations $a$ with the tuples $\{t \in T$ | $t.A=a\}$. Collectively, for a set of given FDs, this construction leads to a two-level bipartite graph connecting FDs and violations to tuples responsible for the violations.  We compare \sys-based approaches with \uguide's implementation\footnote{\small \uguide  proposed novel algorithms for mining and verifying functional dependencies, and implemented a fast version of the data-profiling task using \metanome~\cite{metanome:2015:papenbrock}.  Although latency was not their focus, the system was optimized for performance, so we believe it is a reasonable comparison baseline.}.  Based on correspondence with the authors, it turns out that \uguide internally creates data structures akin to the lineage indexes that \sys captures.  This makes sense because it mirrors a lineage-based description of the problem.

\sstitle{Techniques.} FD violations for $A\rightarrow B$ can be identified by transforming the FD into one or more SQL queries.  We consider two rewrite approaches.  The simple approach (\textbf{\textsc{CD}}) runs the query \texttt{Q$_{cd}$=SELECT A FROM T GROUP BY A HAVING COUNT(DISTINCT B) > 1}; backward and forward lineage indexes correspond to the bipartite graph above. 

\uguide implements an optimization which, \emph{although not modeled as lineage, effectively simulates lineage indexes}.  We thus describe the second approach (\textbf{\textsc{UG}}) in lineage terms.   We first evaluate \texttt{Q$_{ug,attr}$=SELECT DISTINCT attr FROM T} for \texttt{attr}$\in \{A, B\}$, and capture lineage.  We backward trace each $a \in Q_{ug,A}$ to the input $T$, and forward trace each lineage record to $Q_{ug,B}$.  If there are more than one distinct $b$ values in the forward trace output, then the FD is violated.  The lineage indexes also correspond to the desired bipartite graph. The \textbf{\textsc{UG}} implementation is typically faster than \textbf{\textsc{CD}} for FD mining because \textbf{\textsc{UG}} explicitly builds lineage indexes once per attribute and reuses them across FD checks.  Our experiments report the cost of individual FD checks and the cost to construct the bipartite graph, however the relative findings are expected to grow wider for multi-FD checks.
We compare \sys using both approaches (\syscd, \sysug) with \uguide that implements the \textbf{\textsc{UG}} approach (\metanomeuguide).

\begin{figure}[tb]
  \centering
   \includegraphics[width=\columnwidth]{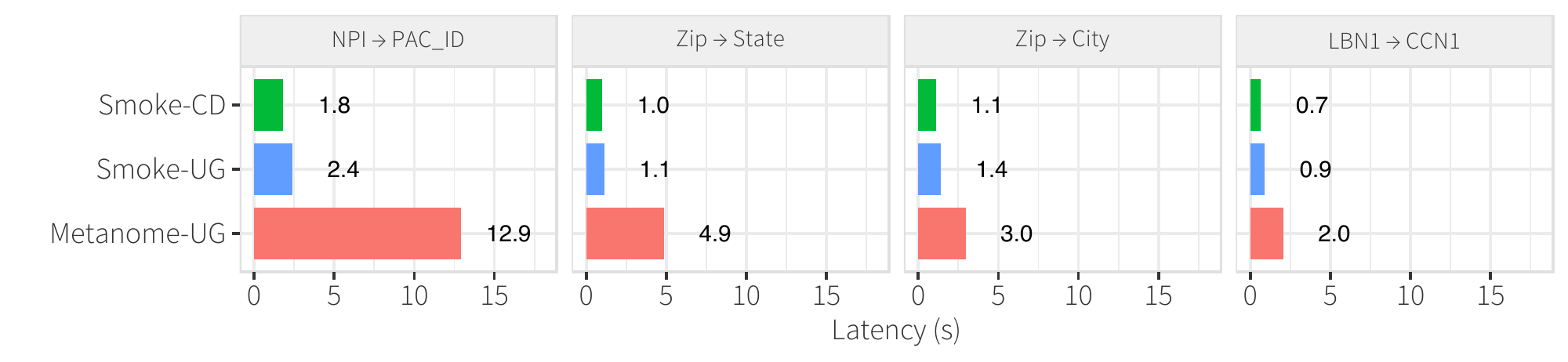}
   \caption{\small Latency of different approaches for FD violation evaluation and bipartite graph construction. \syscd is the minimal overall. \metanomeuguide is affected by virtual function calls for lineage capture, the overheads of JVM, and its data model.  } 
    \label{fig:profiling}
    \vspace{-2em}
\end{figure}

\sstitle{Main Results.}~\Cref{fig:profiling} evaluates the techniques using four functional dependencies over the Physician dataset used in the Holoclean~\cite{holoclean:2017:rekatsinas} paper.  Overall, \sysug outperforms \metanomeuguide by $2-6\times$ while the simpler \syscd approach outperforms both approaches. Both \sys capture overheads are consistent with our microbenchmarks ($<1.2\times$ overhead).  There are several reasons why \sysug outperforms \metanomeuguide.  \metanomeuguide incurs virtual function call costs when constructing its version of lineage indexes ($>2\times$ overhead on \texttt{Q$_{ug,attr}$} that we implemented in \uguide), as well as general JVM overhead even after a warm-up phase to enable JIT optimization. Further, \metanomeuguide models all attribute types as strings, which slows uniqueness checks for integer data types such as \texttt{NPI}. For fairness, the other three FDs are over string attributes (zip is a string). 

{\it \sstitle{Application Takeaways:} Lineage can express many real-world tasks, such as those in visualization and data profiling, that are currently hand-implemented in ad-hoc ways.  We have shown evidence that lineage capture can be fast enough to free developers from implementing lineage-tracing logic without sacrificing, and in many cases, improving performance.}

\section{Conclusions and Future Work}
\label{s:future}
\sys illustrates that it possible to both capture lineage with low overhead and enable fast lineage query performance.  \sys reduces the overhead of fine-grained lineage capture by avoiding shortcomings of logical~\cite{widom:2005:trio,dbnotes2004bhagwat,glavic:2009:perm,gprom,geerts2006mondrian,cui2001lineage,green:2007:orchestra} and physical~\cite{logothetis2013scalable,interlandi2015sparkprovenance,wu2013subzero,ramp,
panda,ikedathesis} approaches in a principal manner, and is competitive or outperforms hand-optimized visualization and data profiling applications. \sys also contributes to the space of physical database design~\cite{chaudhuri:2007:physicaldbdesign,agrawal:2000:automatedselectionmatindexes,dias2005automatic,oracle10g:2003,cliffguard:mozafari:2015,kersten:2005:cracking,idreos:2007:database,amoeba:shanbhag:2016,adaptdb:lu:2017,h20:alagiannis:2014:hha,petraki:2015:holisticindexingcracking,ottertune:2017:vanaken} by being the first engine to consider lineage as a type of information for physical design decisions. Our capture techniques and workload-aware optimization make \sys well-suited for online; adaptive; and offline physical database design settings. Finally, we believe the design principles used in the development of \sys (\textbf{P1-P4} in Introduction) are broadly applicable beyond our design.

There are many areas for future work to explore: 1) leverage modern features such as vectorized and compressed execution, columnar formats, and UDFs~\cite{rheinlander:2017:udfs}, 2) develop cost-based techniques to instrument plans in an application-aware manner (e.g., \typej is best-suited for speculation in-between interactions),  3) model database optimization policies (e.g., statistics computation, cube-construction, key indexes) as lineage queries, and 4) extend support to data cleaning, visualization, machine learning, and what-if~\cite{deutch2013caravan,provisioning2015assadi} applications.

\balance

{
\bibliographystyle{abbrv}
\bibliography{main}
}

\if\techr1
\appendix
Our appendix material covers several details that we did not include in the main body as well as extensions of our techniques, discussion of open problems, additional experiments, and related work. More specifically, we first provide background on the query compilation model that \sys employs (\Cref{s:bgcompile}) and how we tuned  techniques (\Cref{s:tuning}). Furthermore, we provide a more detailed description of (a) the TPC-H Q1 variants, that we used in our experiments with workload-aware optimizations, (\Cref{s:variants}) and (b) the crossfiltering techniques (\Cref{ss:xval}). In addition, we show how our workload optimizations can be used to encode different provenance semantics (\Cref{s:semantics}) and how \sys performs lineage capture for bag and set union; intersection; and difference as well as lineage capture with nested-loop evaluation to support $\theta-$joins and cross products (\Cref{s:physical_algebra}). We conclude with additional experiments (\Cref{ss:exps}) and related work (\Cref{s:related}).

\section{Query Compilation}
\label{s:bgcompile}
One of the main design principles that we realized in \sys is the tight integration of the lineage capture logic with the query execution logic. In this section, we give a brief background on the query compilation and push-based execution model that \sys leverages to realize this principle. (However, note that our techniques are not bound to this execution model. Based on this execution model we better describe tuning techniques of alternative logical and physical approaches for lineage capture in the next section. Readers familiar with query compilation concepts can skip to the next section.) 

Query compilation combines query optimization and execution with low-level compiler-based optimizations (e.g., LLVM, LSM, GCC).  It replaces query interpretation~\cite{hellerstein2007architecture} with a {\it compilation phase} that transpiles queries into intermediate representations (IR) such as C, C++, or LLVM IR, that are further optimized by a standard compiler, and an {\it execution phase} than runs the query as a binary executable. Each operator in the physical plan emits its intermediate representation that implements its logic, and the engine emits glue code to combine operator inputs and outputs.  Many modern database systems~\cite{gebaly2016afterburner,gupta2015amazon,armbrust2015sparksql} have taken this approach, and continues to be an active research area~\cite{klonatoslegobase} with positive results.

Query compilation systems typically implement operators using the {\it producer-consumer} code generation model~\cite{neumann2011efficiently} to derive a push-based execution model that is in contrast to the pull-based iterator~\cite{graefe1993volcano}, batch~\cite{padmanabhan2001block}, or full-column~\cite{manegold2009dae} execution models of traditional query interpreters. Each operator exposes two functions: \texttt{produce} triggers child operators to produce tuples or data structures with the appropriate schema, and \texttt{consume} emits execution logic to process its inputs and hand the result to the parent consume methods.  Borrowing the example from Neumann~\cite{neumann2011efficiently}, the following generates pseudo code for the $\sigma$ and scan operators: 
{\small
\begin{alltt}
\(\sigma\).produce        \(\sigma\).input.produce
\(\sigma\).consume(a,s)   print `if '+\(\sigma\).condition;
                 \(\sigma\).parent.consume(a, \(\sigma\))
scan.produce     print `for t in relation'
                 scan.parent.consume(attrs, scan)
\end{alltt}
}
The compilation phase will call the root operator's \texttt{produce} method to emit IR that generates result tuples.  Consider compiling the physical plan $\sigma_p(scan(T))$.    \texttt{$\sigma$.produce} calls \texttt{scan.produce}, which emits the \texttt{for} loop over $T$ and calls \texttt{$\sigma$.consume} to consume the tuples of the scan. Then, \texttt{$\sigma$.consume} inlines the selection over tuples of $T$ in the for loop.  The final emitted pseudocode will be:
{\small
\begin{alltt}
for t in relation
   if p(t)
     <\(\sigma\)'s parent logic>(t)
\end{alltt}
}

Operators such as hash aggregation and the building side of hash joins are called {\it pipeline breakers} because the entire pipeline (i.e., all operators up to the pipeline breaker) needs to materialize its end result before the next pipeline can start operating.  Each pipeline defines a separate code block in the final emitted program code, where each code block is a separate \texttt{for} loop (similar to the example above). 

Finally, it is important to note that interpretation-based execution models can also enable the tight integration principle by introducing new physical operators similar to the ones we presented in~\Cref{ss:instr}. Designing such a system is an interesting future direction.

\section{Tuning}
\label{s:tuning}

Having provided a basic background on the query compilation and execution model of \sys, in this section we present low-level optimization details that we enabled for  lineage capture techniques. We start with logical approaches, then discuss \sys-based optimizations, and conclude with physical approaches.

\sstitle{Tuning logical techniques.} In our preliminary experiments we used \perm and \gprom as they are implemented over Postgres 8.3 and a commercial database system, respectively. Unfortunately, both systems exhibit increased lineage capture overhead for reasons intrinsic to the underlying DBMSs which are not related to the principals behind the logical rewrite rules of \perm and \gprom. 

In particular, we observed increased lineage capture overhead due to (a) the added complexities from the transactional processing layers of full-fledged database systems~\cite{harizopoulos:2008:oltpglass} and (b) no flexibility in reusing data structures in the same physical query plan; reusing data structures is important for lineage capture, as we also noted for our techniques in~\Cref{s:instr}. For these reasons, we implemented the rewrite rules of \perm and \gprom in \sys because (a) it does not incur the  overhead of the transactional processing layers of a full-fledged database and (b) reuses data structures within the same query plan---hence, enables a fair comparison of only the principals behind \sys and logical lineage capture techniques having fixed the underlying execution engine. Finally, we note that our implementation of logical alternatives in \sys outperforms the available versions of \perm and \gprom by two orders of magnitude because it avoids these caveats. For instance, \lgerids and \lgefull for 1m tuples and 1000 groups in our group-by aggregation microbenchmark take $<200ms$ while \perm and \gprom take $25s-45s$. (depending on optimization knobs.)

\begin{figure}[h]
\centering
\includegraphics[width=\columnwidth]{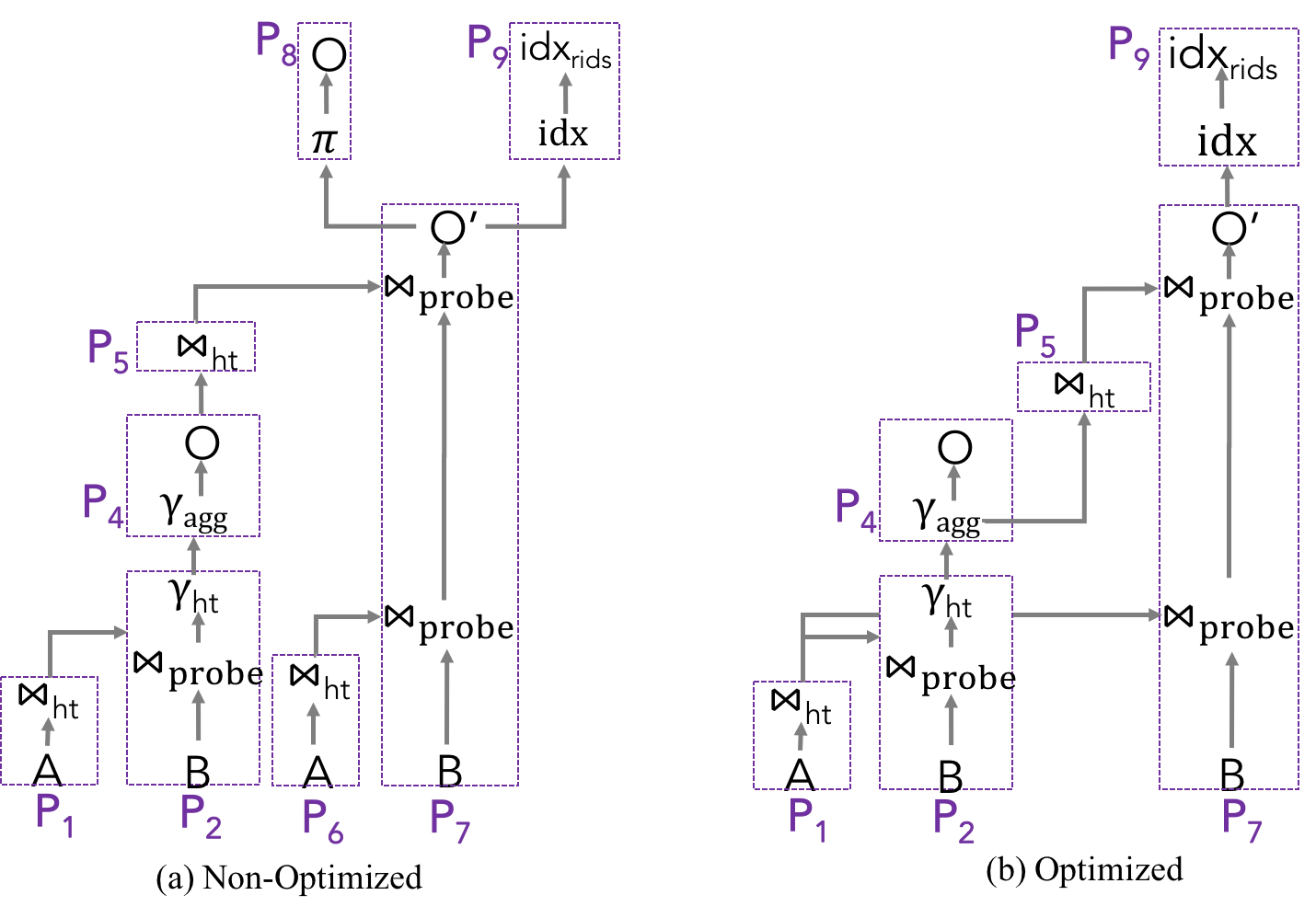}
\caption{\small Physical plans derived for logical approaches for the lineage capture of $\gamma_{a,b,count(1)}(A \bowtie B)$: (a) non-optimized and (b) optimized plan generated by \sys. Boxes \textcolor{magenta}{$P_1$-$P_9$} correspond to individual pipelines per the query compilation model. The index $i$ of a pipeline $P_i$ denotes the order of execution in the compiled plan.}
\label{fig:log_phys_opts}
\end{figure}

To better explain the optimizations that we implemented for logical approaches, consider the following query which has the same structure with the queries that we used for our multi-operator experiments: $\gamma_{a,b,count(1)}(A \bowtie B)$, where $a \in \text{sch}(A)$ and $b \in \text{sch}(B)$; $\text{sch}(X)$ denotes the schema (i.e., set of attributes) of relation $X$. Furthermore, let the join between $A$ and $B$ to be a pk-fk natural join.

\perm's re-rewrite rule for this query results in the following query:  $\rho_{a/a',b/b'}(\gamma_{a,b,count(1)}(\textcolor{red}{A \bowtie B} )) \underset{a=a' \wedge b = b'}{\bowtie} (\textcolor{blue}{A \bowtie B})$; $\rho_{p/p'}$ denotes the relational rename operation of the attribute $p$ to $p'$.~\Cref{fig:log_phys_opts} shows the corresponding non-optimized and optimized physical plans that \sys produces for this query.  Similar non-optimized physical plans, with the same lack of optimizations, are produced by \perm and \gprom. Next, we describe the set of optimizations to derive the optimized plan.   

First, \perm joins the output of the aggregation with the join \textcolor{red}{$A \bowtie B$} on the same attributes that the hash table was built on for the group-by aggregation. This allows us to reuse the hash table to join the output of the join \textcolor{blue}{$A \bowtie B$} with the aggregation result. As such, the built of the hash table at pipeline $P_5$ can be eliminated as we can reuse directly the hash table that was built during pipeline $P_2$. Second, consider the materialization of the output at pipeline $P_8$ and the materialization of the same output at $P_7$. This is the exact output of the base query. $P_8$ is more costly than $P_7$ due to the projection on a, possibly, large $O'$ relation. Hence, we can eliminate $P_8$ and provide the output result $O$ by executing $P_4$. Finally, note that the two natural joins between the tables A and B are the same. Hence, if we materialize the join output there is no need to perform again the join. In our experiments we found that joins are costly to materialize. As an alternative we re-used the hash table built for the inner table A and probed only the outer table B which resulted in the best performance in our experiments with the TPC-H queries. In general, for all n-way joins we materialized the last hash table and probed only the rightmost table in the left-deep plans that \sys produces. 

Note that the set of optimizations that we presented above can be enabled by any DBMS. This means that there is no intrinsic problem with the re-write rules of \perm and \gprom, beyond the ones that we described in~\Cref{s:bg} and experimentally showed in our experiments. Rather, DBMSs, on top of which logical systems are built, need to perform low-level optimizations of physical plans. This is why we did not compare against systems that do not enable these optimizations and directly optimized logical alternatives within \sys. 

\stitle{Tuning \sys techniques.} In~\Cref{ss:instr_multi} we noted that \sys specifically optimizes SPJA queries with pk-fk joins. Note that the set of optimizations presented above are also enabled by \sys for \sysd. The  difference is that \sysd does not need to materialize $O'$ as in $P_7$. Instead, it combines together $P_7$ and $P_9$ to build directly lineage indexes and always outperforms the logical approaches even with the optimizations because it avoids storing the denormalized representation that logical approaches perform. Furthermore, in our experiments we discussed that \sysi oupterforms \sysd in TPC-H queries that we experimented with. For these TPC-H queries, \sysi results in annotating intermediate hash-tables with rids as part of lineage capture for joins and the final aggregation operator uses these rids to perform the \typei method. Hence, \sysi avoids the expensive join of \sysd and outperforms it. 

\sstitle{Tuning physical techniques.} Similarly to logical techniques, we have also implemented \phvfdp and \phbdb within \sys. More specifically, for a compiled query plan as generated by \sysn, we instrument it with virtual function calls to emit lineage (i.e., <output, input> rids). Then, assume that we want to capture backward lineage. For one-to-many relations between output and input, \phvfdp probes a hash table on the output rid. Each entry in the hash table keeps a pointer to an rid index that we use to append the input rid. For one-to-one relations, we use an rid list where we append the input rid. In contrast, \phbdb, adds the <output, input> rids in BerkeleyDB with the key being the output rid. (For forward lineage the process is the same but we probe on the input rids and we append output rids.) For our experiments, we used BerkeleyDB 12.1 that we instructed to (a) be in-memory with cache size 0.5GiB (which is sufficient to avoid spooling to disk) and (b) use B-Tree as its internal indexing structure. Essentially, both approaches are similar to \sysi but \phvfdp implements the capture logic of \sysi in a virtual function instead of having it inline in the compiled plan while \phbdb uses the B-tree indexes of BerkeleyDB instead of the lineage indexes of \sys.

\section{Variants of TPC-H Q1}
\label{s:variants}
In~\Cref{ss:exp:opts} we evaluated the optimizations of \sys on three variants of TPC-H $Q1$, namely, $Q1_a$; $Q1_b$; and $Q1_c$, and compared them with lazy lineage query approaches. Here, we give a more detailed description of these queries and their lazy alternatives. First, we give a brief background on lazy alternatives~\cite{cui2001lineage,ikedathesis}. 

\sstitle{Lazy lineage capture.} Consider a base group-by aggregation query $O=\gamma_{g_1,\ldots, g_n, F}(I)$, where $g_1 \ldots g_n$ are group-by attributes and $F$ is the set of aggregation functions. A backward lineage query $L_B(o \in O, I)$ for a given output $o \in O$ is equivalent to $O_{\text{lazy}}=\sigma_{o.g_1=I.g_1 \wedge \ldots \wedge o.g_n=I.g_n}(I)$.  Furthermore, if the base query $O$ includes selections, these selections need to be added in the selection clause of $O_{\text{lazy}}$. Essentially, instead of accessing lineage through indexes as in the eager approaches of \sys, the lazy approaches access lineage with selection scans on base relations. Similar are the equivalence rules for backward (or forward) lineage queries on top of joins, selections, projections, or general workflows~\cite{cui2001lineage,ikedathesis}.

Now, recall that TPC-H $Q1$ is specified as shown below (\sys's hash-based evaluation precludes sorting and \texttt{ORDER BY} clauses are omitted): 
{\small
\begin{lstlisting}[
	language = SQL,
	showspaces=false,
	basicstyle=\ttfamily,
	commentstyle=\color{gray},
	mathescape=true,
	numbers=none,
	frame = none,
    escapeinside={<}{>},
    label={dl:lsql1},
    upquote=true
 		]
Q1 = SELECT
  l_returnflag, l_linestatus,
  sum(l_quantity) as sum_qty,
  sum(l_extendedprice) as sum_base_price,
  sum(l_extendedprice*(1-l_discount))
                             as sum_disc_price,
  sum(l_extendedprice*(1-l_discount)*(1+l_tax))
                                 as sum_charge,
  avg(l_quantity) as avg_qty,
  avg(l_extendedprice) as avg_price,
  avg(l_discount) as avg_disc,
  count(*) as count_order
FROM lineitem
WHERE l_shipdate <<> '1998-12-01'
GROUP BY
  l_returnflag,
  l_linestatus
\end{lstlisting}
}
\noindent The output of $Q1$ is four groups derived from combinations of l\_returnflag and l\_linestatus (i.e., $\{(A,F), (N,O), (R,F), (N,F)\}$).

$Q1_a$ drills down from $Q1$ on the year and month of l\_shipdate given an output group of $Q1$. Using with the backward lineage query construct this operation can be specified as follows:
{\small
\begin{lstlisting}[
	language = SQL,
	showspaces=false,
	basicstyle=\ttfamily,
	commentstyle=\color{gray},
	mathescape=true,
	numbers=none,
	frame = none,
    escapeinside={<}{>},
    label={dl:lsql1}
 		]
Q1<$_{a}$> = SELECT ..., 
   <\textcolor{purple}{extract(year from l\_shipdate),}>
   <\textcolor{purple}{extract(month from l\_shipdate)}>
FROM <\textcolor{purple}{$L_b(O_i \subseteq \text{Q1}$ ,lineitem)}>
GROUP BY
  <\textcolor{purple}{extract(year from l\_shipdate),}>
  <\textcolor{purple}{extract(month from l\_shipdate)}>
\end{lstlisting}
}
\noindent For the eager approach of \sys, $Q1_a$ is evaluated using only the lineage indexes that we construct during the execution of $Q1$. As a result, we do not need to add the selections specified in $Q1$ in $Q1_a$ because the backward lineage will retrieve tuples of lineitem that satisfy these selections. 

Now, regarding the lazy approach to evaluate $Q1_a$, recall that a backward lineage for a group-by aggregation query $O=\gamma_{g_1,\ldots, g_n, F}(I)$ can be specified lazily as $\sigma_{o.g_1=I.g_1 \wedge \ldots \wedge o.g_n=I.g_n}(I)$. By applying this rule on $Q1_a$ we derive its lazy alternative: 
{\small
\begin{lstlisting}[
	language = SQL,
	showspaces=false,
	basicstyle=\ttfamily,
	commentstyle=\color{gray},
	mathescape=true,
	numbers=none,
	frame = none,
    escapeinside={<}{>},
    label={dl:lsql1}
    caption={Q1a for lazy evaluation}
 		]
Q1<$_{a-lazy}$> = 
   SELECT ..., 
     <\textcolor{purple}{extract(year from l\_shipdate),}>
     <\textcolor{purple}{extract(month from l\_shipdate)}>
   FROM lineitem
   WHERE 
     l_shipdate <<> '1998-12-01' and
     <\textcolor{purple}{l\_linestatus = ? and}>
     <\textcolor{purple}{l\_returnflag = ?}>
   GROUP BY
     <\textcolor{purple}{extract(year from l\_shipdate),}>
     <\textcolor{purple}{extract(month from l\_shipdate)}>
\end{lstlisting}
}
\noindent Given an output group of $Q1$ we can parameterize the selections above and evaluate $Q1_a$ using a selection scan on lineitem (as opposed to the indexed scan using our \sys techniques). Finally, note that in our experiments we considered backward lineage queries to be specified for only one output group of $Q1$. This avoids disjunctive selections on \texttt{l\_linestatus} and \texttt{l\_shipdate} (e.g., change the single selection \texttt{l\_linestatus = ?} above to \texttt{l\_linestatus IN ($\ldots$)}), since we only need to backward trace to tuples that contribute only to one group. However, in the general case, lazy approaches need to consider these expensive disjunctions. In contrast, eager approaches have access to tuples of each group through the lineage indexes and do not require to re-calculate groups with expensive selections.

Following the logic of $Q1_a$ above we similarly derived the eager and lazy alternatives of $Q1_b$ and $Q1_c$. For completeness, we list them below without further discussion.
{\small
\begin{lstlisting}[
	language = SQL,
	showspaces=false,
	basicstyle=\ttfamily,
	commentstyle=\color{gray},
	mathescape=true,
	numbers=none,
	frame = none,
    escapeinside={<}{>},
    label={dl:lsql1}
 		]
Q1<$_{b}$> <$=$> 
   SELECT ... FROM <$\text{L}_\text{b}(\text{Q1}' \subseteq \text{Q1}$,lineitem)>
   WHERE l_shipinstruct = ? and
         l_shipmode = ?
   GROUP BY ... 
\end{lstlisting}
}
{\small
\begin{lstlisting}[
	language = SQL,
	showspaces=false,
	basicstyle=\ttfamily,
	commentstyle=\color{gray},
	mathescape=true,
	numbers=none,
	frame = none,
    escapeinside={<}{>},
    label={dl:lsql1}
 		]
Q1<$_{b-lazy}$> <$=$>
   SELECT ... FROM lineitem
   WHERE l_shipdate <<> '1998-12-01'
         l_shipinstruct = ? and l_shipmode = ? and
         l_linestatus = ? and l_returnflag = ?
   GROUP BY ... 
\end{lstlisting}
}
{\small
\begin{lstlisting}[
	language = SQL,
	showspaces=false,
	basicstyle=\ttfamily,
	commentstyle=\color{gray},
	mathescape=true,
	numbers=none,
	frame = none,
    escapeinside={<}{>},
    label={dl:lsql1}
 		]
Q1<$_{c}$> <$=$> 
   SELECT ... FROM <$\text{L}_\text{b}(\text{$Q1_b$}' \subseteq \text{$Q1_b$}$,lineitem)>
   GROUP BY ..., l_tax
\end{lstlisting}
}
{\small
\begin{lstlisting}[
	language = SQL,
	showspaces=false,
	basicstyle=\ttfamily,
	commentstyle=\color{gray},
	mathescape=true,
	numbers=none,
	frame = none,
    escapeinside={<}{>},
    label={dl:lsql1}
 		]
Q1<$_{c-lazy}$> <$=$> 
   SELECT ... FROM lineitem
   WHERE l_shipdate <<> '1998-12-01' and 
         l_shipinstruct = ? and l_shipmode = ? and
         l_linestatus = ? and l_returnflag = ? and
         extract(year from l_shipdate) = ?
         extract(month from l_shipdate) = ?
    GROUP BY ..., l_tax
\end{lstlisting}
}

\section{Crossfiltering using lineage}
\label{ss:xval}

In~\Cref{sss:exp:xfilter}, we introduced two techniques (i.e., \btxfilter and \btftxfilter) for crossfiltering using lineage that we compared with the \lazyxfilter approach. In this section, we give a more detailed technical discussion.~\Cref{fig:xfilter_eval} shows the proposed physical plans (i.e., \btxfilter and \btftxfilter) for crossfiltering alongside the naive approach (i.e., \lazyxfilter) and drives our discussion. 

Consider a set of queries $\mathcal{Q}=\{\texttt{Q}_x |$ \texttt{Q$_x$ = SELECT G$_x$, F$_x$(J$_x$) FROM T GROUP BY G$_x$}$\}$ for which we seek to support crossfiltering functionality, 

\sstitle{\lazyxfilter.} During the execution of each $Q_x$, \lazyxfilter simply executes the group-by aggregations without capturing lineage. Then, given a selection of a subset of outputs of $Q_{\text{brushed}} \in \mathcal{Q}$ (i.e., $Q'_\text{brushed} \subseteq Q_\text{brushed}$), \lazyxfilter supports crossfiltering by updating each $Q_x \in  \mathcal{Q} \setminus \{Q_{\text{brushed}}\}$ as follows:
{
\begin{lstlisting}[
	language = SQL,
	showspaces=false,
	basicstyle=\ttfamily,
	commentstyle=\color{gray},
	mathescape=true,
	numbers=none,
	frame = none,
    escapeinside={<}{>},
    label={dl:lsql1}
 		]
Q<$_x$'>=SELECT G<$_x$>, F<$_x$>(J<$_x$>)
   FROM T
   WHERE <\textbf{$\bigwedge_{o\in Q'_\text{brushed}} o.G_b = T.G_b$}>
   GROUP BY G<$_x$>
\end{lstlisting}
}
\noindent Essentially, \lazyxfilter supports crossfiltering by performing lazy lineage capture to identify the partitions of the base relation that contributed to the selected outputs. Finally, to evaluate the set of all updated $Q_x'$, \lazyxfilter does not execute each update separately. Rather it uses a shared selection scan of the input relation with the selection being $\bigwedge_{o\in Q'_\text{brushed}} o.G_b = T.G_b$ to avoid multiple, expensive selection scans of the base relation.

\begin{figure}[t]
   \centering
   \includegraphics[width=\columnwidth]{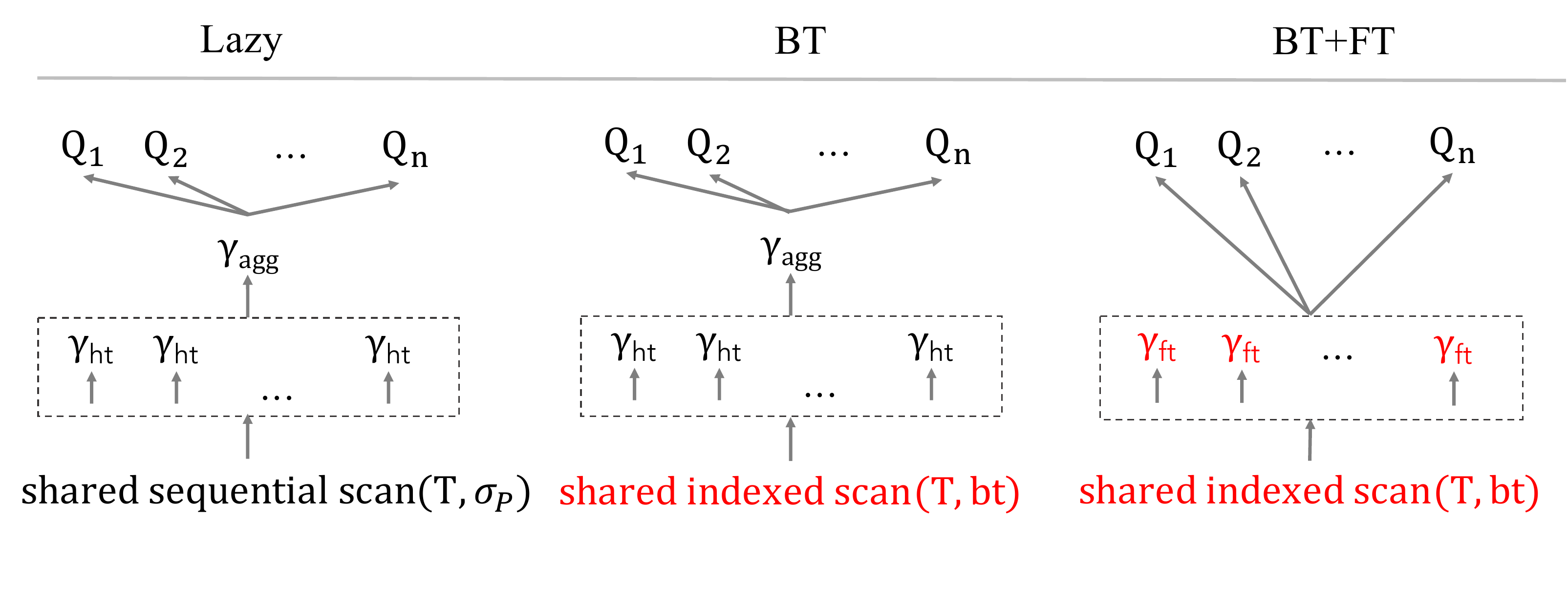}
    \caption{Crossfilter evaluation techniques without using data cubes: (a) Lazy re-evaluates the group-by aggregation queries with a shared selection scan on the base table, (b) \btxfilter uses an index scan on the rids of the backward lineage index of $Q'_{\text{brushed}}$, (c) \btftxfilter performs updates using the fordward indexes that connect each tuple in the base table to each output of aggregation query.}
\label{fig:xfilter_eval}
\end{figure}

During the execution of initial group-by aggregations, \btxfilter and \btftxfilter perform lineage capture in order to to use lineage indexes for crossfiltering.  

\stitle{\btxfilter.} As we noted above, \lazyxfilter supports crossfiltering with lazy lineage capture. This means that $Q_x'$ above is equivalent to the following query.
{
\begin{lstlisting}[
	language = SQL,
	showspaces=false,
	basicstyle=\ttfamily,
	commentstyle=\color{gray},
	mathescape=true,
	numbers=none,
	frame = none,
    escapeinside={<}{>},
    label={dl:lsql1}
 		]
Q<$_x$'>=SELECT G<$_x$>,F<$_x$>(J<$_x$>) 
    FROM backward_trace(<$Q'_{\text{brushed}} \subseteq Q_{\text{brushed}}$>,T)
    GROUP BY G<$_x$>
\end{lstlisting}
}
\noindent Since we have built a backward index bt$_{Q_{\text{brushed}}}$, during the execution of $Q_{\text{brushed}}$, then $Q_x'$ can be evaluated using an indexed-scan on the base table $T$ using only the rid arrays of bt$_{Q'_{\text{brushed}}}$ of bt$_{Q_{\text{brushed}}}$ that correspond to the selected output records $Q'_{\text{brushed}}$. Essentially, with this approach we can avoid the selection scans of \lazyxfilter with indexed scans from lineage. Similarly to \lazyxfilter, \btxfilter uses a shared scan but this time a shared indexed-scan based on bt$_{Q'_{\text{brushed}}}$.

\stitle{\btftxfilter.} Finally, note that \btxfilter still needs to perform group-by aggregations  which is a very costly operation due to the construction of hash tables. Instead, recall the notion of a forward lineage index for a group-by aggregation query: Each tuple in the input is associated with a group in the output. So, forward indexes provide a perfect hashing between tuples in the base table and group-by aggregation results that we have already calculated (i.e., the initial views). Hence, instead of constructing hash tables, \btftxfilter uses the forward indexes as perfect hash tables and performs crossfiltering as follows:
{\small
\begin{lstlisting}[
	language = C++,
	showspaces=false,
	basicstyle=\ttfamily\scriptsize,
	commentstyle=\color{gray},
	mathescape=true,
	numbers=none,
	frame = single,
	escapeinside={<}{>},
	captionpos=b,
	caption={Crossfilter using \btftxfilter.},
	label={xfilter:btft},
	commentstyle=\color{red},
]
Input:  bw[][]       // backward index for <\textcolor{red}{$Q'_{\text{brushed}}$}>
        fw[][]       // forward indexes from each tuple
                     // to groups of each initial
                     // group-by aggregate
        <$Q_1, \ldots , Q_n$>      // outputs of initial views
Output: <$Q'_1, \ldots , Q'_n$>      // crossfiltered views
Init <$Q'_1, \ldots , Q'_n$> using <$Q_1, \ldots , Q_n$>
for i = 0 to bw.size()
  for j = 0 to bw[i].size()
    for z = 0 to n
      agg_update(<$Q'_z$>[fw[<$Q_z$>][bw[i][j]]])
remove_non_affected_groups(<$Q'_1, \ldots , Q'_n$>)
\end{lstlisting}
}
\noindent \texttt{agg\_update()} in the listing above simply updates the aggregation (e.g., for \texttt{COUNT(*)}, \texttt{agg\_update} is simply $Q'_z$\texttt{[fw[}$Q_z$\texttt{][bw[i][j]]]++}). Finally, note the \texttt{remove\_non\_affected\_groups} function at the end of the listing above. This function loops over the groups of each updated group-by and removes the groups that were not affected by the group-by. In the case of \texttt{COUNT(*)} this is simply the groups that have a zero count. For other aggregates, like \texttt{SUM}, we need to track which groups were updated within the \texttt{agg\_update} functions. However, in most interactive visualizations it is more important to maintain the groups even if the have a zero count (or a zero sum). In that case the \texttt{remove\_non\_affected\_groups} can simply by ignored and the \texttt{agg\_update} can perform the update without updating a state of what groups were updated. Our experiments in~\Cref{sss:exp:xfilter} report the latency of~\btftxfilter including the time for this operation.



\section{Lineage Semantics in Smoke}
\label{s:semantics}
In~\Cref{s:bg}, we noted that Smoke uses transformation provenance semantics that can allow us to encode several novel provenance semantics at the will of end-developers. Here, we give a brief discussion on how one can go about encoding new semantics in \sys. 

Consider the following example query and database: \texttt{SELECT COUNT(*), A.cname, B.pname FROM A.cid = B.id GROUP BY A.cname, B.pname}.

\begin{tabular}{l|l|l|}
\cline{2-3}
&\textbf{cid} & \textbf{cname} \\ \cline{2-3}
$a_1$&1 & Bob \\ \cline{2-3}
$a_2$&2 & Alice \\ \cline{2-3}
\end{tabular}
\begin{tabular}{l|l|l|l|l|}
\cline{2-5}
&\textbf{oid} & \textbf{cid} & \textbf{pname} & \textbf{date} \\ \cline{2-5}
$b_1$&1 & 1 & iPhone & 12/25 \\ \cline{2-5}
$b_2$&2 & 1 & iPhone & 12/25\\ \cline{2-5}
$b_3$&3 & 2 & XBox & 12/25 \\ \cline{2-5}
\end{tabular}

\noindent The output of this query is the following:
\begin{center}
\begin{tabular}{l|l|l|l|}
\cline{2-4}
&\textbf{COUNT(*)} & \textbf{A.cname} & \textbf{B.pname} \\ \cline{2-4}
$o_1$&1 & Bob & iPhone\\ \cline{2-4}
$o_2$&2 & Alice & xBox \\ \cline{2-4}
\end{tabular}
\end{center}

According to Smoke's provenance semantics the backward index for $o_1$ with respect to table $A$ contains the tuple rid $a_1$ \emph{twice}. This is important because in this way \sys can encode multiple provenance semantics. The why-provenance of $o_1$ is $\{(a_1, b_1), (a_1, b_2)\}$ and to answer why-provenance queries \sys simply concatenates the backward index rids: rids at the same position in the backward indexes for A and B correspond to the why-provenance witnesses. Now, the which-provenance of $o_1$ is $\{a_1, b_1, b_2\}$  which \sys can derive by performing a set union of the backward indexes. Finally, the how-provenance of $o_1$ is $a_1\cdot(b_1+b_2)$ which \sys can derive by set unioning and concatenating the backward rid indexes, similarly to the operations for which- and why-provenance. 

Now, note that all these operations to derive different provenance semantics are lineage consuming queries whose logic we can push down as workload-aware optimizations similar to the ones in~\Cref{s:optf}. In which case \sys operates as a which-, why-, or how-provenance system for the case illustrated above. However, note that depending on the provenance semiring that how-provenance captures, the lineage consuming logic can be different and expressing semirings as lineage consuming queries is an open question.

These observations, along with the focus of \sys to capture forward lineage, highlight that \sys provides a general architecture for novel provenance semantics.

\begin{figure}[b]
  \centering
  \includegraphics[width=.7\columnwidth]{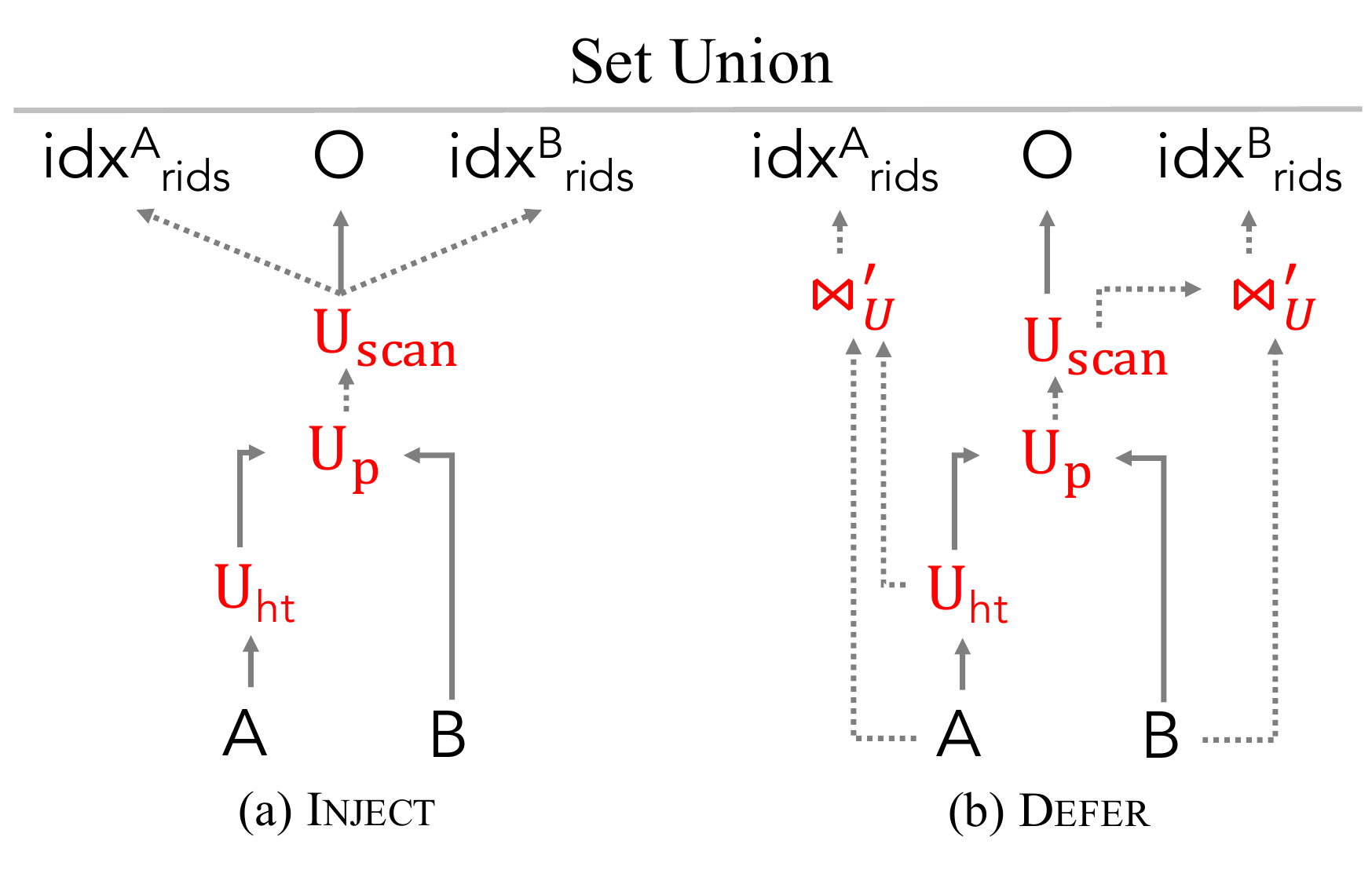}
  \caption{{\small \typei and \typej plans for set union. Dotted arrows are only necessary for lineage capture.}}
  \label{fig:smoke_set_union_inst}
\end{figure}

\section{Instrumentation Algebra}
\label{s:physical_algebra}

In~\Cref{s:instr} we presented the physical algebra of \sys for lineage capture of projection, selections, hash-based group-by aggregations, and hash-based equi-joins. In this section, we extend this algebra for bag and set union, intersection, and difference as well as nested-loop $\theta$-joins and cross products.

\subsection{Set Union}
\label{ss:set_union}

Set union between two relations $A$ and $B$ (i.e., $A \overset{S}{\bigcup}_{\text{uattrs}} B$, where $S$ denotes set union and $\text{uattrs}$ the attributes from $A$ and $B$ to union on) are implemented in a hash-based way with consecutive appends to a hash table: Initially, the operator $\cup_{ht}$ builds a hash table using the relation $A$ with the key being the attributes of the union (i.e., $\text{uattrs}$). Then, $\cup_p$ probes the hash table constructed by $\cup_{ht}$ on the union attributes using relation $B$. If an entry does not already exist for the union attributes, $\cup_p$ appends a new entry in the hash table with the union attributes. Essentially, $\cup_{ht}$ and $\cup_{b}$ are the same operator, that probe and append tuples in a hash table. The only difference is that $\cup_{ht}$ takes as input an empty hash table while $\cup_{p}$ takes as input a pre-built hash table. Finally, $\cup_\text{scan}$ scans the hash table and constructs the output. Next, we discuss \typej and \typei lineage capture approaches for set union;~\Cref{fig:smoke_set_union_inst} shows their corresponding physical plans.

\sstitle{\typei: }~\Cref{dl:set_union_inject} illustrates the \typei lineage capture of \sys for set union. Similarly to group-by aggregation, \typei rewrites $\cup_{ht}$ to append, besides the union attributes, two arrays \texttt{a\_rids} and \texttt{b\_rids} that track which tuples from $A$ and $B$, respectively, contribute to the hash table entry. During $\cup_{ht}$ we populate \texttt{a\_rids} and during $\cup_{p}$ we populate \texttt{b\_rids}. Finally, $\cup_\text{scan}$ outputs the result and the lineage indexes.

\sstitle{\typej: }~\Cref{dl:set_union_defer} illustrates the \typej lineage capture of \sys for set union. Similarly to group-by aggregation, \typej rewrites $\cup_{ht}$ and $\cup_{p}$ to append an \texttt{oid} to each hash table entry, initially set to $-1$, besides the union attributes. Then, $\cup_\text{scan}$ outputs the set union result and assigns the correct oid to each hash table entry. 
{\small
\begin{lstlisting}[
	language = C++,
	showspaces=false,
	basicstyle=\ttfamily\scriptsize,
	commentstyle=\color{gray},
	mathescape=true,
	numbers=none,
	frame = single,
	escapeinside={<}{>},
	captionpos=b,	
	caption={\typei lineage capture for set union $A \overset{S}{\bigcup}_{\text{uattrs}} B$.},
	label={dl:set_union_inject},
	commentstyle=\color{red},
	float=tp,
	floatplacement=tbp,
	belowskip=-1em
]
Input:  A, B
Output: O,                 
        a_fw[A.size()], b_fw[B.size()] // forward indexes
        a_bw[][], b_bw[][]            // backward indexes
Hash Table ht, Hash Function hash
for i = 0 to A.size()  // <\red{$\cup_{ht}$}>: Build phase
  h = hash(A[i].uattrs)
  if(!ht[h]) ht[h]={init_state(A[i].uattrs), 
                    <\texttt{\blue{a\_rids=[], b\_rids=[]}}>}
  <\texttt{\blue{ht[h].a\_rids.insert(i)}}>

for i = 0 to B.size()   // <\red{$\cup_{p}$}>: Probe/Append phase
  h = hash(B[i].uattrs)
  if(!ht[h]) ht[h]={init_state(B[i].uattrs), 
                    <\texttt{\blue{, a\_rids=[], b\_rids=[]}}>}
  <\texttt{\blue{ht[h].b\_rids.insert(i)}}>

oid = -1
a_bw = int[ht.size()][]
b_bw = int[ht.size()][] 
for (state, a_rids, b_rids) in ht   // <\red{$\cup_{\text{scan}}$}>: Scan phase
  O[++oid] = create_output_record(state)
  <\texttt{\blue{a\_bw[oid] = a\_rids}}>
  <\texttt{\blue{for rid in a\_rids}}>
    <\texttt{\blue{a\_fw[rid] = oid}}>
  <\texttt{\blue{b\_bw[oid] = b\_rids}}>
  <\texttt{\blue{for rid in b\_rids}}>
    <\texttt{\blue{b\_fw[rid] = oid}}>
\end{lstlisting}
}
\noindent  To construct the lineage indexes $\bowtie^{'}_{\cup}$ takes as input the relation $A$ and probes the previously constructed hash table to find the \texttt{oid} and properly construct the lineage indexes between the output and input relation $A$. 
\vspace{1em}
{
\begin{lstlisting}[
	language = C++,
	showspaces=false,
	basicstyle=\ttfamily\scriptsize,
	commentstyle=\color{gray},
	mathescape=true,
	numbers=none,
	frame = single,
	escapeinside={<}{>},
	captionpos=b,	
	caption={\typej lineage capture for set union $A \overset{S}{\bigcup}_{\text{uattrs}} B$.},
	label={dl:set_union_defer},
	commentstyle=\color{red},
]
Input:  A, B
Output: O,                 
        a_fw[A.size()], b_fw[B.size()] // forward indexes
        a_bw[][], b_bw[][]            // backward indexes
Hash Table ht, Hash Function hash
for i = 0 to A.size()  // <\red{$\cup_{ht}$}>: Build phase
  h = hash(A[i].uattrs)
  if(!ht[h]) ht[h]={init_state(A[i].uattrs), 
                    <\texttt{\blue{oid=-1}}>}

for i = 0 to B.size()   // <\red{$\cup_{p}$}>: Probe/Append phase
  h = hash(B[i].uattrs)
  if(!ht[h]) ht[h]={init_state(B[i].uattrs), 
                    <\texttt{\blue{, oid=-1}}>}

oid = -1
a_bw = int[ht.size()][]
b_bw = int[ht.size()][] 
for h in ht   // <\red{$\cup_{\text{scan}}$}>: Scan phase
  O[++oid] = create_output_record(h.state)
  h.oid = oid
<\textcolor{MidnightBlue}{for i=0 to A.size()}> // <\red{$\bowtie'_{\cup}$}>: Lineage capture for <\red{$A$}>
  <\textcolor{MidnightBlue}{h = hash(A[i].uattrs)}>
  <\textcolor{MidnightBlue}{a\_bw[ht[h].oid].insert(i)}>
  <\textcolor{MidnightBlue}{a\_fw[i] = ht[h].oid}>
<\textcolor{MidnightBlue}{for i=0 to B.size()}> // <\red{$\bowtie'_{\cup}$}>: Lineage capture for <\red{$B$}>
  <\textcolor{MidnightBlue}{h = hash(B[i].uattrs)}>
  <\textcolor{MidnightBlue}{a\_bw[ht[h].oid].insert(i)}>
  <\textcolor{MidnightBlue}{a\_fw[i] = ht[h].oid}>
\end{lstlisting}
}
\noindent Similar is the process for $\bowtie^{'}_{\cup}$ when taking as input the $B$ to construct lineage indexes between the output and relation $B$.
 
\sstitle{Further optimizations:} An optimization, for both \typei and \typej approaches, is that there is no need to wait to append the right relation $B$ to the hash table to construct the lineage indexes for the relation $A$. This is because the intermediate hash table built for $A$ suffices for the lineage index construction for $A$. For \typej, in particular, this also means that the join $\bowtie_{U}$ for $A$ will not need to probe a hash table that keeps not all entries for $A$ but also $B$. However, this also means that \typej needs to block the output construction until after the $\bowtie_{U}$ for $A$ has been executed, which is a counter-argument to the \typej paradigm (i.e., lineage is constructed without blocking the query execution). To balance this effect we could keep a copy of the intermediate hash table for $A$ and use only that for lineage construction for the $A$ relation at the cost of copying which could be substantial. \sys does not yet support the copy construction but it does support blocking the set union for the lineage construction.

\subsection{Bag Union }
\label{ss:bag_union}
Lineage capture for bag union is simpler than lineage capture for set union. Since for bag union we only concatenate the two input relations, what we only need to maintain is the $rid$ of where one relation ends and the other relation begins in the output of the union. More generally, for bag union of $k$ relations we need $k-1$ such rids. Using these indexes it is sufficient to answer both backward and forward lineage queries. Note, however that this lineage capture relies on the fact that the input relation is a base relation stored in the database. For multi-operator plans the input to the union could be an intermediate relation for which we need to perform lineage capture. For instance, for a query $\sigma_{\theta}(A) \bigcup B$, we need to perform lineage capture for the selection on $A$.

{\small
\begin{lstlisting}[
	language = C++,
	showspaces=false,
	basicstyle=\ttfamily\scriptsize,
	commentstyle=\color{gray},
	mathescape=true,
	numbers=none,
	frame = single,
	escapeinside={<}{>},
	captionpos=b,	
	caption={{\small \typei lineage capture for set intersection $A \overset{S}{\bigcap}_{\text{iattrs}} B$.}},
	label={dl:set_intersect_inject},
	commentstyle=\color{red},
	float=tp,
	floatplacement=tbp
]
Input:  A, B
Output: O,                 
        a_fw[A.size()], b_fw[B.size()] // forward indexes
        a_bw[][], b_bw[][]            // backward indexes
Hash Table ht, Hash Function hash
for i = 0 to A.size()  // <\red{$\cap_{ht}$}>: Build phase
  h = hash(A[i].iattrs)
  if(!ht[h]) ht[h]={init_state(A[i].iattrs), 
                    <\texttt{\blue{a\_rids=[], b\_rids=[]}}>}
  <\texttt{\blue{ht[h].a\_rids.insert(i)}}>

for i = 0 to B.size()   // <\red{$\cap_{p}$}>: Probe phase
  h = hash(B[i].iattrs)
  if(ht[h]) <\texttt{\blue{ht[h].b\_rids.insert(i)}}>

oid = -1
a_bw = int[ht.size()][]
b_bw = int[ht.size()][] 
for (state, a_rids, b_rids) in ht   // <\red{$\cap_{\text{scan}}$}>: Scan phase
  if(b_rids.size()==0) continue;  
  O[++oid] = create_output_record(state)
  <\texttt{\blue{a\_bw[oid] = a\_rids}}>
  <\texttt{\blue{for rid in a\_rids}}>
    <\texttt{\blue{a\_fw[rid] = oid}}>
  <\texttt{\blue{b\_bw[oid] = b\_rids}}>
  <\texttt{\blue{for rid in b\_rids}}>
    <\texttt{\blue{b\_fw[rid] = oid}}>
\end{lstlisting}
}

\pagebreak

\subsection{Set Intersection}
\label{ss:set_intersect}

Set intersection in \sys is broken into three operators. First, $\cap_{ht}$ builds a hash table on the outer relation $A$ with the key being the attributes of the intersection. Each hash table entry, beyond the intersection attributes, also maintains a bit to indicate whether or not it has been matched with a tuple from the inner relation $B$. Then, $\cap_{p}$ probes the hash table and sets the bit if a match was found. Finally, $\cap_{\text{scan}}$ scans the hash table and emits the entries with the bit set to form the output. 

Linage capture for set intersection (see~\Cref{fig:smoke_set_intersect_inst}) follows the logic of set union (see~\Cref{fig:smoke_set_union_inst}). An important difference is that for the \typei approach,  \texttt{a\_rids} that we have kept for non-matched tuples will be discarded. If the fraction of tuples in the outer relation that appear in the intersection is small that could result in the \typej approach to be faster than \typei because it avoids the unnecessary writes in \texttt{a\_rids}. Also, a slight difference from set intersection without lineage capture, is that  \typei does not require a bit indicating whether a hash table entry has been matched with tuples from the outer relation because we maintain \texttt{b\_rids} that provide this information. For completeness, Listings~\ref{dl:set_intersect_inject} and~\ref{dl:set_intersect_defer} show code snippets for \typei and \typej, respectively.

{
\begin{lstlisting}[
	language = C++,
	showspaces=false,
	basicstyle=\ttfamily\scriptsize,
	commentstyle=\color{gray},
	mathescape=true,
	numbers=none,
	frame = single,
	escapeinside={<}{>},
	captionpos=b,	
	caption={{\small \typej lineage capture for set intersection $A \overset{S}{\bigcap}_{\text{iattrs}} B$.}},
	label={dl:set_intersect_defer},
	commentstyle=\color{red},
	float=tp,
	floatplacement=tbp
]
Input:  A, B
Output: O,                 
        a_fw[A.size()], b_fw[B.size()] // forward indexes
        a_bw[][], b_bw[][]            // backward indexes
Hash Table ht, Hash Function hash
for i = 0 to A.size()  // <\red{$\cap_{ht}$}>: Build phase
  h = hash(A[i].iattrs)
  if(!ht[h]) ht[h]={init_state(A[i].iattrs), b_bit = 0
                    <\texttt{\blue{oid=-1}}>}

for i = 0 to B.size()   // <\red{$\cap_{p}$}>: Probe/Append phase
  h = hash(B[i].iattrs)
  if(ht[h]) ht[h].b_bit=1

oid = -1
a_bw = int[ht.size()][]
b_bw = int[ht.size()][] 
for h in ht   // <\red{$\cap_{\text{scan}}$}>: Scan phase
  O[++oid] = create_output_record(h.state)
  h.oid = oid
<\textcolor{MidnightBlue}{for i=0 to A.size()}> // <\red{$\bowtie'_{\cap}$}>: Lineage capture for <\red{$A$}>
  <\textcolor{MidnightBlue}{h = hash(A[i].iattrs)}>
  <\textcolor{MidnightBlue}{if(!h.b\_bit) continue}>
  <\textcolor{MidnightBlue}{a\_bw[ht[h].oid].insert(i)}>
  <\textcolor{MidnightBlue}{a\_fw[i] = ht[h].oid}>
<\textcolor{MidnightBlue}{for i=0 to B.size()}> // <\red{$\bowtie'_{\cap}$}>: Lineage capture for <\red{$B$}>
  <\textcolor{MidnightBlue}{h = hash(B[i].iattrs)}>
  <\textcolor{MidnightBlue}{if(!h) continue}>
  <\textcolor{MidnightBlue}{a\_bw[ht[h].oid].insert(i)}>
  <\textcolor{MidnightBlue}{a\_fw[i] = ht[h].oid}>
\end{lstlisting}
}

\vspace{1em}

\begin{figure}[t]
  \centering
  \includegraphics[width=.69\columnwidth]{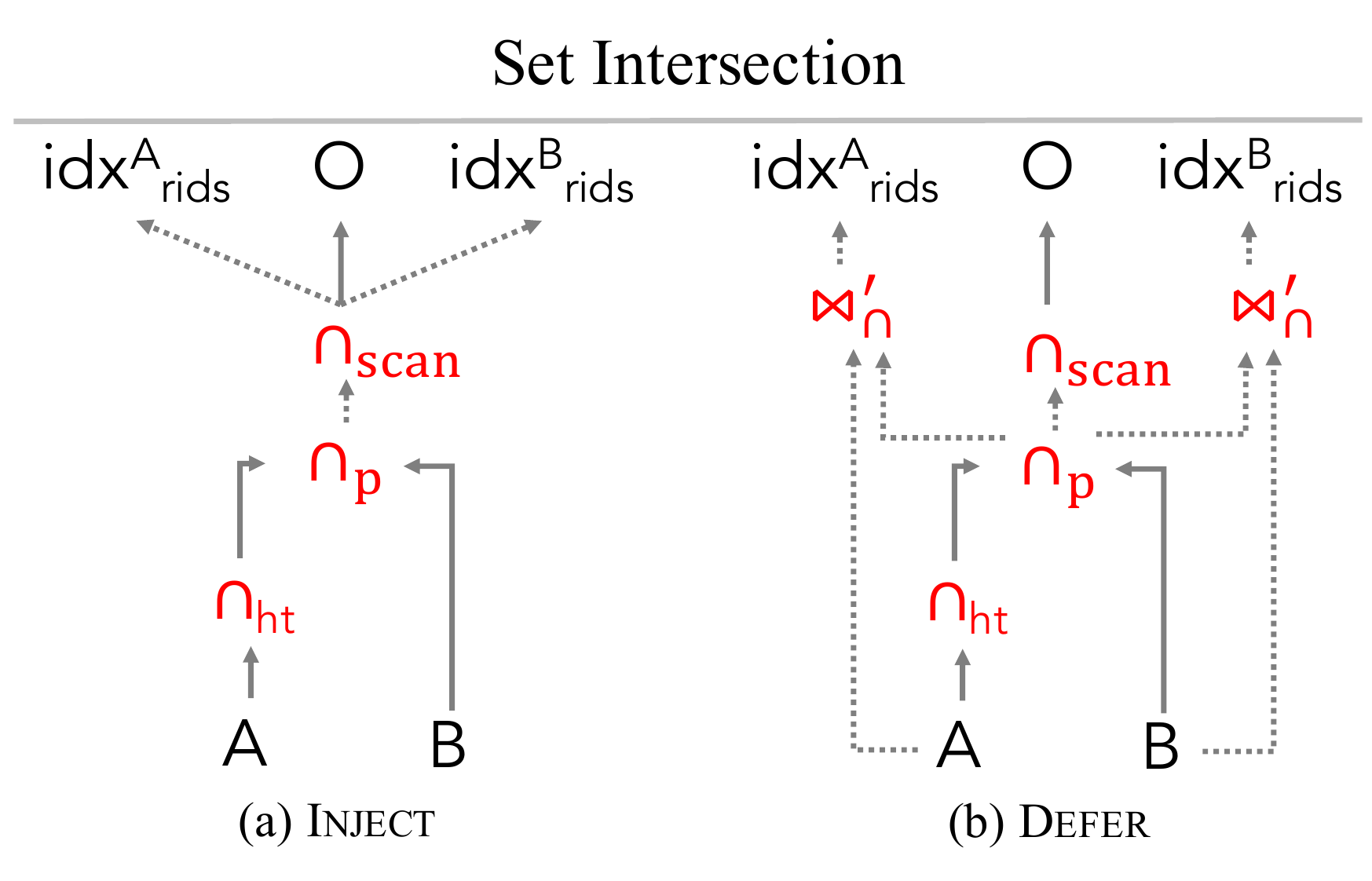}
  \caption{{\small \typei and \typej plans for set intersection. Dotted arrows are only necessary for lineage capture.}}
  \label{fig:smoke_set_intersect_inst}
\end{figure}

\subsection{Bag Intersection}
\label{ss:bag_intersect}

Bag intersection in Smoke follows the same logic as the set intersection. The only difference is that the hash table needs to maintain two more attributes per entry: (a) the number of tuples from the outter relation that are duplicates according to the intersection attributes, and (b) the number of matches with the inner relation. $\cap_{ht}$ adds a hash entry \texttt{$\{$A[i].iattrs, a\_matches=1, b\_matches=0$\}$} if there is no prior entry in the hash table for \texttt{A[i].iattrs}, or updates the matches of $A$ (i.e., \texttt{a\_matches++}) if there was an entry for \texttt{A[i].iattrs}. Then, $\cap_{p}$ probes the hash tables with the tuples from the inner relation $B$ and updates the \texttt{b\_matches}. Finally, $\cap_{scan}$ scans the hash table and outputs each entry \texttt{a\_matches}$\cdot$\texttt{b\_matches} times to provide an output with the correct bag intersection semantics.

\stitle{\typei: } Lineage capture for bag intersection under \typei semantics is straightforward. Instead of keeping \texttt{a\_matches} and \texttt{b\_matches} we maintain two arrays of rids (\texttt{a\_rids} and \texttt{b\_rids}) from where the 
matches have originated. As such, \texttt{a\_matches} = \texttt{a\_rids.size()} and 
\texttt{b\_matches=b\_rids.size()}. Hence, $\cap_{scan}$ can still provide an output with the correct bag intersection semantics. Moreover, $\cap_{scan}$ can provide backward and forward indexes using these rids. Note, however, that while set intersection has 1-to-N backward lineage, bag intersection has 1-to-1.

\stitle{\typej: } Lineage capture for bag intersection under \typej follows the logic of \typej for set intersection. Besides \texttt{a\_matches} and \texttt{b\_matches},
each hash entry maintains an output rid \texttt{oid} of the first tuple in the output for this hash entry. Note that the output will contain tuples related to this hash entry at rids  [\texttt{oid}, \texttt{oid}+\texttt{a\_matches}$\cdot$\texttt{b\_matches}]. Now, the trick is that $\bowtie_{\cap}'$ need to happen in order first with the $A$ relation and then with $B$, and for every match we should increase the \texttt{oid}. For completeness,~\Cref{dl:bag_intersect_defer} provides the corresponding code snippet for \typej.

{
\begin{lstlisting}[
	language = C++,
	showspaces=false,
	basicstyle=\ttfamily\scriptsize,
	commentstyle=\color{gray},
	mathescape=true,
	numbers=none,
	frame = single,
	escapeinside={<}{>},
	captionpos=b,	
	caption={{\small \typej lineage capture for bag intersection $A \overset{B}{\bigcap}_{\text{iattrs}} B$.}},
	label={dl:bag_intersect_defer},
	commentstyle=\color{red},
	belowskip=-1em
]
Input:  A, B
Output: O,                 
        a_fw[A.size()], b_fw[B.size()] // forward indexes
        a_bw[][], b_bw[][]            // backward indexes
Hash Table ht, Hash Function hash
for i = 0 to A.size()  // <\red{$\cap_{ht}$}>: Build phase
  h = hash(A[i].iattrs)
  if(!ht[h]) ht[h]={init_state(A[i].iattrs),
                     a_matches=1, b_matches=0,
                    <\texttt{\blue{oid=-1}}>}
  else ht[h].a_matches++

cnt=0
for i = 0 to B.size()   // <\red{$\cap_{p}$}>: Probe/Append phase
  h = hash(B[i].iattrs)
  if(ht[h])
    ht[h].b_matches++
    cnt+= a_matches 
  
oid = -1
a_bw = int[cnt][]
b_bw = int[cnt][] 
for h in ht   // <\red{$\cap_{\text{scan}}$}>: Scan phase
  O[++oid] = create_output_record(h.state)
  h.oid = oid
<\textcolor{MidnightBlue}{for i=0 to A.size()}> // <\red{$\bowtie'_{\cap}$}>: Lineage capture for <\red{$A$}>
  <\textcolor{MidnightBlue}{h = hash(A[i].iattrs)}>
  <\textcolor{MidnightBlue}{if(!h.b\_matches) continue}>
  <\textcolor{MidnightBlue}{a\_bw[ht[h].oid] = i}>
  <\textcolor{MidnightBlue}{a\_fw[i] = ht[h].oid++}>
<\textcolor{MidnightBlue}{for i=0 to B.size()}> // <\red{$\bowtie'_{\cap}$}>: Lineage capture for <\red{$B$}>
  <\textcolor{MidnightBlue}{h = hash(B[i].iattrs)}>
  <\textcolor{MidnightBlue}{if(!h) continue}>
  <\textcolor{MidnightBlue}{a\_bw[ht[h].oid] =i}>
  <\textcolor{MidnightBlue}{a\_fw[i] = ht[h].oid++}>
\end{lstlisting}
}

\begin{figure}[t]
  \centering
  \includegraphics[width=.7\columnwidth]{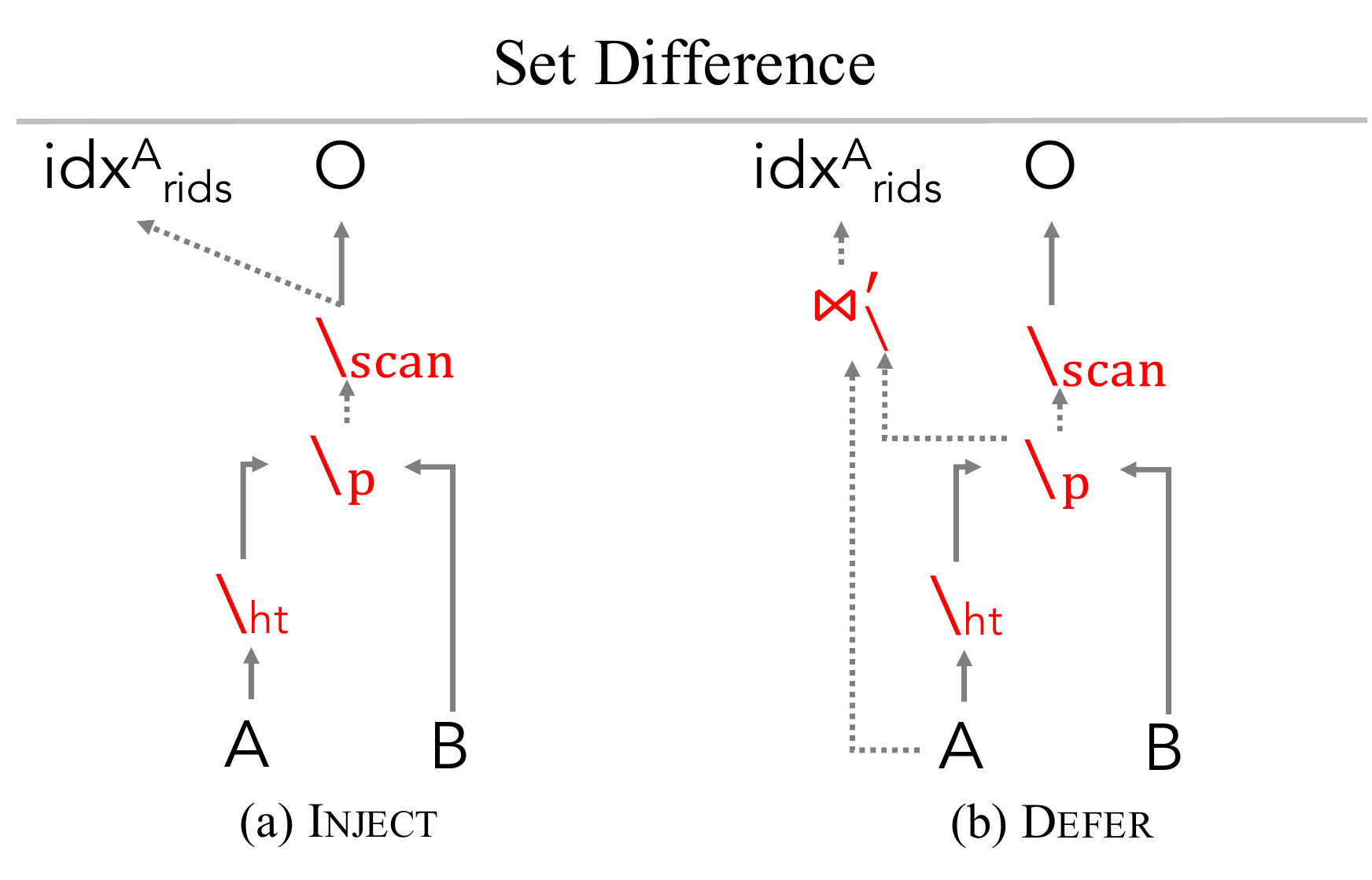}
  \caption{{\small \typei and \typej plans for set difference.  \red{Red} operators are lineage-aware. Dotted arrows are only necessary for lineage capture.}}
  \label{fig:smoke_set_difference_inst}
\end{figure}

\subsection{Set and bag difference}
\label{ss:set_diff}

\sys implements set difference of two relations $A$ and $B$ (i.e., $A \overset{B/S}{\setminus_{dattrs}} B$) in a hash-based way similar to set intersection. The only differences are (a) we set the \texttt{b\_bit} of each hash entry to 1 instead of 0 during the initial build and (b) when we probe the hash table with the inner relation we set the \texttt{b\_bit} to 0 as opposed to 1. The final scan outputs only the hash entries with \texttt{b\_bit}=1 as these are the tuples that appear in the inner relation but do not appear in the outter relation.

Efficient lineage capture for set difference is non-trivial. By definition, the lineage for a tuple $o \in A \setminus B$ depends on (a) the set of tuples in $A$ that it came from and (b) the whole inner relation $B$. Capturing forward indexes for the $A$ tuples follows the lineage capture logic of set intersection and we omit further details. The problem with set difference is that each output depends on the whole outer relation $B$. Our experimental results show that lineage capture is meaningful when lineage has small cardinality. As such, if $B$ is a base relation we do not capture lineage and for backward queries that require access to $B$ we simply scan $B$. Now, if the input relation is an intermediate relation, then \sys performs lineage capture during the execution of the operator whose output is the intermediate relation that is the outer relation to the set difference. Hence, for a backward query on the set difference we can access a base relation that is used to construct the intermediate relation through the backward index of the intermediate relation. More interestingly, a forward query from a tuple of a base relation, that is used in the construction of the intermediate relation that is input to the set difference, is the whole output \emph{times} the amount of tuples it contributes to the intermediate relation. This is because each tuple in the intermediate relation contributes to all the tuples in the output of the set difference. 

As such, \sys captures lineage only for the $A$ relation that follows the logic of lineage capture for the inner relation of set intersection. For completeness,~\Cref{fig:smoke_set_difference_inst} illustrates the corresponding \typei and \typej physical plans.

\subsection{$\theta$-joins and Nested Loops}
\label{ss:nlj}

So far, we have proposed a physical algebra for hash-based implementations of equi-joins, group-by aggregations, unions, intersections, and differences. In this section we give a brief discussion for \typei lineage capture of nested-loop based implementations for $\theta$-joins. Lineage capture with merge-sort approaches and lineage capture based on nested loops for the rest operators are obvious future work.

{
\begin{lstlisting}[
	language = C++,
	showspaces=false,
	basicstyle=\ttfamily\scriptsize,
	commentstyle=\color{gray},
	mathescape=true,
	numbers=none,
	frame = single,
	escapeinside={<}{>},
	captionpos=b,	
	caption={\typei lineage capture for nested loop join $A \bowtie_{\theta} B$.},
	label={dl:nlj_inject},
	commentstyle=\color{red}
]
Input:  A, B
Output: O,                 
    a_fw[A.size()][], b_fw[B.size()][] // forward indexes
    a_bw[], b_bw[]            // backward indexes
oid=-1
for i = 0 to A.size()   
  for j = 0 to B.size()
    if(<$\theta($A[i], B[j]$)$>)
      O[++oid] = create_output_record(A[i], B[j])
      <\texttt{\blue{a\_bw[oid] = i}}>
      <\texttt{\blue{b\_bw[oid] = j}}>
      <\texttt{\blue{a\_fw[i].insert(oid)}}>
      <\texttt{\blue{b\_fw[j].insert(oid)}}>
\end{lstlisting}
} 

\sstitle{\typei: }~\Cref{dl:nlj_inject} illustrates the lineage capture of \sys for nested-loop $\theta$-joins. For each combination of tuples from $A$ and $B$ that satisfy the $\theta$ condition the algorithm emits the record to construct the correct output. Since we write serially the output, we can also write serially the lineage indexes and maintain the alignment between each output record and their corresponding backward lineage index. 

As an optimization, note that the backward index for the $A$ relation can be condensed. All the output records due to $A[i]$ will be consecutive in the output. Hence, instead of keeping the rids for each output \texttt{a\_fw[i].insert(oid)} we can simply store the rid of only the first one.

\subsection{Cross product}
\label{s:crossp}

Regarding cross product, \sys does not perform lineage capture in the general case. Given an input tuple from the outer relation $A$ with rid $a$ we know that its forward lineage is $\{a, a + |B|, \ldots, a + (|A|-1)|B|\}$ due to the semantics of cross product. Similar is the series for the inner relation. Hence, whether we are given an input or output tuple we can directly infer the backward and lineage rids at runtime without a cost. If the input to cross product is intermediate relations, \sys first captures lineage for operators that produce them.

\subsection{Group-By Aggregations and Joins}
\label{s:others}

Finally, we include code snippets for \typej and \typei joins and group-by aggregations that we presented in~\Cref{ss:instr}. Listings~\ref{dl:agg_typei} and~\ref{dl:agg_typej} illustrate the \typei and \typej approaches for group-by aggregation, respectively.~\Cref{dl:join} illustrates the \typei approach of \sys for joins, while~\Cref{dl:join-opt} shows the \typej approach and highlights its differences from the the \typei approach.

{\small
\begin{lstlisting}[
	language = C++,
	showspaces=false,
	basicstyle=\ttfamily\scriptsize,
	commentstyle=\color{gray},
	mathescape=true,
	numbers=none,
	frame = single,
	escapeinside={<}{>},
	captionpos=b,	
	caption={\typei lineage capture for $\gamma_{ht}$ and $\gamma_{agg}$.},
	label={dl:agg_typei},
	commentstyle=\color{red}
]
Input:  A
Output: O,                 
        fw[], bw[][]   // forward, backward index
Hash Table ht, Hash Function hash
for i = 0 to A.size()  // <\red{$\gamma_{ht}$}> Build phase
  h = hash(A[i].gbattr)
  if(!ht[h]) ht[h]={init_agg_state()<\texttt{\blue{, rids=[]}}>}
  ht[h].state.update(A[i])
  <\texttt{\blue{ht[h].rids.insert(i)}}>

fw = int[A.size()]
bw = int[ht.size()][]
oid = -1;               
for (state, rids) in ht // <\red{$\gamma_{agg}$}> Scan phase
  O[++oid] = create_output_record(state)
  <\texttt{\blue{bw[oid] = rids}}>
  <\texttt{\blue{for rid in rids}}>
    <\texttt{\blue{fw[rid] = oid}}>
\end{lstlisting}
}
{\small
\begin{lstlisting}[
	language = C++,
	showspaces=false,
	basicstyle=\ttfamily\scriptsize,
	commentstyle=\color{gray},
	mathescape=true,
	numbers=none,
	frame = single,
	escapeinside={<}{>},
	captionpos=b,
	caption={\typej lineage capture using $L_{\gamma}$.},
	label={dl:agg_typej},
	commentstyle=\color{red}
]
Input:  A
Output: O,
        fw[], bw[][]   // forward, backward index
Hash Table ht, Hash Function hash
for i = 0 to A.size()  // <\red{$\gamma_{ht}$}> Build phase
  h = hash(A[i].gbattr)
  if(!ht[h]) ht[h]={init_agg_state(), <\textcolor{blue}{oid : -1}>}
  ht[h].state.update(A[i])


fw = int[A.size()]
bw = int[ht.size()][]
oid = -1;
for h in ht // <\red{$\gamma_{agg}$}> Scan phase
  O[++oid] = create_output_record(h)
  <\textcolor{blue}{h.oid = oid}>

<\textcolor{MidnightBlue}{for i=0 to A.size()}>
  <\textcolor{MidnightBlue}{h = hash(A[i].gbattr)}>
  <\textcolor{MidnightBlue}{bw[ht[h].oid].insert(i)}>
  <\textcolor{MidnightBlue}{fw[i] = ht[h].oid}>
\end{lstlisting}
}
\vspace{-1em}
{\small
\begin{lstlisting}[
	language = C++,
	showspaces=false,
	basicstyle=\ttfamily\scriptsize,
  %columns=fullflexible,
	commentstyle=\color{gray},
	mathescape=true,
	numbers=none,
	frame = single,
  escapeinside={<}{>},
	captionpos=b,
  caption={\typei lineage capture code for $A\bowtie_{A.a=B.b}B$.},
  label={dl:join},
  commentstyle=\color{red}
]
Input:  relations A, B; 
Output: R                     // A $\red{\bowtie_{A.a=B.b}}$ B
        a_fw[][], b_fw[][]    // Forward indexes 
        a_bw[],   b_bw[]      // Backward indexes 
Hash Table ht, Hash Function hash
for i = 0 to A.size()         // Build Phase
  h = hash(A[i].a)
  if(!ht[h]) ht[h]={records=[]<\texttt{\blue{, i\_rids=[]}}>}
  ht[h].records.insert(A[i])
  <\texttt{\blue{ht[h].i\_rids.insert(i)}}>
  
o = 0;  
for i = 0 to B.size()         // Probe Phase
  h = hash(B[i].b)
  if(!(t = ht.probe(h))) continue;
  for j = 0 to t.i_rids.size()
     R[o] = (t.records[j], B[i])
     <\texttt{\blue{a\_bw[o] = t.i\_rids[j]}}>
     <\texttt{\blue{b\_bw[o] = i}}>
     <\texttt{\blue{a\_fw[j].insert(o)}}>
     <\texttt{\blue{b\_fw[i].insert(o++) }}>
\end{lstlisting}
}
{\small
\begin{lstlisting}[
	language = C++,
	showspaces=false,
	basicstyle=\ttfamily\scriptsize,
	mathescape=true,
	numbers=none,
	frame = single,
	escapeinside={<}{>},
	captionpos=b,
	caption={\typej lineage capture for $A\bowtie_{A.a=B.b}B$.},
	label={dl:join-opt},
	commentstyle=\color{red}
]
...                         // Build Phase
  if(!ht[h])ht[h]={records=[],i_rids=[]<\texttt{\blue{,o\_rids=[]}}>}
...
o = 0;  
for i = 0 to B.size()       // Probe Phase
  h = hash(B[i].b)
  if(!(t = ht.probe(h))) continue;
  <\texttt{\blue{t.o\_rids.insert(o)}}>
  for j = 0 to t.i_rids.size()
    R[o] = (t.records[j], B[i])
    <\texttt{\blue{b\_bw[o] = i}}>
    <\texttt{\blue{b\_fw[i].insert(o++)}}>

<\texttt{\blue{a\_bw = int[o]}}>   	  // Build indexes for left relation
<\texttt{\blue{for h in ht}}>
  <\texttt{\blue{s = 0}}>
  <\texttt{\blue{for r in h.i\_rids}}>
    <\texttt{\blue{a\_fw[r] = int[h.o\_rids.size()])}}>
    <\texttt{\blue{for o in h.o\_rids}}>
      <\texttt{\blue{a\_fw[r].insert(o + s)}}>
      <\texttt{\blue{a\_bw[o+s] = r}}>
    <\texttt{\blue{s++ }}>
\end{lstlisting}
}

\section{More Experiments}
\label{ss:exps}

In this section, we include experiments that did not fit in the main body of the paper due to space limitations.

\begin{figure}[b]
\centering
\includegraphics[width=0.9\columnwidth]{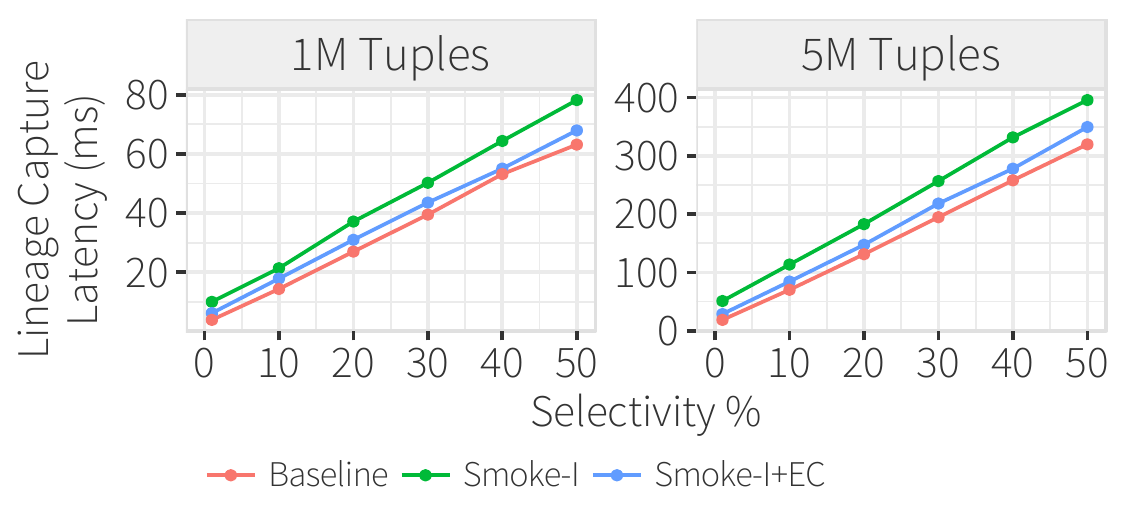}
\caption{\small Instrumented selection latency with estimated predicate selectivity (\sysec) and without (\sysi).  We find that it is better to over estimate, than underestimate and incur resizing costs.}
\label{fig:selection_overhead}
\end{figure}

\subsection{Microbenchmarks with Selection}
\label{sss:exp:micro:selection}
This experiment uses the following base query: \texttt{\lstinline|SELECT * FROM zipf WHERE v < ?|}, where the attribute \texttt{v}$\in [0,100]$ is drawn from a uniform distribution.  Varying the parameter \texttt{?} allows us to vary the query selectivity.  \Cref{fig:selection_overhead} reports the lineage capture costs for two relation sizes ($1,5$ million), and varying the estimated query selectivity between $1\%$ and $50\%$.  We evaluate \sysi, as well as \sysec, which estimates the query selectivity as $\frac{v}{100}$ and, in turn, uses the selectivity estimates to preallocate the lineage indexes.

\sstitle{Comparison of \sys techniques for selection. } \sysi introduces average overhead of 0.38$\times$ and  0.46$\times$, for one and five million records across the varying selectivities.  This is consistent with our finding that the techniques primarily vary by a constant per-tuple overhead.   When using selectivity estimates, \sysec reduces the average overhead to 0.14$\times$ and 0.15$\times$, for the respective relation sizes.  The reason that \sysec fluctuates is that the selectivity estimates may be slightly incorrect. When estimates overestimate the true selectivity, it is typically fine, however if they underestimate then they lead to array resizing overheads.

\subsection{Workload-Aware Optimizations}
\label{sss:exp:wwopts:continued}
This set of experiments evaluate the effects of instrumentation pruning and selection pushdown, which are designed to reduce lineage capture costs.

\begin{figure}[t]
\centering
\includegraphics[width=.8\columnwidth]{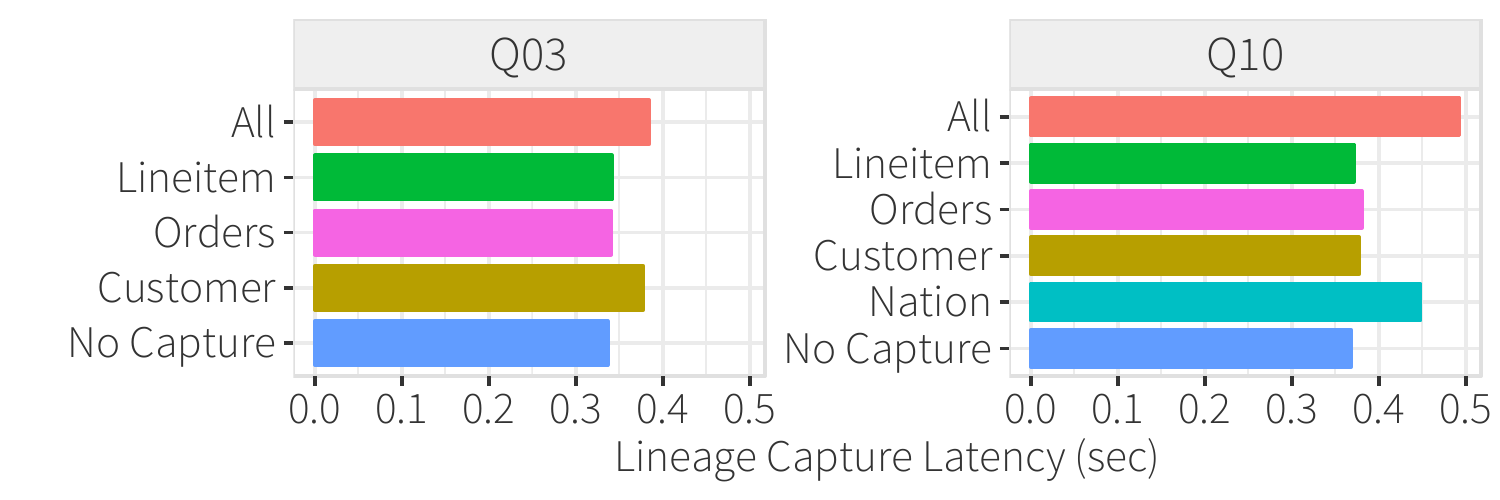}
\caption{\small Lineage capture costs for different table pruning strategies.  \texttt{ALL} refers to lineage capture for all tables.  The lineage indexes for the left-most tables dominate the overhead (\texttt{Q3.Customer}, \texttt{Q10,Nation})}
\label{fig:prune}
\vspace{-1em}
\end{figure}

\stitle{Pruning input relations.}~\Cref{fig:prune} compares the latency of Q3 and Q10, which read three and four relations, respectively, under three sets of conditions: no lineage capture, lineage capture for all input relations (non-optimized \sysi), and \sysi-based lineage capture for a single input relation.   We did not evaluate Q1, which is a single relation query, and Q12 shows the same findings. Although pruning input relations from lineage capture reduces the overall overhead, we find that the main overhead is due to the left-most tables in the join plans (\texttt{Customer} for Q3, \texttt{Nation} for Q10). It tends to be a smaller table, thus the fanout when joined with the other tables is high, and leads to more rid array reallocations.   \texttt{Lineitem} incurs the lowest overhead because it is the right-most join relation, and its join is a primary-foreign key join.   Our pk-fk join optimization uses an rid array rather than an rid index for the forward lineage index, which is much cheaper to populate.

\stitle{Selection pushdown.} To evaluate the impact of the selection pushdown optimization, we used Q1 as the base query, and ran the following lineage consuming query: \\
{\small\texttt{\lstinline|SELECT * FROM | $L_B$\lstinline|(Q1, Lineitem) WHERE l\_tax < ?|}}\\
\Cref{fig:spush} plots the average and standard deviation base query latency when assigning \texttt{?} to 5 distinct \texttt{l\_tax} values, along with the cost of \sysi without selection pushdown, and \sysn.  We find that the effectiveness of selection pushdown depends on the selectivity of the predicate.  The overhead is linear with respect to the predicate selectivity, and there is a cross-over point with \sysi at high selectivities ($>75\%$), where the overhead of evaluating the predicate for every input record outweighs the benefits of building a smaller lineage index.  We expect that increasing the predicate complexity (e.g., string comparisons, more predicate clauses) will likely shift the cross-over point towards lower selectivities.  These results suggest the value of cost-based methods to choose between the two.  

\begin{figure}[t]
\centering
\includegraphics[width=.7\columnwidth]{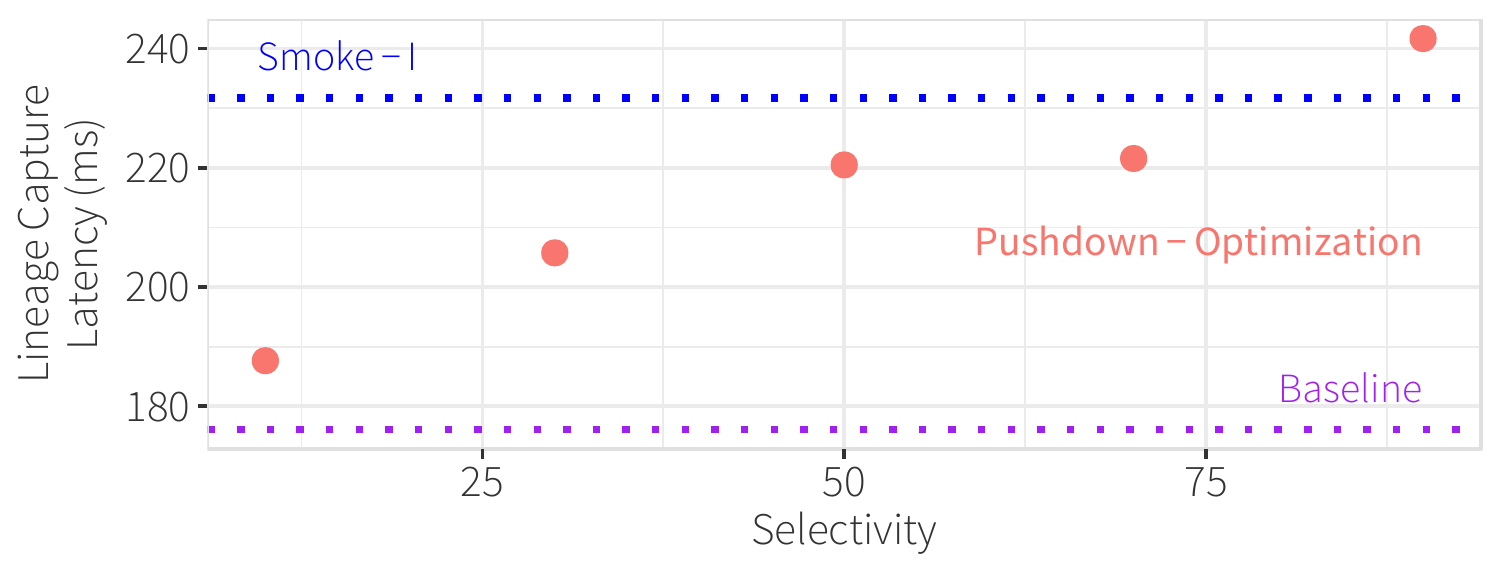}
\caption{\small Lineage capture with selection push-down at varying selectivities of \texttt{l\_tax < ?}.  The crossover point between with and without pushdown is due to the additional cost of predicate evaluation before adding rids to the lineage indexes. }
\label{fig:spush}
\vspace{-1em}
\end{figure}

\section{Related Work}
\label{s:related}

\sstitle{Lineage. } Most related to our work are lineage subsystems for databases that model, capture, index, and query lineage information. In~\Cref{s:bg}, we classified the different subsystems into logical~\cite{widom:2005:trio,dbnotes2004bhagwat,glavic:2009:perm,gprom,geerts2006mondrian,cui2001lineage,green:2007:orchestra} and physical~\cite{logothetis2013scalable,interlandi2015sparkprovenance,wu2013subzero,ramp,panda,ikedathesis}. , explained their differences, and discussed how \sys avoids their drawbacks. Our experiments show how \sys can both provide negligible lineage capture overhead, fast lineage query execution, and outperform state-of-the-art alternatives by multiple orders of magnitude.

\sstitle{Physical Database Design. } The physical database design literature has long studied techniques to create redundant data structures (e.g., indexes and materialized views) and data layouts to minimize the expected execution cost of a possible future query workload~\cite{chaudhuri:2007:physicaldbdesign,agrawal:2000:automatedselectionmatindexes,dias2005automatic,oracle10g:2003,cliffguard:mozafari:2015,kersten:2005:cracking,idreos:2007:database,amoeba:shanbhag:2016,adaptdb:lu:2017,h20:alagiannis:2014:hha,petraki:2015:holisticindexingcracking,ottertune:2017:vanaken}. \sys is the first database engine to consider lineage as a type of information for physical design decisions (i.e., we showed how we can build online lineage indexes and push logic of lineage consuming queries into lineage capture to answer future queries equivalent to SQL queries). Also, \sys does not simply push the physical database design costs into query execution; we both propose write-efficient data structures to minimize construction overheads and {\it overlap lineage index construction costs with query execution logic}.

\sstitle{Lineage Applications.} A core motivation behind \sys is to demonstrate how applications with hand-tuned implementations can leverage lineage systems to express their logic declaratively and enjoy out-of-the(-lineage)-box optimizations. Interactive data visualizations have long materialized specialized data structures---such as data cubes~\cite{liu2013immens,wang2017gaussian,battle2016,kandel2012profiler}, indexes~\cite{pahins2017hashedcubes,lins2013nanocubes}, or precomputed results~\cite{chang2007wirevis}---offline in order to insure sub-150ms response times. In our experiments with crossfiltering we showed how our push-down optimizations can be used to construct such cubes. Moreover, we showed lineage-enabled techniques and workload-aware optimizations that can adequately address the cold-start problem of interactive visualizations~\cite{position:2017:leilani} and the ``Overview first, zoom and filter, and details on demand'' interaction paradigm (by either generating indexes or materializing and partitioning results). Finally, we showed how lineage can express data profiling primitives, of core use across domains~\cite{uguide:2017:thirumuruganathan,holoclean:2017:rekatsinas,profiling:2014:naumann}, and outperform hand-written implementations. These results provide evidence that \emph{\sys does not just provide declarative features for applications to express their logic but it can actually optimize applications holistically}.

\fi

\end{document}